\pdfoutput=1
\documentclass[
  aps,
  prx,
  reprint,
  amsmath,
  superscriptaddress,
  amssymb
]{revtex4-2}

\usepackage[T1]{fontenc}
\usepackage{newtxtext,newtxmath} 

\usepackage{xcolor}
\usepackage{graphicx}
\usepackage{booktabs}
\usepackage{multirow}
\usepackage{tikz}
\usepackage{array}

\usepackage{mathrsfs}
\usepackage{bm}
\usepackage{braket}
\usepackage{wasysym}

\usepackage[nolist,nohyperlinks]{acronym} 
\usepackage{enumitem}
\usepackage{chemformula} 
\usepackage{tcolorbox}
\usepackage{hyperref}
\usepackage[capitalise]{cleveref}

\makeatletter
\g@addto@macro\bfseries{\boldmath}
\makeatother
\makeatletter
\AtBeginDocument{%
  \def\ltx@label#1{\cref@label{#1}}%
  \def\label@in@display@noarg#1{%
    \cref@old@label@in@display{#1}%
  }%
  \def\label@in@mmeasure@noarg#1{%
    \begingroup
      \measuring@false
      \cref@old@label@in@display{#1}%
    \endgroup
  }%
} 
\makeatother
\AddToHook{cmd/appendix/before}{%
  \crefalias{section}{appendix}%
}
\crefname{appendix}{Appendix}{Appendices}

\definecolor{cblue}{RGB}{0,0,255}
\definecolor{cgray}{RGB}{77,77,77}
\hypersetup{
  colorlinks=true,
  linkcolor=cblue,
  citecolor=cblue,
  urlcolor=cblue
}

\newenvironment{cframed}
  {%
    \begin{tcolorbox}[
      colframe=cgray!75!white,
      colback=cgray!3!white,
      boxrule=0.75pt
    ]%
  }
  {%
    \end{tcolorbox}%
    \noindent
  }
\newenvironment{subalign}
  {\subequations\align}
  {\endalign\endsubequations}

\renewcommand{\vec}[1]{\bm{#1}} %
\newcommand{\vhat}[1]{\hat{\vec{#1}}} %
\newcommand{\mat}[1]{\vec{#1}} %
\newcommand{\trp}[1]{{#1}^{\intercal}} %

\newcommand{\del}{\partial}
\newcommand{\h}[1]{{#1}^{\dagger}}

\DeclareMathOperator{\tr}{tr}

\newcommand{\ot}{{\rm O}(3)}

\newcommand{\sot}{{\rm SO}(3)}

\newcommand{\st}{s}

\newcommand{\yes}{\ensuremath{\checkmark}}
\newcommand{\no}{\ensuremath{\times}}
\newcommand{\maybe}{\ensuremath{\ast}}
\newcommand{\sg}[2]{\ensuremath{#1}~(\##2)}
\newcommand{\SSG}{{\rm SSG}}
\DeclareRobustCommand{\CoTaS}{\ch{CoTa3S6}}
\makeatletter
\newcommand{\ssop}{\@ifstar\ssop@big\ssop@small}
\newcommand{\ssop@small}[2]{[#1\Vert #2]}
\newcommand{\ssop@big}[2]{\left[#1\,\middle\Vert\,#2\right]}
\makeatother
\newcommand{\dG}[1]{{}^{#1}\Gamma}

\begin{document}

\title{Spin Space Symmetry Without Spin Space Groups}

\author{Jeffrey G. Rau}
\email{jrau@uwindsor.ca}
\affiliation{Department of Physics, University of Windsor, 401 Sunset Avenue, Windsor, Ontario, N9B 3P4, Canada}

\date{\today}

\begin{abstract}
Spin space groups provide a complete symmetry classification of magnetic structures in the non-relativistic limit and have become an increasingly useful language for organizing magnetic phases and their physical properties. Each spin space group with a given parent space group is specified, up to spin-only symmetries, by a real representation of that space group acting in spin space. We argue that working directly with this representation not only avoids the construction and identification of the group, but also organizes and simplifies the resulting symmetry constraints on allowed responses. Determining the presence or absence of magnetic compensation, piezomagnetism, spin and charge transport, the Edelstein effect, and momentum-space spin textures becomes an ordinary problem in the representation theory of space groups. The resulting conditions reduce to overlaps between the three exterior powers of the defining representation and multipole representations of the point group. Several physically distinct responses then reduce to the same condition on the defining representation. For example, a $p$-wave spin texture and the time-even Edelstein effect are allowed under the same condition, as are piezomagnetism and a $d$-wave spin texture in compensated order. The wave-vector content of the exterior powers gives a further hierarchy of selection rules: depending on the relevant power, a uniform response requires one, two, or three ordering wave vectors to sum to a reciprocal lattice vector. These rules can exclude many responses before point-group symmetry is considered. Among the general consequences, a non-relativistic anomalous Hall effect in compensated order at zero wave vector requires a defining representation built from three one-dimensional irreducible representations, a purely odd-parity spin texture requires a pair of nonzero ordering wave vectors that cancel, and coplanar order at nonzero wave vector in the interior of the Brillouin zone is always a $p$-wave magnet. We illustrate the approach with examples drawn from a wide variety of materials hosting collinear, coplanar and non-coplanar magnetic order. Finally, we discuss extensions of these results to magnetic space groups, connections to Landau theories of collinear order, and the characterization of odd-parity magnets.
\end{abstract}

\maketitle

\section{Introduction}

Symmetry classification is one of the basic organizing principles of
condensed-matter physics. For crystalline systems, space groups provide a finite
atlas of possible spatial symmetries~\cite{ITAnorm,bradley1972mathematical};
once a material has been assigned to a particular group, the corresponding
symmetry constraints follow independent of the microscopic details. In the
presence of magnetic ordering, this group structure can be generalized to
include symmetries involving not only spatial operations but also spatial
operations combined with operations that act in spin
space~\cite{litvin2013magnetic}.

How spatial and spin operations combine depends on the strength of relativistic
effects. The magnetic space
groups~\cite{litvin2013magnetic,bradley1972mathematical} assume spin and spatial
rotations are locked together, as required when spin-orbit coupling or dipolar
interactions are significant. In the opposite, non-relativistic limit, spin
rotations decouple from spatial rotations entirely, and the larger classes of
spin space groups~\cite{brinkman1966theory,litvin1974spin} emerge. Recent work
has revived focus on this limit, partly motivated by the physics of
altermagnetism~\cite{smejkal2022beyond,smejkal2022emerging,jungwirth2026altermagnetic},
culminating in a complete classification and enumeration of the possible spin
space groups~\cite{xiao2024,chen2024enumeration,jiang2024enumeration}.

There is, however, a distinction between two core problems. First is the
\emph{classification
problem}~\cite{xiao2024,chen2024enumeration,jiang2024enumeration}: determine all
possible symmetry groups, identify when two groups are equivalent, and choose a
canonical representative and label for each equivalence class. Second is the
\emph{inference problem}: given the symmetry of a particular ordered state,
determine which observables, couplings, degeneracies, and response functions are
allowed or forbidden. A complete classification provides one route to the second
problem: once a state has been assigned its group, the allowed form of each
quantity follows by imposing invariance under each of its elements in turn, as
in Neumann's principle. For response tensors this inference has been developed
systematically using spin-group transformation
laws~\cite{watanabe2024symmetry,etxebarria2025crystal} and implemented in
automated software tools~\cite{elcoro2026automatic}. The question addressed here is how
to organize these constraints directly through the representation of the space
group, acting in spin space, that specifies the spin space group, making their
common spin-space structure and dependence on translations explicit.

For spin space groups this observation can be made particularly transparent. In
the non-relativistic limit spin rotations and time reversal act independently of
the lattice, and for real quantities they combine into $\mathrm{O}(3)$, so the
paramagnetic phase of a crystal with space group $G_0$ has symmetry
$\mathrm{O}(3)\times G_0$. A magnetically ordered state has a spin space group
$\SSG{} \leq \mathrm{O}(3)\times G$, where the parent space group $G \leq G_0$
may be smaller than $G_0$. For non-coplanar order its
elements are spin-space operations $\ssop{\mat{\Gamma}(g)}{g}$, one for each
$g\in G$, and the group structure makes the matrices $\mat{\Gamma}(g)$ a real
representation, $\Gamma$, of $G$, which we call the defining representation
(\cref{sec:definitions}). The classification of spin space groups can be
formulated as the problem of enumerating all such inequivalent representations
for all parent space groups~\cite{xiao2024,jiang2024enumeration}. Up to an
overall choice of spin frame, this representation contains precisely the
information required to determine how spin-dependent quantities transform.
Collinear, coplanar, and non-coplanar states are described naturally by real
representations of dimension one, two, and three~\cite{xiao2024}, respectively,
with collinear and coplanar orderings enjoying additional spin-only symmetries. 
 
One route to obtaining symmetry constraints on physical quantities may therefore
be sketched as
\begin{equation}
\label{eq:schematic}
\begin{array}{c}
\text{Parent Space Group}\ (G) + \text{Representation}\ (\Gamma),
\\[0.5em]
\downarrow
\\[0.5em]
\text{Spin Space Group}\ (\SSG \leq \ot{} \times G),
\\[0.5em]
\downarrow
\\[0.5em]
\text{Physical Consequences}\ (\vec{M}, \vec{C}_{\mu\nu}, \sigma^-, \ldots),
\end{array}
\end{equation}
where the first step constructs the spin space group, which may then be
identified within a classification, and the second applies its elements to each
quantity of interest. The central point of this work is that the second step can
be framed directly in terms of $\Gamma$. Doing so avoids the construction and
identification of the group, and has the further advantage that it organizes the
resulting constraints: the transformation laws of many observables are fixed by
a few objects built from $\Gamma$, so that the constraints on different
observables are related, and their origin in the ordered state is visible. 

The relevant objects are the exterior powers $\Lambda^1\Gamma$,
$\Lambda^2\Gamma$ and $\Lambda^3\Gamma$ of the defining representation. For the
quantities in \cref{tab:summary}, the spin transformation law is fixed by the
presence or absence of a spin index and by time-reversal parity. A time-odd spin
vector transforms under $\Lambda^1\Gamma$, a time-even spin vector under
$\Lambda^2\Gamma$, and a time-odd spin scalar under $\Lambda^3\Gamma$, with the
constraints of the spin-only symmetries included for collinear and coplanar
order. The spatial tensor character then determines which multipole
representation the selected power must overlap. Consequently, observables
selecting the same power share constraints, and a power that vanishes, as
$\Lambda^2\Gamma$ and $\Lambda^3\Gamma$ do for collinear order, removes all of
them at once. 

\begin{figure}[t]
\includegraphics[width=\columnwidth]{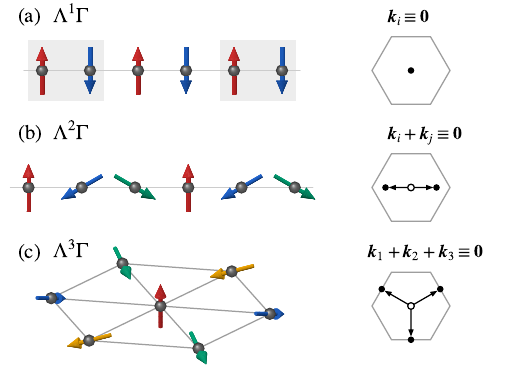}
\caption{Orders whose wave vectors give $\Lambda^p\Gamma$ a uniform component,
\cref{eq:wavevector-rules}, each beside its wave vectors in the Brillouin zone.
(a)~A two-sublattice order at $\vec{k}=\vec{0}$, the same in every unit cell
(shaded). (b)~A $120^\circ$ spiral, with wave vectors $\pm\vec{k}$. (c)~A
tetrahedral triple-$Q$ order, with wave vectors at the three $M$ points.}
\label{fig:hierarchy}
\end{figure}

An important consequence is a hierarchy of selection rules arising from
translational symmetry, independent of the detailed point-group representation
content. For a three-dimensional representation, a spatially uniform response
controlled by $\Lambda^1\Gamma$ requires an ordering wave vector at zero, one
controlled by $\Lambda^2\Gamma$ requires two wave vectors that sum to zero, and
one controlled by $\Lambda^3\Gamma$ requires all three to sum to zero, in each
case modulo a reciprocal lattice vector (\cref{fig:hierarchy}). The symmetry
analysis therefore proceeds in two steps: first identifying whether the relevant
exterior power contains a translation-invariant sector, and then determining
whether that sector overlaps the spatial representation of the observable. In
some cases the first step alone is enough to rule out a response. For example, a
non-coplanar triple-$Q$ state with three nonzero ordering wave vectors whose sum
vanishes can support an anomalous Hall response even when the spin
conductivities and momentum-space spin texture are forbidden. We demonstrate
this separation explicitly for the tetrahedral order in \CoTaS{} in
\cref{sec:co13tas2}.

Applied to the net spin magnetization, piezomagnetism~\cite{dzyaloshinskii1958piezomagnetism,%
moriya1959piezomagnetism,borovik1960piezomagnetism}, spin and charge
transport~\cite{naka2019spin,gonzalez2021efficient}, momentum-space spin
textures~\cite{liu2022spin,smejkal2022beyond,hellenes2023p,song2026}, and the
Edelstein effect~\cite{edelstein1990spin}, the exterior powers give the
conditions summarized in \cref{tab:summary}. Several responses share identical
symmetry criteria: for compensated order, piezomagnetism and a $d$-wave spin
texture are allowed under the same conditions, as are a $p$-wave spin texture
and the time-even Edelstein effect. Relations of this kind are not as apparent
when each response is constrained separately by the elements of the group.

\begin{table}[t]
\centering
\footnotesize
\setlength{\tabcolsep}{3pt}
\renewcommand{\arraystretch}{1.25}
\begin{tabular}{@{}l l c l l@{}}
\toprule
\multicolumn{2}{@{}l}{Response} & Symbol & Required overlap & Section \\
\midrule
\multicolumn{2}{@{}l}{Net spin magnetization} & $\vec{M}$
  & $\Gamma \cap \Gamma_0^+$ & \ref{sec:magnetization} \\
\multicolumn{2}{@{}l}{Piezomagnetism} & $\vec{C}_{\mu\nu}$
  & $\Gamma \cap \Gamma_2^+$ & \ref{sec:piezomagnetism} \\
\multirow{2}{*}{Spin conductivity}
  & Time-odd & $\Sigma^-$
  & $\Gamma \cap (\Gamma_0^+ \oplus \Gamma_1^+ \oplus \Gamma_2^+)$
  & \ref{sec:spin-conductivity} \\
  & Time-even & $\Sigma^+$
  & $\Lambda^2\Gamma \cap (\Gamma_0^+ \oplus \Gamma_1^+ \oplus \Gamma_2^+)$
  & \ref{sec:spin-conductivity} \\
\multicolumn{2}{@{}l}{Anomalous Hall conductivity} & $\sigma^-$
  & $\Lambda^3\Gamma \cap \Gamma_1^+$ & \ref{sec:hall} \\
\multirow{2}{*}{Edelstein effect}
  & Time-odd & $\alpha^-$ & $\Gamma \cap \Gamma_1^-$
  & \ref{sec:edelstein} \\
  & Time-even & $\alpha^+$ & $\Lambda^2\Gamma \cap \Gamma_1^-$
  & \ref{sec:edelstein} \\
\midrule
\multirow{2}{*}{Spin texture}
  & Even rank & $\vec{\st}(\vec{q})$ & $\Gamma \cap \Gamma_\ell^+$
  & \ref{sec:spintexture} \\
  & Odd rank & $\vec{\st}(\vec{q})$ & $\Lambda^2\Gamma \cap \Gamma_\ell^-$
  & \ref{sec:spintexture} \\
\bottomrule
\end{tabular}
\caption{Summary of the symmetry conditions for the non-relativistic responses
derived in this work. Each is expressed as a nonzero overlap of an exterior
power of the defining representation $\Gamma$ with a multipole representation
$\Gamma_\ell^\pm$ of the parent space group $G$. The section in which each
condition is derived is indicated.
For collinear order $\Lambda^2\Gamma = \Lambda^3\Gamma = 0$, and for coplanar
order $\Lambda^3\Gamma = 0$, so the responses selecting these powers vanish. The
piezomagnetic condition assumes compensated order, and a rank-$\ell$
contribution to the spin texture requires the overlap shown for $\ell$ of the
stated parity.}
\label{tab:summary}
\end{table}

We stress that our goal is not to replace spin space groups, nor to argue that
their classification is unnecessary. A group classification remains essential
when one wishes to enumerate possible symmetry types, compare structures across
different crystallographic settings, construct
databases~\cite{gallego2016magndata,aroyo2006bilbao,ssgdatabase,yu2026findspingroup},
or assign canonical symmetry labels~\cite{perezmato2015symmetry}. The full spin
space group is also the natural object for questions about individual states,
such as band degeneracies, band connectivity and symmetry-diagnosed topology,
rather than the allowed form of a response (\cref{sec:scope}). Imposed element
by element, the spin space group gives the same allowed forms. Organizing these
constraints through the defining representation makes their common origin
explicit.

Beyond the general methods noted above, several individual results have
already been reported. \Citet{watanabe2024symmetry} analyzed the spin-group conditions
for the non-relativistic Hall response, and noted that in \CoTaS{} it occurs
without non-relativistic spin--momentum locking. The symmetry equivalence of
piezomagnetism and a $d$-wave spin texture is known for collinear
order~\cite{mcclarty2024landau} and was discussed for magnetic point
groups by \citet{radaelli2024tensorial}, and the spin-texture parity relations
appear in the classification of \citet{song2026}. Criteria for odd-parity
magnets have been given in terms of spin-group
operations~\cite{hellenes2023p,luo2025odd,luo2026spin,bai2026time}, for coplanar
order at a time-reversal invariant momentum (TRIM)~\cite{yu2025odd}, and for
coplanar spirals at any wave vector~\cite{larsen2026odd}.

Working with the exterior powers, these and more general connections can be exposed
systematically. We use the exterior powers to derive a wave-vector
selection hierarchy of \cref{sec:exterior-prelim}, which can settle a response
before any point-group character is computed. The same construction restricts
the non-relativistic anomalous Hall response to non-cubic parent point groups
and, for compensated order at $\vec{k}=0$, to defining representations built
from three real one-dimensional irreps (\cref{sec:hall}). We also
characterize odd-parity magnets systematically, fixing their wave-vector
content, inversion parity and wave character directly from $\Gamma$
(\cref{sec:oddparity}). For example, an interior non-TRIM pair
$\{\vec{k},-\vec{k}\}$ in coplanar order always gives a $p$-wave texture and a
time-even Edelstein response. Combining the
representation conditions with Onsager reciprocity, we further show that any
non-relativistic time-odd spin conductivity in the collinear orders of
$\alpha$-\ch{Fe2O3} or \ch{CuFeS2} is antisymmetric in its spatial indices and
requires interactions coupling the spin channels (\cref{sec:spin-conductivity}).

The outline of the paper is as follows. We first review in
\cref{sec:definitions} how collinear, coplanar, and non-coplanar spin-space
symmetries are encoded by real representations of the parent space group.
\Cref{sec:exterior-prelim} then collects the properties of the exterior powers
of that representation, and of their wave-vector content, that are used
throughout, and \cref{sec:observables} derives from them the conditions of
\cref{tab:summary}. We next apply these in \cref{sec:examples} to eight
materials spanning collinear, coplanar and non-coplanar order at both zero and
finite ordering wave vector, including two, \ch{MnO} and
\ch{Gd3(Ru_{0.95}Rh_{0.05})4Al12}, in which the order selects part of a wave-vector star, so that its parent space
group is smaller than $G_0$, and several in which the dimension
and wave-vector content of $\Gamma$ alone settle most or all of the responses. The
same analysis of the commensurate magnetic structures in the MAGNDATA database,
using code published with this paper~\cite{zenodo-code}, is summarized in
\cref{sec:magndata} and tabulated in the Supplemental
Material~\cite{supplemental}. We close with
a discussion of how the same construction relates to
conventional magnetic space groups, each of which is likewise encoded in a real
one-dimensional representation of its parent space group (\cref{sec:msg}), to
the Landau theory of collinear magnetic transitions (\cref{sec:landau}), to
odd-parity magnets (\cref{sec:oddparity}), and to existing
spin space group classifications of the momentum-space spin textures
(\cref{sec:textureclass}).

\section{Definitions}
\label{sec:definitions}

Consider the paramagnetic phase of a crystal, with space group $G_0$ and full spin rotation symmetry,
$\mathbb{Z}^T_2 \times \sot{} \times G_0$, where $\mathbb{Z}^T_2$ is time
reversal. Since we work exclusively with real quantities, time reversal and
spin rotations may be combined as
\begin{equation}
\label{eq:z2-so3-to-o3}
\mathbb{Z}^T_2 \times \sot{} \rightarrow \ot{},
\end{equation}
interpreting any $\mat{U} \in \ot{}$ with $\det \mat{U} = -1$ as a spin rotation
followed by time reversal, which for real quantities is a spin inversion. The
paramagnetic symmetry group is then $\ot{} \times G_0$ where $G_0$ is the space group of the paramagnetic phase.\footnote{In some contexts, including the magCIF format and MAGNDATA, the parent space group refers to $G_0$. Here we reserve the term for the group $G$ introduced below, formed by the spatial operations of the spin space group.}

At a magnetic phase transition this is broken to a spin space group $\SSG{} \leq
\ot{} \times G$ where $G \leq G_0$ is the parent space group of the spin space group. This parent space group may be smaller than $G_0$: it contains those spatial operations that preserve the magnetic
ordering pattern up to a global transformation in spin space. An explicit method for construction of $G$
and $\Gamma$ for a commensurate magnetic structure is given in
\cref{app:constructing-gamma}. The elements of
the spin space group may be written as the spin-space operations
\begin{equation}
\label{eq:spin-space-element}
\ssop{\mat{\Gamma}(g)}{g},
\end{equation}
with $g \in G$ and $\mat{\Gamma}(g) \in \ot{}$. For
non-coplanar order, the group structure of $\SSG{}$ requires
$\mat{\Gamma}(g_1)\mat{\Gamma}(g_2) = \mat{\Gamma}(g_1g_2)$ for all $g_1$, $g_2
\in G$, so the $\mat{\Gamma}$ form a
real three-dimensional representation of the parent space group
$G$~\cite{brinkman1966theory,litvin1974spin,xiao2024}, which need not be
irreducible. A choice of basis for this representation fixes a (right-handed)
frame in spin space: a global spin rotation $\tilde{\mat{\Gamma}}(g) \equiv
\mat{R} \mat{\Gamma}(g) \mat{R}^{-1}$, with $\mat{R} \in \sot{}$ arbitrary,
yields an equivalent subgroup. We treat spin space groups related by such a
rotation as equivalent. If the order is collinear or coplanar, the group also
contains the elements of the spin-only group.

The representation $\Gamma$ is a representation of the full space group. Its
action on lattice translations encodes the wave-vector content of the magnetic
order. In particular, the relevant wave vectors can be organized into stars
under the point-group action: if $\vec{k}$ occurs in the magnetic structure,
then symmetry relates it to the other wave vectors in the star of $\vec{k}$,
modulo reciprocal lattice vectors. The representation $\Gamma$ therefore
reflects not only the propagation wave vector itself, but also how the different
arms of its star are related by the spatial symmetries (not just a little group
representation). 

For collinear and coplanar orders the magnetic structure spans only a one- or
two-dimensional subspace of spin space, and after factoring out the additional
spin-only symmetries the corresponding defining representation is one-
or two-dimensional~\cite{xiao2024}. The remaining spin directions are acted on
by additional spin-only symmetries. For collinear order the representation is a
real one-dimensional irrep, $\Gamma(g) = \pm 1$. Denoting the collinear axis by
$\vhat{N}$, convenient three-dimensional
representatives of the full spin-space
operations can be chosen to be of the form
\begin{equation}
\label{eq:collinear-gamma}
\ssop{\Gamma(g)\mat{1}}{g},
\end{equation}
where $\Gamma(g) = +1$ has no spin operation and $\Gamma(g) = -1$ indicates
time reversal. The group is extended to include all rotations
$\mat{R}_{\vhat{N}}(\theta)$ about the collinear direction, together with the
two-fold rotations perpendicular to $\vhat{N}$ followed by time
reversal.~\footnote{These three-dimensional representatives are not unique: taking
$\vhat{N}$ along $\vhat{z}$ and combining an element $g \in G$ with elements of
the spin-only group, we can
change the action on the $\vhat{x}$ and $\vhat{y}$ spin directions. For example,
if we combine each element with $\Gamma(g) = -1$ with a two-fold rotation about
$\vhat{z}$ then $\mat{\Gamma}(g) = {\rm diag}(+1,+1,\Gamma(g))$.}
 
For coplanar order the representation is a real two-dimensional representation
$\Gamma$. Taking the preferred plane, spanned by $\vhat{N}_1$ and $\vhat{N}_2$,
perpendicular to $\vhat{z}$, convenient three-dimensional representatives of the full spin-space operations can be
written
\begin{equation}
\label{eq:coplanar-gamma}
\ssop*{
\begin{pmatrix}
\mat{\Gamma}(g) & 0 \\
0 & \det \mat{\Gamma}(g)
\end{pmatrix}
}{g}.
\end{equation}
The group is again extended to include spin-only operations, here by a two-fold
rotation $\mat{R}_{\vhat{z}}(\pi)$ combined with time reversal~\footnote{Again
the three-dimensional representative is not unique: as in the collinear case,
acting with $\mat{R}_{\vhat{z}}(\pi) T = {\rm diag}(+1,+1,-1)$ maps a chosen
three-dimensional representative of determinant $+1$ to one of determinant $-1$,
leaving the action on $\vhat{z}$ arbitrary.}.

Groups obtained this way are not necessarily distinct. To carry out a
classification we would further need to remove groups that are equivalent under
changes of axes or origin, so-called affine equivalences~\cite{ITAnorm}. This
is the program carried out in \citet{xiao2024} and in the companion enumerations
of \citet{chen2024enumeration} and \citet{jiang2024enumeration} using different
techniques.\footnote{The classification of \citet{xiao2024} additionally
identifies representations induced from wave vectors with the same little group
when they can be continuously deformed into one another as the wave vectors vary
within the corresponding fixed-point manifold.}

Note that magnetic orders whose spin space group has a parent space group identical to
the paramagnetic space group ($G=G_0$) are called \emph{regular magnetic
orders}~\cite{messio2011}. These
orders have the property that they are stationary points of Heisenberg spin
Hamiltonians (absent free continuous parameters in the ansatz).
\Citet{messio2011} used an algebraic approach to classify all spin wallpaper
groups with parent groups of $p6m$ or $p4m$ symmetry, and constructed examples
of the associated regular magnetic orders.

\section{Exterior Powers}
\label{sec:exterior-prelim}
Before turning to specific observables it will be helpful to discuss some
properties of the defining representation that are used repeatedly below. Let
$V$ denote the spin space on which the real representation $\Gamma$ acts. The natural representations on antisymmetric products of spin
vectors are the exterior powers $\Lambda^p\Gamma$ acting on $\Lambda^p
V$~\cite{fulton1991representation,bradley1972mathematical}. Here $\Lambda^p V$
is the space of totally antisymmetric $p$-fold products, spanned by the wedge
products $\vec{v}_1\wedge\cdots\wedge\vec{v}_p$, which change sign under the
exchange of any two factors; it has dimension $\binom{n}{p}$ for $\dim V = n$
and vanishes for $p>n$. Acting with $\mat{\Gamma}(g)$ on each factor preserves
the antisymmetry and so defines a representation,
\begin{equation}
\label{eq:exterior-action}
\Lambda^p\mat{\Gamma}(g)\,
{\left(\vec{v}_1 \wedge \cdots \wedge \vec{v}_p\right)}
\equiv
{\left(\mat{\Gamma}(g)\vec{v}_1\right)} \wedge \cdots \wedge
{\left(\mat{\Gamma}(g)\vec{v}_p\right)}.
\end{equation}
For non-coplanar order, where $\Gamma$ is three-dimensional, these satisfy
\begin{subequations}
\label{eq:exterior-powers}
\begin{align}
\Lambda^1\Gamma &\equiv \Gamma, \\
\Lambda^2\Gamma &\equiv \Gamma\otimes\det\Gamma, \\
\Lambda^3\Gamma &\equiv \det\Gamma.
\end{align}
\end{subequations}
The third holds because $\mat{\Gamma}(g)$ rescales the oriented volume element
spanning the one-dimensional $\Lambda^3 V$ by its determinant. The second
isomorphism follows from $\Lambda^2 V \equiv V^*\otimes\Lambda^3 V$ together
with the identification $V^*\equiv V$; this is a generalization of the familiar
identification of planes in three dimensions with axial vectors.

Geometrically, these correspond to
\begin{equation*}
\label{eq:exterior-power-hierarchy}
\begin{array}{ccl}
\Lambda^1\Gamma &:& \text{spin vector}, \\
\Lambda^2\Gamma &:& \text{spin pseudovector}, \\
\Lambda^3\Gamma &:& \text{spin pseudoscalar}.
\end{array}
\end{equation*}
The second can be seen from the cross product. For any $\mat{\Gamma}(g) \in
\ot{}$,
\begin{equation}
\label{eq:cross-product-exterior}
{\left(\mat{\Gamma}(g)\vec{S}_1\right)} \times
{\left(\mat{\Gamma}(g)\vec{S}_2\right)}
=
\det \mat{\Gamma}(g) \mat{\Gamma}(g)
{\left(\vec{S}_1\times\vec{S}_2\right)},
\end{equation}
so an antisymmetric spin bilinear, such as the vector spin chirality
$\vec{S}_1\times\vec{S}_2$ of two moments, transforms under $\Lambda^2\Gamma$
and, unlike a spin vector, is even under spin inversion. The third is the
oriented volume in spin space, realized most simply by the scalar spin chirality
\begin{equation}
\label{eq:scalar-chirality}
\chi_{123}
=
\vec{S}_1 \cdot {\left(\vec{S}_2\times\vec{S}_3\right)},
\end{equation}
which transforms as
\begin{equation}
\label{eq:chirality-det-gamma}
\chi_{123}
\rightarrow
\det\mat{\Gamma}(g) \chi_{123}.
\end{equation}
Both reappear in \cref{sec:transport}, in the time-even spin conductivity
and the anomalous Hall conductivity.

For a two-dimensional representation $\Gamma$ the exterior powers are
\begin{subalign}
\label{eq:exterior-powers-2d}
\Lambda^1\Gamma &\equiv \Gamma, \\
\Lambda^2\Gamma &\equiv \det\Gamma,
\end{subalign}
where $\Lambda^3\Gamma = 0$. Here $\Lambda^2\Gamma=\det\Gamma$ is the oriented
area of the spin plane or, equivalently, the representation carried by its
normal $\vhat{N}_1\times\vhat{N}_2$. For a one-dimensional representation $\Gamma$ the
higher exterior powers vanish, with 
\begin{equation}
\Lambda^1\Gamma \equiv \Gamma,
\end{equation}
and $\Lambda^2 \Gamma = \Lambda^3 \Gamma = 0$. For an irreducible $\Gamma$ of
dimension two or three, $\det\Gamma$ can be read from \cref{tab:det-reps}. For
a reducible $\Gamma$ the exterior powers follow from those of its irreducible
summands, as given in \cref{app:exterior-sum}.

\begin{table}[t]
\centering
\footnotesize
\setlength{\tabcolsep}{5pt}
\renewcommand{\arraystretch}{1.25}
\begin{tabular}{l >{$}l<{$} @{\hspace{1.5em}} l >{$}l<{$}}
\toprule
$P$ & \Gamma \rightarrow \det\Gamma & $P$ & \Gamma \rightarrow \det\Gamma \\
\midrule
$C_3$    & E \rightarrow A                  & $C_6$    & E_1,E_2 \rightarrow A \\
$S_6$    & E_g,E_u \rightarrow A_g          & $C_{6h}$ & E_{1g},E_{2g},E_{1u},E_{2u} \rightarrow A_g \\
$D_3$    & E \rightarrow A_2                & $D_6$    & E_1,E_2 \rightarrow A_2 \\
$C_{3v}$ & E \rightarrow A_2                & $C_{6v}$ & E_1,E_2 \rightarrow A_2 \\
$D_{3d}$ & E_g,E_u \rightarrow A_{2g}       & $D_{6h}$ & E_{1g},E_{2g},E_{1u},E_{2u} \rightarrow A_{2g} \\
$C_{3h}$ & E',E'' \rightarrow A'            &          & \\
$D_{3h}$ & E',E'' \rightarrow A_2'          & $T$      & E,T \rightarrow A \\
         &                                  & \multirow[t]{2}{*}{$T_h$} & E_g,E_u,T_g \rightarrow A_g \\
$C_4$    & E \rightarrow A                  &          & T_u \rightarrow A_u \\
$S_4$    & E \rightarrow A                  & \multirow[t]{2}{*}{$O$}   & E,T_2 \rightarrow A_2 \\
$C_{4h}$ & E_g,E_u \rightarrow A_g          &          & T_1 \rightarrow A_1 \\
$D_4$    & E \rightarrow A_2                & \multirow[t]{2}{*}{$T_d$} & E,T_2 \rightarrow A_2 \\
$C_{4v}$ & E \rightarrow A_2                &          & T_1 \rightarrow A_1 \\
$D_{2d}$ & E \rightarrow A_2                & \multirow[t]{4}{*}{$O_h$} & E_g,E_u,T_{2g} \rightarrow A_{2g} \\
$D_{4h}$ & E_g,E_u \rightarrow A_{2g}       &          & T_{1g} \rightarrow A_{1g} \\
         &                                  &          & T_{1u} \rightarrow A_{1u} \\
         &                                  &          & T_{2u} \rightarrow A_{2u} \\
\bottomrule
\end{tabular}
\caption{Determinant representations for real irreducible representations of
dimension two or three in the crystallographic point groups $P$. Point groups
with only one-dimensional real irreducible representations are omitted.}
\label{tab:det-reps}
\end{table}

We can make general statements about the wave-vector content of these exterior powers.
Restricted to the translations, a real $n$-dimensional representation $\Gamma$
becomes, over $\mathbb{C}$, a sum of characters $[\mat{1}|\vec{t}] \mapsto
e^{i\vec{k}_i\cdot\vec{t}}$, and we call $\{\vec{k}_1,\dots,\vec{k}_n\}$ its
wave-vector content. Wedge
products of translation eigenvectors diagonalize the translations on
$\Lambda^p V$, with characters
$e^{i(\vec{k}_{i_1}+\cdots+\vec{k}_{i_p})\cdot\vec{t}}$ for $i_1<\cdots<i_p$,
so $\Lambda^p\Gamma$ contains a component at $\vec{k}=0$ if and only if
\begin{subequations}
\label{eq:wavevector-rules}
\begin{align}
\Lambda^1 \Gamma
&:
\vec{k}_i\equiv\vec{0}\text{ for some }i,
\\
\Lambda^2 \Gamma
&:
\vec{k}_i+\vec{k}_j\equiv\vec{0}\text{ for some }i\neq j,
\\
\Lambda^3 \Gamma
&: 
\sum_i \vec{k}_i \equiv\vec{0},
\end{align}
\end{subequations}
in each case modulo a reciprocal lattice vector. The three exterior powers
therefore test whether one ordering wave vector vanishes, whether two cancel, or
whether all three sum to zero (\cref{fig:hierarchy}), and identical conditions apply to the non-trivial
powers of one- and two-dimensional representations.

Reality closes the wave-vector content of $\Gamma$ under
$\vec{k}\rightarrow-\vec{k}$,
so any unpaired entry must be a TRIM $\vec{K}$ with
$2\vec{K}\equiv 0$. A three-dimensional $\Gamma$ thus has content
$\{\vec{k},-\vec{k},\vec{K}\}$ or $\{\vec{K}_1,\vec{K}_2,\vec{K}_3\}$, and a
two-dimensional one $\{\vec{k},-\vec{k}\}$ or $\{\vec{K}_1,\vec{K}_2\}$. For the content
$\{\vec{k},-\vec{k},\vec{K}\}$, $\Lambda^2\Gamma$ always reaches $\vec{k}=0$
and $\Lambda^3\Gamma$ does so only for $\vec{K}\equiv 0$; for three TRIMs,
$\Lambda^2\Gamma$ requires two of them to coincide, since $2\vec{K}\equiv 0$
turns $\vec{K}_j+\vec{K}_k\equiv 0$ into $\vec{K}_j\equiv\vec{K}_k$, and
$\Lambda^3\Gamma$ requires them to sum to zero.

\section{Observables}
\label{sec:observables}
We now apply this representation hierarchy to several observables, expressing
the symmetry constraints directly in terms of the representation $\Gamma$. For
each observable the spin transformation selects one of $\Lambda^1\Gamma$,
$\Lambda^2\Gamma$, or $\Lambda^3\Gamma$, while its spatial tensor character
determines the representation with which that exterior power must overlap.
Here two real representations are said to overlap when they share a real
irreducible component, or equivalently when their tensor product contains the
trivial irrep.~\footnote{Only whether an overlap is nonzero matters for the
criteria below. The trivial irrep can occur more than once per shared
component, as for a real two-dimensional irrep built from a complex-conjugate
pair, such as $E$ of $C_3$, whose square contains it twice.}

Throughout we use $\Gamma_\ell^\pm$ to denote the rank-$\ell$ irreducible
representation of $\mathrm{O}(3)$ with inversion parity $\pm$, regarded as a
$\vec{k}=0$ representation of $G$, or equivalently of its point group
$P \equiv G/T$, where $T$ is the translation subgroup of $G$.
Thus $\Gamma_0^+$ is the trivial irrep, $\Gamma_1^-$ and $\Gamma_1^+$ are the
polar- and axial-vector representations, and $\Gamma_2^+$ is the quadrupole
representation. Note that this superscript is an inversion parity, whereas on
$\Sigma^\pm$, $\alpha^\pm$ and $\sigma^\pm$ below it denotes time-reversal
parity.

\subsection{Magnetization}
\label{sec:magnetization}
We start by formulating a basic question in this representation-first picture:
which of these spin space groups allow a net spin magnetization? Consider the spin space group
corresponding to the representation $\Gamma$. The net (spin) magnetization
$\vec{M}$~\footnote{Here and below, $\vec{M}$ denotes the spin magnetization.
For non-coplanar order the orbital magnetization $\vec{M}_{\rm orb}$ transforms
as $\det\Gamma \otimes \Gamma_1^+$, as discussed in \cref{sec:hall}; for
collinear and coplanar order it is forbidden in the strict non-relativistic
limit by the spin-only symmetry.} is a uniform spin vector, so the spatial part
of an operation leaves it unchanged and it transforms under
$\Lambda^1\Gamma=\Gamma$. For a non-coplanar order,
\begin{equation}
\label{eq:magnetization-transform}
\vec{M} \xrightarrow{\ssop{\mat{\Gamma}(g)}{g}} \mat{\Gamma}(g) \vec{M},
\end{equation}
where $g \in G$. A magnetization allowed in this phase must then be invariant,
\begin{equation}
\label{eq:magnetization-invariance}
\vec{M} = \mat{\Gamma}(g)\vec{M},
\end{equation}
for all $g \in G$, and so spans a one-dimensional subspace left invariant by
the group; that is, $\Gamma$ contains a copy of the trivial irrep.

For collinear or coplanar orders the same condition applies to the one- or
two-dimensional representations. The spin-only group in the collinear case
further restricts $\vec{M}$ to $M\vhat{N}$, and in the coplanar case to
$M_1 \vhat{N}_1 + M_2\vhat{N}_2$, where $\vhat{N}_1$, $\vhat{N}_2$ define the
preferred plane. Together, we can say:
\begin{cframed}
    \textbf{Net Spin Magnetization:} A spin space group defined by a real
    representation $\Gamma$ allows a net spin magnetization if and only if
    $\Gamma$ contains the trivial irrep $\Gamma^+_0$.
\end{cframed}
We call a state \emph{compensated} when $\Gamma$ does not contain $\Gamma^+_0$, so that
its net moment vanishes by symmetry. If all parts of the representation
$\Gamma$ are at finite wave vector it is
automatically compensated, as in a conventional antiferromagnet, since the
trivial irrep is at $\vec{k}=0$. The same argument applies to every observable
below that is controlled by $\Lambda^1\Gamma$, since each must overlap a
representation at $\vec{k}=0$: for such an order piezomagnetism, the time-odd
spin conductivity $\Sigma^-$, the time-odd Edelstein response $\alpha^-$ and the
even-rank spin texture all vanish as well
(\cref{sec:piezomagnetism,sec:transport,sec:edelstein,sec:spintexture}).

The magnetization may still vanish despite being allowed by symmetry. An example
of recent interest is the class of so-called ``Luttinger-compensated''
magnets~\cite{vanleuken1995half,mazin2024altermagnetism}, collinear insulators
or half-metals in which at least one spin channel is gapped. Due to the
constraints from Luttinger's theorem~\cite{luttinger1960fermi} (or,
nonperturbatively, Oshikawa's
argument~\cite{oshikawa2000commensurability,oshikawa2000topological}), the net
magnetization can only take on quantized values, and thus may exactly vanish
even when the symmetry allows a finite magnetization. 

Potential examples are the
double perovskite \ch{LaMn2SbO6}, identified as a compensated magnetic
semiconductor~\cite{solanamadruga2018anisotropic,hou2025unconventional}, the
$A$ phase of the semiconducting calcium ferrite
\ch{CaFe2O4}~\cite{corliss1967magnetic,songvilay2020disorder,ida2010preparation},
and the polar insulator
\ch{Mn2Mo3O8}~\cite{bertrand1975etude,stanislavchuk2020spectroscopic,liao2025magnetic}.
In each, $\Gamma$ is the trivial irrep ($A_g$ of $P4_2/n$ for \ch{LaMn2SbO6},
$A_g$ of $Pnma$ for \ch{CaFe2O4} and $A_1$ of $P6_3mc$ for \ch{Mn2Mo3O8}), so a
net moment is allowed by symmetry, yet the experimentally refined
structures have zero net magnetization, the two inequivalent sublattices of
\ch{Mn^{2+}} or \ch{Fe^{3+}} carrying equal and opposite
moments~\cite{gallego2016magndata}.%
~\footnote{This quantization relies on conserved spin and exact stoichiometry.
Spin-orbit coupling and defects lift it, consistent with the minute net moment
reported in the $A$ phase of epitaxial \ch{CaFe2O4} films, part of which was
attributed to defective \ch{Fe^{2+}}~\cite{ueda2022unusual}.} For \ch{Mn2Mo3O8}
this zero is a ground-state property: it is an L-type
ferrimagnet,\footnote{The designation ``L-type ferrimagnet'' originates from
Louis N\'eel's 1948 classification of the possible temperature dependences of
ferrimagnetic magnetization~\cite{neel1948proprietes}. In Sec.~16 (Fig.~8L),
N\'eel uses ``L'' simply as a label for the unusual case in which two
inequivalent magnetic sublattices compensate at zero temperature but develop a
net magnetization upon warming; the same paper also introduced the term
``ferrimagnetism.'' Interestingly, \ch{Mn2Mo3O8} has more recently been
proposed as a Luttinger-compensated magnet~\cite{gao2025ferroelectric}. Thus,
the ``L'' has nothing to do with Luttinger compensation, although, rather
curiously, the two concepts meet in this very material.} whose two
sublattices are unequally polarized at finite temperature, giving a net moment
below $T_C \approx 41.5$~K that vanishes only as $T \to 0$~\cite{liao2025magnetic}.
We will encounter a
similar non-symmetry-based constraint in \cref{sec:edelstein} for the Edelstein
effect in collinear magnets.

\subsection{Piezomagnetism}
\label{sec:piezomagnetism}
Next we consider piezomagnetism, defined as the generation of a net
magnetic moment in the presence of a finite
strain~\cite{dzyaloshinskii1958piezomagnetism,moriya1959piezomagnetism,%
borovik1960piezomagnetism,etxebarria2025crystal}.
Explicitly,
\begin{equation}
\label{eq:piezomagnetism-def}
\vec{M} = \sum_{\mu\nu} \vec{C}_{\mu\nu} \epsilon_{\mu\nu} + {\dots}.
\end{equation}
For this to be allowed in a spin space group the coefficient $\vec{C}_{\mu\nu}$
must transform trivially. Like the magnetization, $\vec{C}_{\mu\nu}$ carries a
single spin index and so transforms under $\Lambda^1\Gamma = \Gamma$ in spin
space. We assume that $\Gamma$ does not contain any copies of the trivial irrep,
so that the state has no intrinsic magnetization in the absence of strain.

The symmetric strain tensor, $\mat{\epsilon}$, transforms as the symmetric
square of the polar-vector representation $\Gamma_1^-$ of the point group $P$. We denote this symmetric square by ${\rm sym}^2(\Gamma_1^-)$. For
non-coplanar order, under a transformation $\ssop{\mat{\Gamma}(g)}{g}$ with $g
\equiv [\mat{R}|\vec{t}]$ we have
\begin{subalign}
    \label{eq:strain-transform-M}
    \vec{M} &\rightarrow \mat{\Gamma}(g) \vec{M}, \\
    \label{eq:strain-transform-eps}
    \mat{\epsilon} &\rightarrow \mat{R} \mat{\epsilon} \trp{\mat{R}}.
\end{subalign}
Matching these two requires that ${\rm sym}^2(\Gamma_1^-)
\otimes \Gamma$ contains the trivial irrep. We can write the symmetric square in
a different way as
\begin{equation}
\label{eq:sym-square}
{\rm sym}^2(\Gamma_1^-) = \Gamma_0^+ \oplus \Gamma_2^+,
\end{equation}
where $\Gamma_0^+$ is the trivial irrep and $\Gamma_2^+$ is the quadrupole
representation. Thus piezomagnetism is allowed by symmetry in a compensated spin
space group when $\Gamma$ overlaps with $\Gamma_2^+$. Without compensation the
trivial part of the strain, a uniform dilation, can also change the
magnetization, and the condition becomes an overlap with
$\Gamma_0^+\oplus\Gamma_2^+$. An identical argument
applies to collinear and coplanar orders, with the spin-only groups only
restricting the allowed direction of $\vec{C}_{\mu\nu}$ in spin space.
\begin{cframed} 
    \textbf{Piezomagnetism:} A compensated spin space group defined by a real
    representation $\Gamma$ allows (linear) piezomagnetism if and only if $\Gamma$
    overlaps with the quadrupole representation $\Gamma_2^+$.
\end{cframed}
We give a listing of the polar-vector, axial-vector and quadrupole
representations $\Gamma_1^-$, $\Gamma_1^+$ and $\Gamma_2^+$ for each of the
crystallographic point groups in \cref{tab:vector-quadrupole}.

\begin{table}[t]
\centering
\begin{tabular}{>{$}l<{$} >{$}l<{$} >{$}l<{$} >{$}l<{$}}
\toprule
P & \Gamma_1^- & \Gamma_1^+ & \Gamma_2^+ \\
\midrule
C_{1}   & 3A
        & 3A
        & 5A \\
C_{i}   & 3A_{u}
        & 3A_{g}
        & 5A_{g} \\
C_{2}   & A \oplus 2B
        & A \oplus 2B
        & 3A \oplus 2B \\
C_{s}   & 2A' \oplus A''
        & A' \oplus 2A''
        & 3A' \oplus 2A'' \\
C_{2h}  & A_{u} \oplus 2B_{u}
        & A_{g} \oplus 2B_{g}
        & 3A_{g} \oplus 2B_{g} \\
D_{2}   & B_{1} \oplus B_{2} \oplus B_{3}
        & B_{1} \oplus B_{2} \oplus B_{3}
        & 2A \oplus B_{1} \oplus B_{2} \oplus B_{3} \\
C_{2v}  & A_{1} \oplus B_{1} \oplus B_{2}
        & A_{2} \oplus B_{1} \oplus B_{2}
        & 2A_{1} \oplus A_{2} \oplus B_{1} \oplus B_{2} \\
D_{2h}  & B_{1u} \oplus B_{2u} \oplus B_{3u}
        & B_{1g} \oplus B_{2g} \oplus B_{3g}
        & 2A_{g} \oplus B_{1g} \oplus B_{2g} \oplus B_{3g} \\
C_{4}   & A \oplus E
        & A \oplus E
        & A \oplus 2B \oplus E \\
S_{4}   & B \oplus E
        & A \oplus E
        & A \oplus 2B \oplus E \\
C_{4h}  & A_{u} \oplus E_{u}
        & A_{g} \oplus E_{g}
        & A_{g} \oplus 2B_{g} \oplus E_{g} \\
D_{4}   & A_{2} \oplus E
        & A_{2} \oplus E
        & A_{1} \oplus B_{1} \oplus B_{2} \oplus E \\
C_{4v}  & A_{1} \oplus E
        & A_{2} \oplus E
        & A_{1} \oplus B_{1} \oplus B_{2} \oplus E \\
D_{2d}  & B_{2} \oplus E
        & A_{2} \oplus E
        & A_{1} \oplus B_{1} \oplus B_{2} \oplus E \\
D_{4h}  & A_{2u} \oplus E_{u}
        & A_{2g} \oplus E_{g}
        & A_{1g} \oplus B_{1g} \oplus B_{2g} \oplus E_{g} \\
C_{3}   & A \oplus E
        & A \oplus E
        & A \oplus 2E \\
S_{6}   & A_{u} \oplus E_{u}
        & A_{g} \oplus E_{g}
        & A_{g} \oplus 2E_{g} \\
D_{3}   & A_{2} \oplus E
        & A_{2} \oplus E
        & A_{1} \oplus 2E \\
C_{3v}  & A_{1} \oplus E
        & A_{2} \oplus E
        & A_{1} \oplus 2E \\
D_{3d}  & A_{2u} \oplus E_{u}
        & A_{2g} \oplus E_{g}
        & A_{1g} \oplus 2E_{g} \\
C_{6}   & A \oplus E_{1}
        & A \oplus E_{1}
        & A \oplus E_{1} \oplus E_{2} \\
C_{3h}  & A'' \oplus E'
        & A' \oplus E''
        & A' \oplus E' \oplus E'' \\
C_{6h}  & A_{u} \oplus E_{1u}
        & A_{g} \oplus E_{1g}
        & A_{g} \oplus E_{1g} \oplus E_{2g} \\
D_{6}   & A_{2} \oplus E_{1}
        & A_{2} \oplus E_{1}
        & A_{1} \oplus E_{1} \oplus E_{2} \\
C_{6v}  & A_{1} \oplus E_{1}
        & A_{2} \oplus E_{1}
        & A_{1} \oplus E_{1} \oplus E_{2} \\
D_{3h}  & A_{2}'' \oplus E'
        & A_{2}' \oplus E''
        & A_{1}' \oplus E' \oplus E'' \\
D_{6h}  & A_{2u} \oplus E_{1u}
        & A_{2g} \oplus E_{1g}
        & A_{1g} \oplus E_{1g} \oplus E_{2g} \\
T       & T
        & T
        & E \oplus T \\
T_{h}   & T_{u}
        & T_{g}
        & E_{g} \oplus T_{g} \\
O       & T_{1}
        & T_{1}
        & E \oplus T_{2} \\
T_{d}   & T_{2}
        & T_{1}
        & E \oplus T_{2} \\
O_{h}   & T_{1u}
        & T_{1g}
        & E_{g} \oplus T_{2g} \\
\bottomrule
\end{tabular}
\caption{Polar-vector representation $\Gamma_1^-$, axial-vector representation
$\Gamma_1^+$ and quadrupole representation $\Gamma_2^+$ for each
crystallographic point group $P$. The quadrupole representation $\Gamma_2^+$ is
the traceless part of the symmetric square ${\rm
sym}^2(\Gamma_1^-)=\Gamma_0^+\oplus\Gamma_2^+$, the trivial irrep $\Gamma_0^+$
carrying the trace. Note that $\Gamma_2^+$ may itself contain copies of
$\Gamma_0^+$ in some lower-symmetry groups.}
\label{tab:vector-quadrupole}
\end{table}

\subsection{Spin and Charge Transport}
\label{sec:transport}
We next consider constraints on charge and spin transport. The symmetry-imposed
form of transport tensors has been determined for magnetic point
groups~\cite{seeman2015symmetry,gonzalez2021efficient}, while spin transport in
the non-relativistic limit has been discussed in specific magnetic
systems~\cite{zel2017spin,zhang2018spin} and more generally using spin-group
symmetry~\cite{watanabe2024symmetry,etxebarria2025crystal}. Define a charge
current $j_{\nu}$ and a spin current $J^\mu_\nu$,~\footnote{We do not require a
microscopic definition of the spin-current operator here. We assume a
spin-current response $J^\mu_\nu$ transforming as a spatial current carrying a
spin index, together with an appropriately defined Onsager-reciprocal response.
Different microscopic definitions, such as the conventional and
torque-dipole-corrected currents~\cite{shi2006proper}, may share the same tensor
transformation properties, although the identification of the conjugate force
and reciprocal response can depend on that definition.} where $\mu$ is a spin
index and $\nu$ is a spatial index. The spin and charge currents are related to
the electric field $E_{\nu}$ and the field gradient $\del_{\nu} B^{\mu}$ by
conductivity tensors
\begin{subalign}
\label{eq:charge-current}
j_{\nu} &= \sum_\rho \sigma_{\nu\rho} E_\rho
+ \sum_{\mu\rho} \tilde{\Sigma}^\mu_{\nu\rho} \del_{\rho} B^\mu, \\
\label{eq:spin-current}
J^\mu_\nu &= \sum_\rho \Sigma^\mu_{\nu\rho} E_\rho +
\sum_{\mu'\rho} S^{\mu\mu'}_{\nu\rho} \del_{\rho} B^{\mu'}.
\end{subalign}
\Cref{eq:charge-current,eq:spin-current} contain the usual charge conductivity
$\sigma_{\nu\rho}$, the spin conductivities $\Sigma^\mu_{\nu\rho}$ and
$\tilde{\Sigma}^\mu_{\nu\rho}$, and the spin-spin conductivity
$S^{\mu\mu'}_{\nu\rho}$, which we do not consider further. Spin currents of this kind underlie the forward and
inverse spin Hall
effects~\cite{dyakonov1971possibility,dyakonov1971current,hirsch1999spin,%
sinova2015spin}, and can be generated in compensated antiferromagnets even in
the absence of spin-orbit coupling~\cite{naka2019spin}.

\subsubsection{Spin Conductivity}
\label{sec:spin-conductivity}
The spin conductivity $\Sigma^\mu_{\nu\rho}$ describes the generation of a spin
current in response to an electric field. Its partner
$\tilde{\Sigma}^\mu_{\nu\rho}$ describes the reciprocal response of a charge
current to a field conjugate to the spin density. Onsager--Casimir
reciprocity~\cite{onsager1931reciprocal,casimir1945onsager} gives
\begin{equation}
\label{eq:onsager}
\Sigma^\mu_{\nu\rho}(H) 
=
-\tilde{\Sigma}^\mu_{\rho\nu}(THT^{-1}),
\end{equation}
where $H$ is the system Hamiltonian and $T$ is the (antiunitary) time reversal
operator. Since $\Sigma$ and $\tilde{\Sigma}$ are exchanged by time reversal,
neither has a definite time-reversal parity on its own. We therefore define
\begin{equation}
\label{eq:sigma-even-odd-def}
\Sigma^{\mu,\pm }_{\nu\rho}
\equiv
\frac{1}{2}
{\left(
\Sigma^\mu_{\nu\rho}
\mp
\tilde{\Sigma}^\mu_{\rho\nu}
\right)}, 
\end{equation}
where $\Sigma^-$ is time odd and $\Sigma^+$ is time even.

These combinations have definite transformation laws under the spin-space
symmetries. Consider a non-coplanar spin space group defined by a real three-dimensional
representation $\Gamma$ of the parent space group $G$. Taking into account the
sign of $\det \mat{\Gamma}(g)$, which indicates time reversal, we have
\begin{subalign}
\label{eq:sigma-odd-transform}
\vec{\Sigma}^{-}_{\nu\rho} &\xrightarrow{\ssop{\mat{\Gamma}(g)}{g}}
\sum_{\nu'\rho'} \mat{\Gamma}(g)\vec{\Sigma}^{-}_{\nu'\rho'} R_{\rho\rho'}(g) R_{\nu\nu'}(g), \\
\label{eq:sigma-even-transform}
\vec{\Sigma}^{+}_{\nu\rho} &\xrightarrow{\ssop{\mat{\Gamma}(g)}{g}}
\det\mat{\Gamma}(g)\!\sum_{\nu'\rho'}\!
\mat{\Gamma}(g)\vec{\Sigma}^{+}_{\nu'\rho'} R_{\rho\rho'}(g) R_{\nu\nu'}(g),
\end{subalign}
where $\mat{R}(g)$ is the rotational part of the space group operation $g =
[\mat{R}|\vec{t}]$. The spin conductivity therefore involves $\Gamma$ for the
time-odd component and the determinant twist $\Gamma \otimes \det\Gamma =
\Lambda^2\Gamma$ of \cref{eq:exterior-powers} for the time-even component,
while both spatial indices transform as ordinary vectors. The spin index of
$\Sigma^{+}$ is thus a spin pseudovector, like the cross product in
\cref{eq:cross-product-exterior}: it is even under spin inversion, as a
time-even quantity must be, while that of $\Sigma^{-}$ is odd.

Explicitly, the time-odd spin conductivity transforms as
\begin{equation}
\label{eq:time-odd-spin-cond}
\Gamma \otimes
\left(
\Gamma_1^- \otimes \Gamma_1^-
\right)
=
\Gamma \otimes
{\left(
\Gamma_0^+ \oplus \Gamma_1^+ \oplus \Gamma_2^+
\right)}.
\end{equation}
Here $\Gamma_0^+ \oplus \Gamma_2^+ = {\rm sym}^2(\Gamma_1^-)$ is the part
symmetric in the spatial indices [\cref{eq:sym-square}] and $\Gamma_1^+ =
\Lambda^2\Gamma_1^-$ the antisymmetric part.
We thus require that $\Gamma$ have nonzero overlap with $\Gamma_0^+$,
$\Gamma_1^+$ or $\Gamma_2^+$. The time-even part follows from the same argument,
with the spin index now transforming as $\Lambda^2 \Gamma$ rather than in
$\Gamma$ itself, giving
\begin{equation}
\label{eq:time-even-spin-cond}
\Lambda^2 \Gamma
\otimes
{\left(
\Gamma_0^+ \oplus \Gamma_1^+ \oplus \Gamma_2^+
\right)}.
\end{equation}
Therefore the time-even spin conductivity is allowed if $\Lambda^2 \Gamma$ has
nonzero overlap with $\Gamma_0^+$, $\Gamma_1^+$ or $\Gamma_2^+$.

The spin-only groups impose additional constraints on collinear and coplanar
orders. For a collinear magnet, the residual spin rotation symmetry about the
ordering axis $\vhat{N}$ requires
\begin{equation}
\label{eq:collinear-spin-only}
\Sigma^\mu_{\nu\rho}
=
\Sigma_{\nu\rho} \hat{N}^\mu.
\end{equation}
Time reversal combined with a two-fold spin rotation perpendicular to $\vhat{N}$
then forces the time-even component to vanish,
\begin{equation}
\label{eq:collinear-even-vanishes}
\Sigma^{\mu,+}_{\nu\rho}=0.
\end{equation}
We thus require the one-dimensional irrep $\Gamma$ defining the collinear order
to have nonzero overlap with $\Gamma_0^+$, $\Gamma_1^+$ or $\Gamma_2^+$ for the
time-odd part to be allowed.

For coplanar order, let $\vhat{N}_1$ and $\vhat{N}_2$ span the preferred plane.
The spin-only symmetry instead requires
\begin{subalign}
\label{eq:coplanar-odd}
\vec{\Sigma}^{-}_{\nu\rho}
&=
\Sigma_{1,\nu\rho}\vhat{N}_1
+
\Sigma_{2,\nu\rho}\vhat{N}_2, \\
\label{eq:coplanar-even}
\vec{\Sigma}^{+}_{\nu\rho}
&=
\Sigma_{3,\nu\rho}
{\left(
\vhat{N}_1 \times \vhat{N}_2
\right)}.
\end{subalign}
Thus the time-odd response lies in the ordering plane while the time-even
response is perpendicular to it. In terms of the two-dimensional representation
$\Gamma$, the in-plane parts require $\Gamma$ to overlap with $\Gamma_0^+$,
$\Gamma_1^+$ or $\Gamma_2^+$, while the out-of-plane part requires overlap of $\det \Gamma =
\Lambda^2 \Gamma$ with the same.

These conditions for collinear, coplanar and non-coplanar orders can be
encapsulated in the following unified statements:
\begin{cframed}
\textbf{Time-Odd Spin Conductivity:}
A spin space group defined by a real 
representation $\Gamma$ allows time-odd spin conductivity if and only if
$\Gamma$ has nonzero overlap with $\Gamma_0^+$, $\Gamma_1^+$ or $\Gamma_2^+$.
\end{cframed}
\begin{cframed}
\textbf{Time-Even Spin Conductivity:}
A spin space group defined by a real representation $\Gamma$ allows
time-even spin conductivity if and only if $\Lambda^2 \Gamma$ has nonzero
overlap with $\Gamma_0^+$, $\Gamma_1^+$, or $\Gamma_2^+$.
\end{cframed}
Note that for a one-dimensional representation $\Lambda^2 \Gamma = 0$, so this
also encapsulates that the time-even response vanishes for collinear order.

Because $\Gamma_0^+$, $\Gamma_1^+$ and $\Gamma_2^+$ all live at $\vec{k}=0$,
they also constrain which wave vectors may appear in $\Gamma$, and these
restrictions follow from \cref{sec:exterior-prelim}. The time-odd response is
controlled by $\Gamma \equiv \Lambda^1\Gamma$ and therefore requires a
$\vec{k}=0$ component to be present in $\Gamma$, while the time-even response is
controlled by $\Lambda^2\Gamma$ and requires only a pair of wave vectors that
sum to zero. For coplanar order this is a pair $\vec{k}$, $-\vec{k}$; for
non-coplanar order $\Gamma$ must contain such a pair together with a TRIM
$\vec{K}$.

For collinear order the conventional band conductivity can obey a further
restriction that does not follow from the transformation laws alone. The
spin-only symmetry combining time reversal with a two-fold spin rotation
perpendicular to $\vhat{N}$ acts within each spin sector as spinless time
reversal. If the two sectors conduct independently, as for non-interacting
electrons, $\Sigma_{\nu\rho} \propto \sigma^\uparrow_{\nu\rho} -
\sigma^\downarrow_{\nu\rho}$ [\cref{eq:collinear-spin-only}] is a difference of spin-resolved charge
conductivities that are each symmetric by Onsager reciprocity under this
spinless time reversal. Then $\Sigma^-$ is symmetric in its spatial indices and
thus requires $\Gamma \cap (\Gamma_0^+ \oplus
\Gamma_2^+) \neq 0$; for compensated order this is identical to the
piezomagnetic condition. This symmetric response is the spin-splitter effect of
\citet{gonzalez2021efficient}.

However, an antisymmetric part is not excluded once the sectors are coupled, for example by spin Coulomb
drag~\cite{damico2000theory}, which violates the assumption of independent spin
channels. The purely antisymmetric channel is the only one allowed, for
example, in the reported 300~K structure of
$\alpha$-\ch{Fe2O3}~\cite{hill2008neutron,gallego2016magndata}, above the Morin
transition, whose $\Gamma = A_{2g}$ of $R\bar{3}c$ overlaps with
$\Gamma_1^+ = A_{2g} \oplus E_g$ but not with $\Gamma_0^+$ nor
$\Gamma_2^+ = A_{1g} \oplus 2E_g$, so that piezomagnetism and the symmetric
part of $\Sigma^-$ are also forbidden.~\footnote{Since the spin direction does not enter $\Gamma$ for
collinear order, the same holds below the transition, where the moments lie
along $c$.} The same holds for the chalcopyrite
\ch{CuFeS2}~\cite{donnay1958symmetry,gallego2016magndata}, whose $\Gamma = A_2$
of $I\bar{4}2d$ likewise overlaps $\Gamma_1^+ = A_2 \oplus E$ but not
$\Gamma_0^+$ nor $\Gamma_2^+$. Any antisymmetric spin conductivity arising in
these materials therefore requires interactions that couple the spin channels.

\subsubsection{Anomalous Hall Conductivity}
\label{sec:hall}
The charge conductivity follows from the same construction with the spin index
removed, so that only the determinant of $\mat{\Gamma}(g)$ enters. Define the
symmetric and antisymmetric parts of the charge conductivity as
\begin{equation}
\label{eq:sigma-sym-antisym-def}
\sigma^{\pm}_{\nu\rho} \equiv \frac{1}{2}{\left(\sigma_{\nu\rho} \pm \sigma_{\rho\nu}\right)}.
\end{equation}
The symmetric part is time-even while the antisymmetric part is time-odd, so the
superscripts again denote time-reversal parity. For a non-coplanar order a
spin-space operation $\ssop{\mat{\Gamma}(g)}{g}$ acts as
\begin{subalign}
\label{eq:sigma-sym-transform}
\sigma^+_{\nu\rho} &\xrightarrow{\ssop{\mat{\Gamma}(g)}{g}}
\sum_{\nu'\rho'}\sigma^+_{\nu'\rho'} R_{\rho\rho'}(g) R_{\nu\nu'}(g),\\
\label{eq:sigma-antisym-transform}
\sigma^-_{\nu\rho} &\xrightarrow{\ssop{\mat{\Gamma}(g)}{g}}
\det \mat{\Gamma}(g) \sum_{\nu'\rho'} \sigma^-_{\nu'\rho'} R_{\rho\rho'}(g) R_{\nu\nu'}(g).
\end{subalign}
In representation theory language this states that
\begin{subalign}
\label{eq:sigma-sym-rep}
\sigma^+ &\sim \Gamma_0^+ \oplus \Gamma_2^+, \\
\label{eq:sigma-antisym-rep}
\sigma^- &\sim \det\Gamma \otimes \Gamma_1^+.
\end{subalign}
The symmetric conductivity is therefore always allowed. The antisymmetric part
is equivalent to a Hall pseudovector, an axial vector twisted by the
determinant representation. That representation is the top exterior power
$\Lambda^3\Gamma$, and can be read directly from \cref{tab:det-reps} when
$\Gamma$ is irreducible.

Collinear and coplanar states possess spin-only symmetries that force
$\sigma^-=0$, while a non-coplanar state can exhibit a finite anomalous Hall
conductivity even without relativistic
interactions~\cite{nagaosa2010ahe,martin2008itinerant,smejkal2022anomalous}.
\begin{cframed}
\textbf{Anomalous Hall Conductivity:}
A spin space group defined by a real representation $\Gamma$ allows
anomalous Hall conductivity in the non-relativistic limit if and only if
$\Lambda^3 \Gamma$ has nonzero overlap with the axial-vector representation
$\Gamma_1^+$.
\end{cframed} 
Since $\Lambda^3 \Gamma=0$ for collinear and coplanar orders this encodes that
they cannot support an anomalous Hall effect. The same symmetry analysis applies
unchanged to the thermal conductivity and the thermal Hall conductivity.

The Hall pseudovector transforms exactly as the orbital magnetization, an axial
spatial vector odd under time reversal, rather than as the spin magnetization,
which transforms under $\Gamma$, so in the non-relativistic limit a Hall
response neither requires nor implies a net spin magnetization. This
non-relativistic Hall response was termed the \emph{geometrical Hall effect} in
\citet{watanabe2024symmetry}; related discussions are given in
Refs.~\cite{etxebarria2025crystal,liu2026symmetry}. Its determinant factor
matches that of the scalar spin chirality, \cref{eq:scalar-chirality}, whose microscopic
counterpart is the Berry phase accumulated by an electron hopping through a
chiral spin background~\cite{ohgushi2000spin,taguchi2001spin,nagaosa2010ahe}.
This correspondence is purely a statement about symmetry. It does not require a
particular microscopic chirality to be nonzero, nor does it identify the
mechanism responsible for an allowed Hall response.

Since the Hall response is controlled by $\Lambda^3\Gamma$, its translation
selection rule is that the three wave vectors sum to zero, so $\Gamma$ need not
contain a $\vec{k}=0$ component: its content is either $\vec{0}$, $\vec{k}$,
$-\vec{k}$ with $\vec{k}$ arbitrary, or three TRIM points $\vec{K}_1$,
$\vec{K}_2$, $\vec{K}_3$ with $2\vec{K}_n \equiv 0$ and $\vec{K}_1 + \vec{K}_2 +
\vec{K}_3 \equiv 0$. 
 
Two general restrictions on the non-relativistic anomalous Hall response follow
from the fact that $\Lambda^3\Gamma=\det\Gamma$ is one-dimensional.

First, for the cubic point groups $T$, $T_h$, $O$, $T_d$ and $O_h$ the
axial-vector representation $\Gamma_1^+$ is a single three-dimensional irrep, so
the overlap vanishes for every $\Gamma$ and a cubic parent point group forbids
the anomalous Hall effect outright, as for pyrochlore \ch{FeF3} in
\cref{sec:fef3}.
The vanishing Hall conductivity of the triple-$Q$ state on an undistorted fcc
lattice discussed in Ref.~\cite{shindou2001orbital} is an early instance of this rule.

Second, at $\vec{k}=0$ the representation $\Gamma$ is one of the point group,
and outside the cubic groups it is either a sum of three one-dimensional irreps
or a one-dimensional irrep together with a two-dimensional one. Writing $\dG{d}$
for a $d$-dimensional irreducible representation, the second case has
$\det\Gamma = \dG{1} \otimes \det(\dG{2})$ (\cref{app:exterior-sum}), and comparing
\cref{tab:det-reps,tab:vector-quadrupole} shows that $\det(\dG{2})$ is the
one-dimensional part of $\Gamma_1^+$ in every such group. Overlap then requires
$\dG{1} = \Gamma_0^+$ and so the order is uncompensated. A compensated anomalous
Hall response from $\vec{k}=0$ order therefore requires $\Gamma$ to be a sum of
three one-dimensional irreps; orders built from layers or sublattices that are
each collinear, with moments along three non-coplanar axes, may be a natural
place to realize this. The reported
structure of the polar magnet
\ch{DyCrWO6}~\cite{ghara2018synthesis,gallego2016magndata} illustrates a
version of this case: its compensated non-coplanar $\vec{k}=0$ order, combining a non-collinear
\ch{Dy^{3+}} sublattice with a collinear \ch{Cr^{3+}} one, has
$\Gamma = A_2 \oplus 2B_1$ of $Pna2_1$, so that $\det\Gamma = A_2$ overlaps
$\Gamma_1^+ = A_2 \oplus B_1 \oplus B_2$. Since the \ch{Dy^{3+}} moments likely make
the magnetism of this compound relativistic, we use its structure
only as an illustration.

These conditions for the spin and charge transport tensors, in terms of the
representations and the wave-vector content, are summarized in
\cref{tab:transport}.

\begin{table}[t]
\centering
\footnotesize
\setlength{\tabcolsep}{4pt}
\renewcommand{\arraystretch}{1.25}
\begin{tabular}{l l l}
\toprule
Response & Required overlap & Wave vectors in $\Gamma$ \\
\midrule
\multicolumn{3}{l}{Collinear order} \\
$\Sigma^-$ & $\Lambda^1\Gamma \cap (\Gamma_0^+ \oplus \Gamma_1^+ \oplus \Gamma_2^+)$ & $\vec{0}$ \\
$\Sigma^+$ & --- & --- \\
$\sigma^-$ & --- & --- \\
\addlinespace
\multicolumn{3}{l}{Coplanar order, $\Lambda^2\Gamma = \det\Gamma$} \\
$\Sigma^-$ & $\Lambda^1\Gamma \cap (\Gamma_0^+ \oplus \Gamma_1^+ \oplus \Gamma_2^+)$ & $\vec{0}$ \\
$\Sigma^+$ & $\Lambda^2\Gamma \cap (\Gamma_0^+ \oplus \Gamma_1^+ \oplus \Gamma_2^+)$ & $\vec{k}$, $-\vec{k}$ \\
$\sigma^-$ & --- & --- \\
\addlinespace
\multicolumn{3}{l}{Non-coplanar order, $\Lambda^2\Gamma = \Gamma\otimes\det\Gamma$, $\Lambda^3\Gamma = \det\Gamma$} \\
$\Sigma^-$ & $\Lambda^1\Gamma \cap (\Gamma_0^+ \oplus \Gamma_1^+ \oplus \Gamma_2^+)$ & $\vec{0}$ \\
$\Sigma^+$ & $\Lambda^2\Gamma \cap (\Gamma_0^+ \oplus \Gamma_1^+ \oplus \Gamma_2^+)$ & $\vec{k}$, $-\vec{k}$, $\vec{K}$ \\
$\sigma^-$ & $\Lambda^3\Gamma \cap \Gamma_1^+$ & $\vec{0}$, $\vec{k}$, $-\vec{k}$ or \\
 & & $\vec{K}_1$, $\vec{K}_2$, $\vec{K}_3$ ($\sum_n \vec{K}_n \equiv \vec{0}$) \\
\bottomrule
\end{tabular}
\caption{Conditions for the time-odd and time-even spin conductivities and the
anomalous Hall conductivity, by type of magnetic order, where $\cap$ denotes a
nonzero overlap of representations and a dash a response forbidden by spin-only
symmetry. The last column gives the
wave vectors the representation must contain: $\vec{k}$ is arbitrary, while
$\vec{K}$ always denotes a time-reversal invariant momentum, $2\vec{K}\equiv 0$.
The wave-vector conditions are derived in \cref{sec:exterior-prelim}.}
\label{tab:transport}
\end{table}

\subsection{Momentum-Space Spin Textures}
\label{sec:spintexture} 
\begin{figure*}[t]
\includegraphics[width=\textwidth]{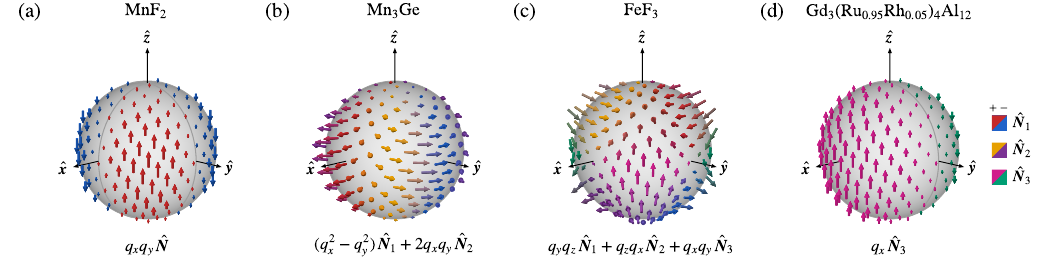}
\caption{Momentum-space spin textures $\vec{\st}(\vec{q})$ of four of the
examples of \cref{sec:examples}, drawn on the sphere of directions $\vhat{q}$, with the spin axes laid along the lattice axes. Each arrow
is colored by the spin axes it lies along, with a pair of colors for the two
signs along each axis, and $\vhat{N}$ in (a) colored as $\vhat{N}_1$. Gray lines mark where the texture vanishes.
(a)~\ch{MnF2} (\cref{sec:mnf2}), a collinear $d$-wave texture. (b)~\ch{Mn3Ge}
(\cref{sec:mn3ge}), an in-plane $d$-wave texture. (c)~Pyrochlore \ch{FeF3}
(\cref{sec:fef3}), a three-component $d$-wave texture.
(d)~\ch{Gd3(Ru_{0.95}Rh_{0.05})4Al12} (\cref{sec:gdru}), a $p$-wave texture
along $\vhat{N}_3=\vhat{N}_1\times\vhat{N}_2$, normal to the spin plane and odd
in $\vec{q}$.}
\label{fig:textures}
\end{figure*}
The same representation-first viewpoint can be applied to the
momentum-space spin texture of the electronic bands~\footnote{This is closely
related to the momentum-space spin-\emph{splitting} for collinear orders, but it
generalizes more naturally to coplanar and non-coplanar orders where they
differ.}. Let $\vec{\st}(\vec{q})$ denote the energy-resolved spin spectral
density at a fixed energy $\epsilon$, where $\vec{q}$ is the crystal momentum,
distinguished here from the ordering wave vector $\vec{k}$. This quantity is a
distribution supported on the constant-energy surfaces
$\epsilon=\epsilon_n(\vec{q})$. Explicitly, we define 
\begin{equation}
    \label{eq:spin-texture}
    \vec{\st}(\vec{q}) \equiv \tr\left[\vec{\st} \delta(\epsilon - H(\vec{q}))\right]
\end{equation}
where $H(\vec{q})$ is the Bloch Hamiltonian at wave vector $\vec{q}$ and
$\vec{\st}$ the electronic spin operator. The trace over electronic states makes
this definition independent of the choice of basis within any degenerate subspaces.

Under a spin-space operation $\ssop{\mat{\Gamma}(g)}{g}$ with $g \equiv
[\mat{R}|\vec{t}]$ the spin is rotated by $\mat{\Gamma}(g)$ while the momentum
is rotated by $\mat{R}(g)$ and, when the operation includes time reversal,
reversed. For a non-coplanar order the momentum-space spin texture is thus
transformed as
\begin{equation}
\label{eq:texture-transform}
\vec{\st}(\vec{q})
\xrightarrow{\ssop{\mat{\Gamma}(g)}{g}}
\mat{\Gamma}(g)
\vec{\st}{\left(
\det \mat{\Gamma}(g) \trp{\mat{R}}(g) \vec{q}
\right)}.
\end{equation}
The factor of $\det \mat{\Gamma}(g)$ is the momentum reversal accompanying time
reversal. Invariance within the ordered phase requires the right-hand side to
equal $\vec{\st}(\vec{q})$ for every $g \in G$. 

More concretely, we can characterize the texture $\vec{\st}(\vec{q})$ by its
projections onto momentum-space lattice harmonics. Since $\vec{q}$ transforms as
the polar-vector representation $\Gamma_1^-$, its traceless rank-$\ell$
multipoles transform as $\Gamma_\ell^{\smash{(-1)^\ell}}$. Let $f_{\ell
a}(\vec{q})$ be a basis of real periodic functions on the Brillouin zone
transforming in this representation, where $a$ labels its distinct components.
The moments
\begin{equation}
    \label{eq:spin-texture-multipoles} 
    \int_{\rm BZ} d^3q\,
    \vec{\st}(\vec{q}) f_{\ell a}(\vec{q})
\end{equation}
then identify the corresponding rank-$\ell$ components of the momentum-space
spin texture. Near the zone center, the $f_{\ell a}(\vec{q})$ may be chosen to
reduce to the usual homogeneous traceless polynomials of degree $\ell$. The
familiar $s$-, $p$-, $d$-, $f$-, and $g$-wave terminology therefore corresponds
respectively to the multipole representations $\Gamma_0^+$, $\Gamma_1^-$,
$\Gamma_2^+$, $\Gamma_3^-$, and $\Gamma_4^+$ (\cref{fig:textures}).

The spin-only symmetries of collinear and coplanar order constrain the parity of
$\vec{\st}(\vec{q})$ under $\vec{q} \rightarrow -\vec{q}$, and thus which of
these multipoles may appear. We write $\st_i(\vec{q}) \equiv \vhat{N}_i \cdot
\vec{\st}(\vec{q})$ for its components, with $\vhat{N}_1 \equiv \vhat{N}$ for
collinear order and $\vhat{N}_3 \equiv \vhat{N}_1 \times \vhat{N}_2$ for coplanar
order. For collinear order, the continuous rotations about $\vhat{N}$ require
\begin{equation}
\label{eq:texture-collinear}
\vec{\st}(\vec{q}) = \st_1(\vec{q})\vhat{N}_1,
\end{equation}
while the two-fold spin rotation perpendicular to $\vhat{N}$ combined with time
reversal further requires
\begin{equation}
\label{eq:texture-collinear-parity}
\st_1(\vec{q}) = +\st_1(-\vec{q}).
\end{equation}
The spin texture therefore contains only even momentum multipoles, and a
contribution of even rank $\ell$ is allowed when $\Gamma$ overlaps with
$\Gamma_\ell^+$. For a compensated collinear order the scalar contribution is
absent, so the lowest possible term is the quadrupolar ($d$-wave) contribution,
present if $\Gamma$ overlaps with $\Gamma_2^+$ [\cref{fig:textures}(a)]. If this
overlap also vanishes
the leading spin-texture multipole occurs at still higher even rank, as in the
$g$-wave texture found in \ch{MnTe} discussed in \cref{sec:mnte}.

For coplanar order, the spin-only symmetry, a two-fold rotation
$\mat{R}_{\vhat{N}_3}(\pi)$ combined with time reversal, instead requires
\begin{subalign}
\label{eq:texture-coplanar-x}
\st_1(\vec{q}) &= +\st_1(-\vec{q}), \\
\label{eq:texture-coplanar-y}
\st_2(\vec{q}) &= +\st_2(-\vec{q}), \\
\label{eq:texture-coplanar-z}
\st_3(\vec{q}) &= -\st_3(-\vec{q}).
\end{subalign}
The in-plane spin texture is thus even in momentum and transforms under
$\Gamma$ [\cref{fig:textures}(b)], while the component along $\vhat{N}_3$ is odd
and transforms under $\det \Gamma$ [\cref{fig:textures}(d)]. An even-rank contribution in $\st_1(\vec{q})$ or $\st_2(\vec{q})$
therefore requires $\Gamma$ to overlap with $\Gamma_\ell^+$, and an odd-rank
contribution appears in $\st_3(\vec{q})$ if $\det \Gamma$ overlaps with
$\Gamma_\ell^-$. In particular, a $p$-wave spin texture perpendicular to the
ordering plane is possible when $\det \Gamma$ overlaps with the polar-vector
representation $\Gamma_1^-$.

For non-coplanar order there is no residual spin-only symmetry fixing the parity
of the spin texture, and both even- and odd-rank contributions may occur. The
momentum reversal in \cref{eq:texture-transform} contributes a factor of $\det
\Gamma$ for each power of $\vec{q}$, so an even-rank contribution requires
$\Gamma$ to overlap with $\Gamma_\ell^+$ [\cref{fig:textures}(c)], while an
odd-rank contribution
requires the determinant twist $\Gamma \otimes \det\Gamma$ to overlap with
$\Gamma_\ell^-$. This can be stated more compactly: $\Gamma \otimes (\det
\Gamma)^\ell$ must overlap with $\Gamma_\ell^{\smash{(-1)^\ell}}$. The even- and
odd-rank components therefore alternate between $\Gamma$ and $\Lambda^2\Gamma$,
matching the conditions found for collinear and coplanar orders. The common
condition for collinear, coplanar and non-coplanar orders can thus be stated
compactly as
\begin{cframed}
\textbf{Momentum-Space Spin Texture:}
A spin space group defined by a real representation $\Gamma$ allows a
rank-$\ell$ contribution to the momentum-space spin texture if and only if
$\Gamma$ overlaps with $\Gamma_\ell^{+}$ when $\ell$ is even, and $\Lambda^2
\Gamma$ overlaps with $\Gamma_\ell^{-}$ when $\ell$ is odd.
\end{cframed}

All even-rank contributions require that the representation has components that
live at $\vec{k}=0$, while the odd-rank contributions may live entirely at
finite wave vector. The conditions on the wave-vector content of odd-rank parts
are identical to those that appear for the time-even spin conductivity (see
\cref{sec:spin-conductivity}). For coplanar order, the odd-rank contributions are only
possible when $\vec{k}$ and $-\vec{k}$ are present in $\Gamma$. If $\vec{k} \neq
0$ here the even-rank contributions vanish and the state can only support an
odd-parity spin texture perpendicular to the ordering plane, as for the
odd-parity magnets discussed in \cref{sec:oddparity}. For non-coplanar order, the odd-rank
contributions are only possible when $\vec{k}$, $-\vec{k}$, and $\vec{K}$ are
present in $\Gamma$ where $2\vec{K} \equiv \vec{0}$, as shown in
\cref{sec:exterior-prelim}.

The parity conditions on $\vec{\st}(\vec{q})$, the required representation
overlaps and the possible wave-vector content are collected in
\cref{tab:spintexture} for collinear, coplanar and non-coplanar order.

Note that an identical set of symmetry conditions applies to the magnetic moment
carried by a magnon excitation, which is closely related to the magnon chirality
probed in polarized inelastic neutron scattering. If we define the single magnon
state
\begin{equation}
\label{eq:single-magnon}
\ket{\vec{q},n} \equiv \h{a}_{\vec{q},n}\ket{0},
\end{equation}
where $\h{a}_{\vec{q},n}$ is the creation operator for a magnon in band $n$ at
momentum $\vec{q}$, then the magnon magnetic moment for that state is
\begin{equation}
\label{eq:magnon-moment}
\vec{m}_n(\vec{q})
\equiv
\braket{\vec{q},n|\vec{M}|\vec{q},n} - \braket{0|\vec{M}|0}.
\end{equation}
The quantity analogous to the spin texture is then energy resolved and summed
over all magnon bands
\begin{equation}
    \vec{m}(\vec{q}) \equiv \sum_n \vec{m}_n(\vec{q})\delta(\omega - \omega_n(\vec{q}))
\end{equation}
at some fixed energy $\omega$, where $\omega_n(\vec{q})$ are the associated
magnon energies. This transforms exactly as the spin texture
$\vec{\st}(\vec{q})$ does in \cref{eq:texture-transform}, so the conditions on
its appearance and symmetry are those already derived: its parity under $\vec{q}
\rightarrow -\vec{q}$ is fixed by the spin-only symmetry of the underlying
magnetic order, and the multipoles it may carry are fixed by
$\Gamma$.~\footnote{In a conventional two-sublattice collinear
antiferromagnet, additional symmetries can enforce degeneracy
between magnon branches carrying opposite spin at the same
momentum. The conserved spin component along the ordering
axis distinguishes these branches. This degeneracy can
be lifted in altermagnets, producing a momentum-dependent
splitting.}

The appearance of chiral magnons is well known in collinear ferromagnets and
ferrimagnets, with the opposite polarizations of distinct magnon branches first
directly measured in \ch{Y3Fe5O12}~\cite{nambu2020observation} and magnon
chirality subsequently measured in the ferromagnet
\ch{SrRuO3}~\cite{jenni2022chirality}, while a momentum-dependent splitting of
opposite-chirality branches is a characteristic signature of
altermagnets~\cite{smejkal2023chiral,liu2024chiral}. Magnon chiralities can be
accessed with polarized inelastic neutron
scattering~\cite{mcclarty2025polarized}, as has been demonstrated more recently
in \ch{MnF2}~\cite{faure2025altermagnetism} and
\ch{MnTe}~\cite{liu2026switchable}.

\begin{table}[t]
\centering
\footnotesize
\setlength{\tabcolsep}{4pt}
\renewcommand{\arraystretch}{1.25}
\begin{tabular}{l l l l}
\toprule
Component & Parity in $\vec{q}$ & Required overlap & Wave vectors in $\Gamma$ \\
\midrule
\multicolumn{4}{l}{Collinear order} \\
$\st_1$ & even & $\Lambda^1\Gamma \cap \Gamma_\ell^+$ & $\vec{0}$ \\
$\vec{\st}\perp\vhat{N}_1$ & --- & --- & --- \\
\addlinespace
\multicolumn{4}{l}{Coplanar order, $\Lambda^2\Gamma = \det\Gamma$} \\
$\st_1$, $\st_2$ & even & $\Lambda^1\Gamma \cap \Gamma_\ell^+$ & $\vec{0}$ \\
$\st_3$ & odd & $\Lambda^2\Gamma \cap \Gamma_\ell^-$ & $\vec{k}$, $-\vec{k}$ \\
\addlinespace
\multicolumn{4}{l}{Non-coplanar order, $\Lambda^2\Gamma = \Gamma\otimes\det\Gamma$} \\
\multirow{2}{*}{$\vec{\st}$}
 & even & $\Lambda^1\Gamma \cap \Gamma_\ell^+$ & $\vec{0}$ \\
 & odd & $\Lambda^2\Gamma \cap \Gamma_\ell^-$ & $\vec{k}$, $-\vec{k}$, $\vec{K}$ \\
\bottomrule
\end{tabular}
\caption{Conditions for a rank-$\ell$ contribution to the momentum-space spin
texture $\vec{\st}(\vec{q})$, by type of magnetic order, with components
$\st_i = \vhat{N}_i\cdot\vec{\st}$ as defined in \cref{sec:spintexture}. The parity of
a component under $\vec{q} \rightarrow -\vec{q}$ fixes which multipoles it may
carry: an even component only even $\ell$, transforming as $\Gamma_\ell^+$, and
an odd component only odd $\ell$, transforming as $\Gamma_\ell^-$. Here $\cap$
denotes a nonzero overlap of representations and a dash a component that
vanishes identically by spin-only symmetry. The last column gives
the wave vectors the representation must contain, derived in
\cref{sec:exterior-prelim}: $\vec{k}$ is arbitrary, while $\vec{K}$ denotes a
time-reversal invariant momentum, $2\vec{K}\equiv 0$.}
\label{tab:spintexture}
\end{table}

\subsection{Edelstein Effect}
\label{sec:edelstein}

A closely related consequence of the momentum-space spin texture is the
Edelstein effect~\cite{edelstein1990spin,johansson2024theory}, or
current-induced spin polarization.~\footnote{Here we consider the uniform,
unit-cell-averaged nonequilibrium spin polarization of the current-carrying
electrons. This need not equal the change in the total spin of the magnet, which
also includes that of the ordered moments. In the non-relativistic limit the
total spin is conserved, so any such response is accompanied by an equal and
opposite change in the spin of the ordered moments. Local or sublattice-resolved
Edelstein responses may obey different symmetry constraints and are not
considered here.} We write its linear response as
\begin{equation}
\label{eq:edelstein}
M^\mu
=
\sum_\nu \alpha^\mu_\nu E_\nu .
\end{equation}
The Edelstein tensor and its Onsager partner, the inverse Edelstein
effect~\cite{shen2014microscopic}, may be combined into components that are even
and odd under time reversal, $\alpha^{\pm}$, as for the spin conductivity in
\cref{sec:spin-conductivity}. Although the Edelstein response has the same spatial
tensor structure as the equilibrium linear magnetoelectric
effect~\cite{dzyaloshinskii1960magnetoelectrical,astrov1960magnetoelectric,%
fiebig2005revival,spaldin2008toroidal}, whose conventional symbol $\alpha$ we
borrow for it, it is a nonequilibrium, or kinetic, magnetoelectric
response~\cite{levitov1985magnetoelectric}.~\footnote{Where symmetry permits a
uniform field-induced torque, the rate of change of the electronic spin entering
the spin continuity equation, that torque is closely related to the Edelstein
response. In a DC steady state this source of spin angular momentum is balanced
by relaxation, with the Edelstein spin density representing the resulting
stationary accumulation. A spatially uniform bulk source cannot be absorbed into
a torque-dipole redefinition of the spin current: such a redefinition requires
the bulk-averaged torque to vanish~\cite{shi2006proper}. Thus a homogeneous
Edelstein polarization may reflect local spin generation rather than the
accumulation of spin current flowing in from elsewhere.} In particular, its
time-even component may remain finite in a time-reversal-symmetric
system~\cite{levitov1985magnetoelectric,edelstein1990spin,%
gonzalez2024nonrelativistic}, whereas an
equilibrium magnetoelectric coefficient relating a time-even electric field
to a time-odd magnetization is necessarily time odd.

The symmetry conditions follow directly from the defining representation. For
non-coplanar order described by $\Gamma$, the time-odd and time-even components
transform with spin indices in $\Gamma$ and $\Gamma\otimes\det\Gamma$, and are
therefore controlled by $\Lambda^1\Gamma$ and $\Lambda^2\Gamma$ respectively.
Since the electric field transforms in the polar-vector representation
$\Gamma_1^-$, the time-odd component is allowed by symmetry when
\begin{equation}
    \alpha^-:\ \Gamma \cap \Gamma_1^- \neq 0.
\end{equation}
The time-even component is allowed when
\begin{equation}
    \alpha^+:\ \left(\Gamma\otimes\det\Gamma\right)
    \cap \Gamma_1^- \neq 0.
\end{equation}

For coplanar order the spin-only symmetry separates the two responses. The
time-odd induced spin lies within the ordering plane and is controlled by
$\Gamma$, whereas the time-even induced spin is perpendicular to the ordering
plane and is controlled by $\det\Gamma$~\cite{gonzalez2024nonrelativistic}. Thus
\begin{subalign}
    \alpha^-
    &: \qquad
    \Gamma \cap \Gamma_1^- \neq 0,
    \\
    \alpha^+
    &: \qquad
    \det\Gamma \cap \Gamma_1^- \neq 0.
\end{subalign}
For collinear order the spin-only symmetry forbids the time-even component
entirely, so only the time-odd component survives, with the condition $\Gamma
\cap \Gamma_1^- \neq 0$. In each case the condition on the time-odd component is
the same as that for the equilibrium linear magnetoelectric effect, which shares
its spatial tensor structure.

We summarize the conditions for collinear, coplanar and non-coplanar order as
\begin{cframed}
\textbf{Time-Odd Edelstein Effect:}
A spin space group defined by a real representation $\Gamma$ allows a
time-odd Edelstein response if and only if $\Gamma$ has nonzero overlap with the
polar-vector representation $\Gamma_1^-$.
\end{cframed}
\begin{cframed}
\textbf{Time-Even Edelstein Effect:}
A spin space group defined by a real representation $\Gamma$ allows a
time-even Edelstein response if and only if $\Lambda^2 \Gamma$ has nonzero
overlap with the polar-vector representation $\Gamma_1^-$.
\end{cframed}

The wave-vector content enters exactly as it does for the spin conductivities of
\cref{sec:spin-conductivity}. Since $\Gamma_1^-$ lives at $\vec{k}=0$, the time-odd
component is controlled by $\Lambda^1\Gamma$ and requires a $\vec{k}=0$
component of $\Gamma$, so a representation built entirely from finite
wave vectors permits no time-odd Edelstein response. The time-even
component is controlled by $\Lambda^2\Gamma$ and requires only a pair of
wave vectors that cancel.

The condition for the time-even Edelstein response is precisely the rank-one
case of the odd-in-momentum spin-texture criterion derived in
\cref{sec:spintexture}. For coplanar or non-coplanar order both require
$\Lambda^2 \Gamma$ to overlap with $\Gamma_1^-$. The symmetry conditions for the
two effects are therefore identical: a $p$-wave momentum-space spin texture is
allowed if and only if a time-even linear Edelstein response is allowed. The
forward direction of this equivalence underlies the large non-relativistic
Edelstein response found in nodal $p$-wave magnets~\cite{chakraborty2025highly}.

This gives a direct transport interpretation of the lowest-rank odd-parity spin
texture. Of the odd-rank multipoles of $\vec{\st}(\vec{q})$, only the $p$-wave
one has the momentum dependence of a polar vector, the representation
$\Gamma_1^-$ of the electric field driving the Edelstein response. Symmetry thus
ties a time-even spin polarization at linear order to the $p$-wave part of the
texture, and not to its $f$-wave or higher components. Odd-parity magnets
therefore divide into those with a $p$-wave component and those whose lowest
odd rank is higher, a distinction we return to in \cref{sec:oddparity}.

For collinear order the spin-only symmetry confines the time-odd response to the
ordering axis and forbids the time-even one. As for the spin conductivity, the
remaining response obeys a further restriction in simple models that does not
follow from the transformation laws. In the usual Bloch description the carriers
move in a fixed collinear exchange field, so their spin along $\vhat{N}$ is
conserved,
\begin{equation}
\label{eq:collinear-conservation}
[H,\vec{\st}\cdot\vhat{N}] = [H_E,\vec{\st}\cdot\vhat{N}] = 0,
\end{equation}
where $\vec{\st}$ is the carrier spin and $H_E$ the coupling to the electric
field. Its expectation value then cannot change to any order in the applied
field, and the uniform Edelstein response vanishes~\cite{leeb2026collinear}.
This conservation is not protected by symmetry. The residual spin rotations
about $\vhat{N}$ conserve only the total spin of the carriers and the ordered
moments together. Once the carriers can exchange spin angular momentum with the
ordered moments, for example through electron--magnon scattering, the carrier
spin is no longer conserved. A collinear magnet with $\Gamma\cap\Gamma_1^-\neq0$
should then generically show a time-odd Edelstein response, just as an
antisymmetric spin conductivity appears once the spin channels are coupled.

\section{Examples}
\label{sec:examples} 
We now illustrate the representation-first approach with several magnetic
materials hosting collinear, coplanar, and non-coplanar orders. In each case we
identify the parent space group and the real representation describing the
magnetic structure, and then infer the allowed physical responses directly from
its representation content. We include examples where the parent space group of
the spin space group differs from the paramagnetic space group $G_0$, as occurs when the order selects only part of a wave-vector star, so that each
$\vec{k}$-domain has its own parent group. We note that these conditions are symmetry statements about the allowed
tensor components and apply irrespective of the magnitude or microscopic origin
of the corresponding observables. For example, in insulating materials they may
be strongly suppressed, particularly at low temperature, but are subject to the
same symmetry constraints. A summary of these examples and the observables
allowed in each is given in \cref{tab:examples}.

\begin{table*}[t]
\centering
\footnotesize
\setlength{\tabcolsep}{2.7pt}
\begin{tabular}{l l l >{$}l<{$} l >{$}l<{$} l l l ccccc c}
\toprule
& & \multicolumn{2}{c}{Paramagnetic} & \multicolumn{2}{c}{Parent}
& \multicolumn{3}{c}{Exterior powers}
& \multicolumn{5}{c}{Allowed observables} & Texture \\
\cmidrule(lr){3-4} \cmidrule(lr){5-6} \cmidrule(lr){7-9} \cmidrule(lr){10-14}
\cmidrule(lr){15-15}
Compound & Order & Space group ($G_0$) & \vec{k} & Space group ($G$) & \vec{k}
& $\Lambda^1\Gamma$ & $\Lambda^2\Gamma$ & $\Lambda^3\Gamma$
& $\vec{M}$ & $\vec{C}$ & $\Sigma^{-/+}$ & $\sigma^-$
& $\alpha^{-/+}$ & $\vec{\st}(\vec{q})$ \\
\midrule
\ch{MnF2}
  & collinear & \sg{P4_2/mnm}{136} & \vec{0} & \sg{P4_2/mnm}{136} & \vec{0}
  & $B_{2g}$ & $0$ & $0$
  & \no & \yes & \yes/\no & \no & \no/\no & $d$ \\
\ch{MnTe}
  & collinear & \sg{P6_3/mmc}{194} & \vec{0} & \sg{P6_3/mmc}{194} & \vec{0}
  & $B_{2g}$ & $0$ & $0$
  & \no & \no & \no/\no & \no & \no/\no & $g$ \\
\ch{Mn3Ge}$^{\ddagger}$
  & coplanar & \sg{P6_3/mmc}{194} & \vec{0} & \sg{P6_3/mmc}{194} & \vec{0}
  & $E_{2g}$ & $A_{2g}$ & $0$
  & \no & \yes & \yes/\yes & \no & \no/\no & $d$ \\
\ch{FeF3}
  & non-coplanar & \sg{Fd\bar{3}m}{227} & \vec{0} & \sg{Fd\bar{3}m}{227} & \vec{0}
  & $T_{2g}$ & $T_{1g}$ & $A_{2g}$
  & \no & \yes & \yes/\yes & \no & \no/\no & $d$ \\
\CoTaS{}$^{\ddagger}$
  & non-coplanar & \sg{P6_322}{182} & M & \sg{P6_322}{182} & M
  & ${}^*M_1$ & ${}^*M_2$ & $A_2$
  & \no & \no & \no/\no & \yes & \no/\no & --- \\
\ch{Cr2O3}
  & collinear & \sg{R\bar{3}c}{167} & \vec{0} & \sg{R\bar{3}c}{167} & \vec{0}
  & $A_{2u}$ & $0$ & $0$
  & \no & \no & \no/\no & \no & \yes$^{\dagger}$/\no & --- \\
\ch{MnO}
  & collinear & \sg{Fm\bar{3}m}{225} & L & \sg{R\bar{3}m}{166} & T
  & $T_1^+$ & $0$ & $0$
  & \no & \no & \no/\no & \no & \no/\no & --- \\
\ch{Gd3(Ru_{0.95}Rh_{0.05})4Al12}$^{\ddagger}$
  & coplanar & \sg{P6_3/mmc}{194} & (\tfrac{1}{6},\tfrac{1}{6},0) & \sg{Cmcm}{63} & (\tfrac{1}{3},0,0)
  & ${}^*\Sigma_1$ & $B_{3u}$ & $0$
  & \no & \no & \no/\no & \no & \no/\yes & $p$ \\
\bottomrule
\end{tabular}
\caption{Summary of the examples of \cref{sec:mnf2,sec:mnte,sec:mn3ge,%
sec:fef3,sec:co13tas2,sec:cr2o3,sec:mno,sec:gdru}, listed in the order in which
they are discussed. A double dagger marks compounds treated in an idealized form
of the reported magnetic structure, as described in the corresponding section. For each compound we list the type of magnetic order, the
space group $G_0$ of the paramagnetic phase, the parent space group $G$ of the spin space
group, the propagation vector $\vec{k}$ referred to each of the two groups in
turn, and the exterior powers of the real representation $\Gamma=\Lambda^1\Gamma$ of $G$
that defines the spin space group, with a zero where a power vanishes by the
spin-only symmetry. The next five columns indicate whether the net
magnetization $\vec{M}$, the piezomagnetic coefficient $\vec{C}_{\mu\nu}$, the
spin conductivities $\Sigma^{-/+}$, the anomalous Hall conductivity $\sigma^-$,
and the Edelstein responses $\alpha^{-/+}$ are allowed (\yes) or forbidden (\no)
by symmetry in the non-relativistic limit. The paired $\Sigma^{-/+}$ and
$\alpha^{-/+}$ entries give the time-odd and time-even components in that order.
A dagger marks an entry allowed by the representation overlap that nonetheless
vanishes when the carrier spin is conserved, \cref{eq:collinear-conservation},
as for Bloch electrons in a fixed collinear exchange field. The last column gives the leading momentum-space
spin texture, labeled by the lowest multipole rank allowed by the conditions of
\cref{tab:spintexture}, with a dash where the texture vanishes identically. The finite-$\vec{k}$
examples are labeled in the Miller--Love notation~\cite{miller1967tables}.
Propagation vectors are given in the conventional reciprocal basis of the group
heading their column, and are quoted by their Brillouin-zone label where one
applies.}
\label{tab:examples}
\end{table*}

\subsection{\ch{MnF2}} 
\label{sec:mnf2}
The material \ch{MnF2} is a canonical example of an
altermagnet~\cite{smejkal2022beyond,smejkal2022emerging}. It crystallizes in the
tetragonal rutile structure with space group $P4_2/mnm$
(\#136)~\cite{stout1954crystal}, and has a collinear antiferromagnetic
order~\cite{erickson1953neutron} that appears below the N\'eel temperature $T_N
\approx 67$~K~\cite{heller1966nuclear}.

\begin{figure}[t]
\includegraphics[width=\columnwidth]{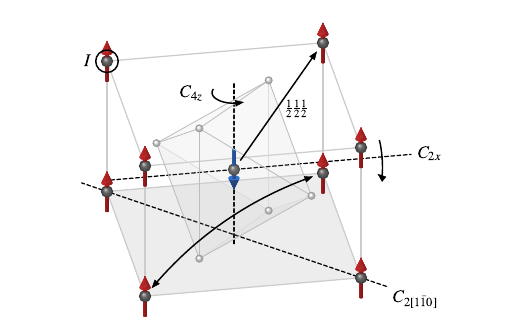}
\caption{The unit cell of \ch{MnF2}, with the \ch{MnF6} octahedron about the
center shaded. The corner (red) and center (blue) Mn carry opposite moments
along $c$. The screws $[C_{4z}|\tfrac{1}{2}\tfrac{1}{2}\tfrac{1}{2}]$ and
$[C_{2x}|\tfrac{1}{2}\tfrac{1}{2}\tfrac{1}{2}]$ take the central Mn to a
corner, so $\Gamma=-1$. The two-fold $C_{2[1\bar{1}0]}$ about the base diagonal
exchanges two corners, and the inversion $I$ through the ringed corner Mn fixes
it, so $\Gamma=+1$ for both.}
\label{fig:mnf2}
\end{figure}

Each symmetry element in the space group is associated with a spin-space
operation that either preserves or flips the spin direction and thus the parent
space group of the spin space group is the same as the paramagnetic space group $G_0$. The two Mn ions occupy the two sites of the $2a$ orbit, $(0,0,0)$ and
$(\tfrac{1}{2},\tfrac{1}{2},\tfrac{1}{2})$, so the one-dimensional
representation follows from whether each generator preserves or exchanges them.
Since $P4_2/mnm$ is non-symmorphic these must be specified as full Seitz
elements: the $4_2$ screw $[C_{4z}|\tfrac{1}{2}\tfrac{1}{2}\tfrac{1}{2}]$ and
the two-fold screws $[C_{2x}|\tfrac{1}{2}\tfrac{1}{2}\tfrac{1}{2}]$ about the
in-plane $x$ and $y$ axes exchange the two sublattices, whereas the two-fold
rotations $[C_{2[110]}|\vec{0}]$ about the in-plane diagonals and the inversion
$[I|\vec{0}]$ preserve them (\cref{fig:mnf2}). Hence
\begin{subalign}
\label{eq:mnf2-screw}
\Gamma([C_{4z}|\tfrac{1}{2}\tfrac{1}{2}\tfrac{1}{2}]) &= -1, \\
\label{eq:mnf2-c2x}
\Gamma([C_{2x}|\tfrac{1}{2}\tfrac{1}{2}\tfrac{1}{2}]) &= -1, \\
\label{eq:mnf2-c2d}
\Gamma([C_{2[110]}|\vec{0}]) &= +1, \\
\label{eq:mnf2-inversion}
\Gamma([I|\vec{0}]) &= +1,
\end{subalign}
which identify the representation as
\begin{equation}
\label{eq:mnf2-rep}
\Gamma = B_{2g}
\end{equation}
of $D_{4h}$, using the convention in which $B_{1g}$ and $B_{2g}$ transform as
$x^2-y^2$ and $xy$, respectively. Alternatively, we can obtain the same result
using the permutation representation of the two Mn sites, which is $A_{1g}
\oplus B_{2g}$~\cite{mcclarty2024landau,schiff2025collinear}, the $A_{1g}$ being
the trivial ferromagnet.

Since $\Gamma = B_{2g}$ does not contain the trivial irrep $A_{1g}$, the order
is compensated. The quadrupole representation in the $D_{4h}$ point group is
$\Gamma_2^+ = A_{1g} \oplus B_{1g} \oplus B_{2g} \oplus E_g$, so that $\Gamma$
overlaps with $\Gamma_2^+$ and piezomagnetism is allowed in the non-relativistic
limit~\cite{etxebarria2025crystal}. The overlap is with the single $B_{2g}$
component of $\Gamma_2^+$, corresponding to the shear strain $\epsilon_{xy}$, so
that the allowed coupling is $\vec{M} \propto \epsilon_{xy}\vhat{N}$: a
basal-plane shear generates a moment along the N\'eel axis. This component was
observed in \ch{MnF2} by \citet{baruchel1988piezomagnetism} indirectly through
domain selection in the presence of uniaxial strain. Earlier measurements by
Borovik-Romanov observed a different transverse piezomagnetic component in
\ch{MnF2}, which is absent in the strict non-relativistic
limit~\cite{borovik1960piezomagnetism}. The $d$-wave spin splitting and the
relativistic transverse piezomagnetic response of \ch{MnF2} have also been
discussed in terms of ferroically ordered magnetic
octupoles~\cite{bhowal2024ferroically}. The same $B_{2g}$ invariant, read as a
coupling between strain and magnetic field, underlies the elastocaloric response
recently used to probe the multipolar character of the order in
\ch{MnF2}~\cite{ohlendorf2026probing}. The time-odd spin conductivity is allowed
since $\Gamma$ overlaps with $\Gamma_2^+$, as for the isostructural
\ch{RuO2}~\cite{gonzalez2021efficient}. The time-even spin conductivity and the
anomalous Hall effect vanish since the order is collinear. The same
representation fixes the momentum-space spin texture. Since $\Gamma=B_{2g}$
occurs in the quadrupole representation, the leading texture is $d$-wave, with
$\vec{\st}(\vec{q})$ having the same symmetry as $q_xq_y\vhat{N}$ in the
convention used here [\cref{fig:textures}(a)]~\footnote{Equivalently, the momentum-space multipole
moment
$
\int_{\rm BZ} d^3 q\ f(\vec{q})\, \vec{\st}(\vec{q})\cdot\vhat{N}
$ is finite, where $f(\vec{q})$ is a lattice harmonic reducing to $q_xq_y$ near
the zone center.
}. 

\subsection{\ch{MnTe}}
\label{sec:mnte}
\ch{MnTe} is also a prominent candidate altermagnetic material. It crystallizes
in the NiAs structure with space group $P6_3/mmc$ (\#194)~\cite{oftedal1927uber},
and has a collinear antiferromagnetic order in which the moments are
ferromagnetically aligned within each triangular Mn layer and alternate in sign
between neighboring layers along the $c$
axis~\cite{greenwald1953antiferromagnetic,kunitomi1964neutron}, appearing below
the N\'eel temperature $T_N \approx 307$~K~\cite{zanmarchi1967optical}. It is a
leading example of a $g$-wave altermagnet~\cite{krempasky2024altermagnetic}. The
Mn ions occupy the two sites of the $2a$ orbit, at $z=0$ and $z=1/2$, so the
one-dimensional representation describing the order can be identified directly
from how the space-group operations act on these two magnetic sublattices.

\begin{figure}[t]
\includegraphics[width=\columnwidth]{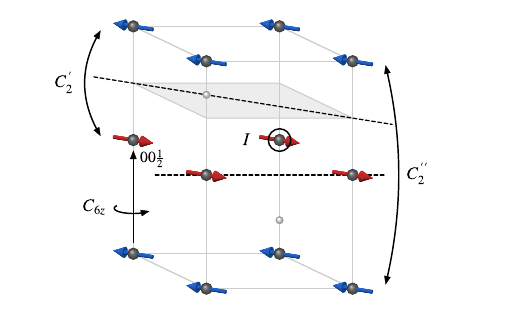}
\caption{The unit cell of \ch{MnTe} from $z=-\tfrac{1}{2}$ to $\tfrac{1}{2}$,
with Mn dark and Te light. The Mn layers are ferromagnetic and alternate
between the two sublattices, red at $z=0$ and blue at $z=\pm\tfrac{1}{2}$. The
two-fold $C_2''$ about $a$ in the red layer maps each layer onto itself, so
$\Gamma=+1$. The two-fold $C_2'$ about $[1\bar{1}0]$ in the shaded plane
$z=\tfrac{1}{4}$ exchanges red and blue layers, so $\Gamma=-1$. The screw
$[C_{6z}|00\tfrac{1}{2}]$ about the lower-left Mn column takes the blue Mn to
the red one above it, so $\Gamma=-1$. The inversion $I$ through the ringed Mn
fixes the red layer and exchanges the two blue ones, so $\Gamma=+1$.}
\label{fig:mnte}
\end{figure}

As for $P4_2/mnm$, the space group $P6_3/mmc$ is non-symmorphic and its elements
must be specified as full Seitz elements. The $6_3$ screw
$[C_{6z}|00\tfrac{1}{2}]$ maps the Mn layer at $z=0$ onto the layer at $z=1/2$
and therefore exchanges the two oppositely polarized sublattices. Consequently,
\begin{equation}
\label{eq:mnte-sixfold}
\Gamma([C_{6z}|00\tfrac{1}{2}])=-1.
\end{equation}
The inversion $[I|\vec{0}]$, on the other hand, maps each Mn layer back onto
itself, since $z=0$ is fixed and $z=1/2$ is mapped to $-1/2\equiv 1/2$, and
hence
\begin{equation}
\label{eq:mnte-inversion}
\Gamma([I|\vec{0}])=+1.
\end{equation}
Finally, the in-plane two-folds fall into two classes distinguished by their
fractional translations (\cref{fig:mnte}). Those about the $a$ axes, $[C_{2[100]}|\vec{0}]$, carry
no translation and map $z=0$ and $z=1/2$ each onto itself, whereas those about
the intervening in-plane axes, $[C_{2[1\bar{1}0]}|00\tfrac{1}{2}]$, rotated by
$30^\circ$ from the first set, exchange the two sublattices. Taking the latter
as the $C_2'$ class~\footnote{This choice follows the convention in the Bilbao
Crystallographic Server~\cite{aroyo2006bilbao}, which is also the convention
used by~\citet{schiff2025collinear}. \Citet{mcclarty2024landau} use the opposite
convention, in which $\Gamma=B_{1g}$. Switching conventions swaps $C_2'$ and
$C_2''$ and interchanges $B_{1g}$ and $B_{2g}$.}, we have
\begin{equation}
\label{eq:mnte-c2}
\Gamma([C_{2[1\bar{1}0]}|00\tfrac{1}{2}])=-1.
\end{equation}
These characters identify the one-dimensional representation of $D_{6h}$ as
\begin{equation}
\label{eq:mnte-rep}
\Gamma=B_{2g}.
\end{equation}
Alternatively, we can obtain the same result using the permutation
representation of the two Mn sites, which is $A_{1g} \oplus
B_{2g}$~\cite{mcclarty2024landau,schiff2025collinear}.

Since $\Gamma=B_{2g}$ does not contain the trivial irrep $A_{1g}$, the order is
compensated. In the $D_{6h}$ point group the axial-vector and quadrupole
representations are
\begin{subequations}
\label{eq:mnte-reps}
\begin{align}
\Gamma_1^+ &= A_{2g}\oplus E_{1g}, \\
\Gamma_2^+ &= A_{1g}\oplus E_{1g}\oplus E_{2g}.
\end{align}
\end{subequations}
Thus $B_{2g}$ has no overlap with either $\Gamma_1^+$ or $\Gamma_2^+$, and there
is no linear piezomagnetism or time-odd spin conductivity in the
non-relativistic limit. The time-even spin conductivity and anomalous Hall
conductivity vanish because the order is collinear; the spontaneous anomalous
Hall effect measured in \ch{MnTe}~\cite{betancourt2023spontaneous} thus
requires relativistic interactions or extrinsic perturbations that break the
protecting spin-only symmetry. The absence of overlap with the scalar,
axial-vector, and quadrupole representations is the representation-theory
signature of the $g$-wave character of \ch{MnTe}: the order first transforms
like a higher-rank spatial multipole. By the conditions of
\cref{sec:spintexture} the momentum-space texture is correspondingly $g$-wave.
With the present choice of crystallographic axes, the leading form of
$\vec{\st}(\vec{q})$ transforms as $q_yq_z(3q_x^2-q_y^2)\vhat{N}$, the $B_{2g}$
representation defining the order.

\subsection{\ch{Mn3Ge}} 
\label{sec:mn3ge}

The material \ch{Mn3Ge}, in its hexagonal $D0_{19}$ form, crystallizes in space
group $P6_3/mmc$ (\#194)~\cite{ohoyama1961xray,chen2020antichiral} and hosts a
coplanar $120^\circ$ antiferromagnetic order at
$\vec{k}=0$~\cite{nagamiya1982triangular,tomiyoshi1983triangular,soh2020ground},
appearing below the N\'eel temperature $T_N \approx
365$~K~\cite{chen2020antichiral}. The parent space group of its spin space group is
the same as the paramagnetic space group $G_0$. We have assumed an idealized
coplanar $120^\circ$ structure with equal moments; the observed state carries in
addition a small weak-ferromagnetic moment, an inevitable consequence of its
symmetry once relativistic effects are included~\cite{soh2020ground}.

\begin{figure}[t]
\includegraphics[width=\columnwidth]{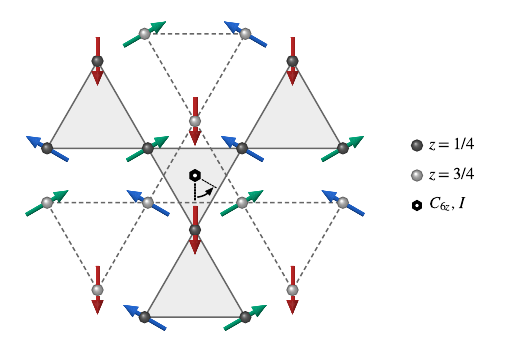}
\caption{The $120^\circ$ order of \ch{Mn3Ge} viewed along $c$. Inversion through the
center maps each site onto one with the same moment, so
$\mat{\Gamma}([I|\vec{0}])=\mat{1}$. The screw $[C_{6z}|00\tfrac{1}{2}]$ carries
each site onto the next one around the center (arc), whose moment is rotated by
$120^\circ$, so $\mat{\Gamma}([C_{6z}|00\tfrac{1}{2}])$ is a $120^\circ$
spin rotation.}
\label{fig:mn3ge}
\end{figure}

The two-dimensional representation describing the (idealized) order can be identified
directly from the action of the space-group generators on the magnetic
structure. The inversion $[I|\vec{0}]$ exchanges the two kagome layers and
relates sites with parallel moments, so that $\mat{\Gamma}([I|\vec{0}])=\mat{1}$
and the representation is even under inversion. The three-fold rotation
$[C_{3z}|\vec{0}]$ cyclically permutes the three Mn sublattices within each
layer, and since the moments on successive sublattices differ by an angle of
$2\pi/3$, $\mat{\Gamma}([C_{3z}|\vec{0}])$ is a spin rotation through
$\mp 2\pi/3$, with trace $-1$. The $6_3$ screw $[C_{6z}|00\tfrac{1}{2}]$ maps
the kagome layer at $z=1/4$ onto that at $z=3/4$ and permutes the three moment
directions in the opposite cyclic order (\cref{fig:mn3ge}), so
$\mat{\Gamma}([C_{6z}|00\tfrac{1}{2}]) = \mat{\Gamma}([C_{3z}|\vec{0}])^{-1}$ is
a spin rotation through $\pm 2\pi/3$, and its trace is also $-1$. Together with
the even inversion parity, the character under the six-fold operation
distinguishes the representation as
\begin{equation}
\label{eq:mn3ge-rep}
\Gamma=E_{2g},
\end{equation}
rather than $E_{1g}$, for which
$\tr\mat{\Gamma}([C_{6z}|00\tfrac{1}{2}])=+1$.~\footnote{This $E_{2g}$ is the
representation $\Gamma$ that defines the spin space group, and describes how the
spin-space operations act. It should not be confused with the $E_{1g}$ symmetry
reported for the ordered structure in \citet{soh2020ground}, which labels the
full transformation of the magnetic moments themselves, spin part included, as
in the conventional magnetic representation used for \ch{FeF3} in
\cref{sec:fef3}.} We note that the ``triangular'' and ``inverse triangular''
$120^\circ$ states (distinguished by the sign of their vector spin chirality),
which carry distinct magnetic irreps once relativistic effects are included,
differ only by a global spin rotation and therefore share the same $\Gamma$.

Since $\Gamma=E_{2g}$ does not contain the trivial irrep $A_{1g}$, the order is
compensated. The quadrupole representation in the $D_{6h}$ point group is
\begin{equation}
\label{eq:mn3ge-quad}
\Gamma_2^+ = A_{1g} \oplus E_{1g} \oplus E_{2g},
\end{equation}
so that $\Gamma$ overlaps with $\Gamma_2^+$ and the order exhibits
piezomagnetism. This has been observed in the isostructural \ch{Mn3Sn}, where
the strain-induced moment has been used to switch the anomalous Hall effect at
room temperature~\cite{ikhlas2022piezomagnetic}. The time-odd spin conductivity
is allowed for the same reason, consistent with previous analyses of
spin-polarized currents in non-collinear antiferromagnets~\cite{zel2017spin}. For the time-even spin conductivity, the
relevant representation is
\begin{equation}
\label{eq:mn3ge-det}
\Lambda^2\Gamma = \det \Gamma = A_{2g}.
\end{equation}
Since $A_{2g}$ overlaps with the axial-vector representation $\Gamma_1^+$, the
time-even spin conductivity is also allowed. Related time-even spin-current
responses have been discussed previously in non-collinear
antiferromagnets~\cite{zhang2018spin}. The anomalous Hall
conductivity vanishes in the non-relativistic limit because the order is
coplanar; the large anomalous Hall effect observed in
\ch{Mn3Ge}~\cite{nayak2016large,kiyohara2016giant} is thus necessarily of
relativistic origin. The spin texture lies in the ordering plane and is $d$-wave.
Choosing a
suitable basis $\vhat{N}_1,\vhat{N}_2$ of the ordering plane, the leading form
of $\vec{\st}(\vec{q})$ transforms as
$(q_x^2-q_y^2)\vhat{N}_1+2q_xq_y\vhat{N}_2$ [\cref{fig:textures}(b)]; the perpendicular odd component
vanishes since $\det\Gamma=A_{2g}$ is even under inversion.

\subsection{Pyrochlore \ch{FeF3}}
\label{sec:fef3}
The material \ch{FeF3}, in its metastable pyrochlore form, crystallizes in space
group $Fd\bar{3}m$ (\#227), in which the \ch{Fe^3+} ions form a
three-dimensional network of corner-sharing tetrahedra~\cite{depape1986new}.
Geometrical frustration suppresses long-range order to $T_N \approx
15.5$~K~\cite{reimers1991short}, below which the moments adopt a $\vec{k}=0$
non-collinear all-in--all-out (AIAO) structure~\cite{ferey1986ordered}; the
selection of this state by a positive biquadratic term has been studied
microscopically~\cite{sadeghi2015spin}. The four moments on each tetrahedron
point along the four local $\langle 111\rangle$ axes,
\begin{equation}
\label{eq:fef3-local-axes}
\begin{split}
\vhat{n}_0 &= \frac{1}{\sqrt{3}}(+1,+1,+1), \qquad
\vhat{n}_1 = \frac{1}{\sqrt{3}}(+1,-1,-1),\\
\vhat{n}_2 &= \frac{1}{\sqrt{3}}(-1,+1,-1), \qquad
\vhat{n}_3 = \frac{1}{\sqrt{3}}(-1,-1,+1),
\end{split}
\end{equation}
with the two magnetic domains corresponding to
$\vec{S}_a=\pm S\vhat{n}_a$.

\begin{figure}[t]
\includegraphics[width=\columnwidth]{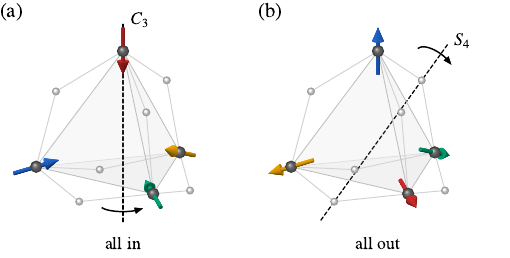}
\caption{One \ch{Fe4} tetrahedron of pyrochlore \ch{FeF3}, with an F bridging
each edge. (a)~The all-in order is unchanged by the three-fold rotation about
the axis through one Fe, so $C_3$ has character $+1$. (b)~Its image under the
improper rotation $S_4$ about $z$, each moment moved with its site and turned as
an axial vector (colors follow the sites), is all-out.
Since $S_4=IC_4^{-1}$ and inversion leaves the order unchanged, $C_4$ has
character $-1$.}
\label{fig:fef3}
\end{figure}

It is useful first to identify the conventional magnetic representation of this
state, in which spatial and spin rotations act together. A three-fold rotation
about a $\langle 111\rangle$ axis preserves the AIAO pattern
[\cref{fig:fef3}(a)]: it fixes one local axis and cyclically permutes the other
three. By contrast, a four-fold rotation about a cubic axis reverses the
pattern [\cref{fig:fef3}(b)]. For example, under $C_{4z}$,
\begin{equation*}
\vhat{n}_0 \rightarrow -\vhat{n}_1,\qquad
\vhat{n}_1 \rightarrow -\vhat{n}_3,\qquad
\vhat{n}_3 \rightarrow -\vhat{n}_2,\qquad
\vhat{n}_2 \rightarrow -\vhat{n}_0,
\end{equation*}
so that, together with the corresponding permutation of the four Fe sublattices,
the two AIAO domains are exchanged. Inversion leaves the magnetic pattern
unchanged, since the Fe magnetic sublattices are mapped onto themselves modulo a
Bravais translation and the magnetic moments are axial vectors. The
corresponding one-dimensional order parameter therefore has characters
\begin{equation}
\label{eq:fef3-a2g-characters}
\chi(C_3)=+1,\qquad
\chi(C_4)=-1,\qquad
\chi(I)=+1,
\end{equation}
which identify the conventional AIAO magnetic representation as $A_{2g}$ of
$O_h$.

The representation relevant for the spin space group is not this one-dimensional
magnetic representation. In the conventional description the spin transforms as
an axial vector, which carries the representation
\begin{equation}
\label{eq:fef3-axial}
\Gamma_1^+=T_{1g}.
\end{equation}
Interpreting this as a spin space group, as for any magnetic space group
(\cref{sec:msg}), we have
\begin{equation}
\label{eq:fef3-gamma}
\Gamma
=
A_{2g}\otimes T_{1g}
=
T_{2g}.
\end{equation}
 Since $\Gamma=T_{2g}$ contains no copy of the trivial irrep $A_{1g}$, the order
is compensated. The quadrupole representation in the $O_h$ point group is
\begin{equation}
\label{eq:fef3-quad}
\Gamma_2^+=E_g\oplus T_{2g},
\end{equation}
so that $\Gamma$ overlaps with $\Gamma_2^+$ and the order exhibits
piezomagnetism. The time-odd spin conductivity is allowed for the same reason.

The higher exterior powers follow from $\Gamma = A_{2g}\otimes T_{1g}$. The
axial-vector representation has unit determinant for every spatial operation,
and multiplying a three-dimensional representation by the one-dimensional
$A_{2g}$ multiplies its determinant by the cube of that character, and hence by
the character itself, so that
\begin{subalign}
\label{eq:fef3-det}
\Lambda^3\Gamma &= \det\Gamma = A_{2g}, \\
\label{eq:fef3-twist}
\Lambda^2\Gamma &= \Gamma\otimes\det\Gamma = T_{2g}\otimes A_{2g} = T_{1g}.
\end{subalign}
The time-even spin conductivity is therefore allowed, since
$\Lambda^2\Gamma = T_{1g}$ overlaps with the axial-vector representation
$\Gamma_1^+$. The anomalous Hall conductivity vanishes in the non-relativistic
limit, since $\Lambda^3\Gamma = A_{2g}$ does not, as required for any cubic
parent point group (\cref{sec:hall}). In the spin frame used above,
the conductivity tensors take the particularly simple forms
\begin{align*}
\Sigma^{\mu,-}_{\nu\rho}&\propto |\varepsilon_{\mu\nu\rho}|,&
\Sigma^{\mu,+}_{\nu\rho}&\propto \varepsilon_{\mu\nu\rho},
\end{align*}
where we adopt $\varepsilon_{xyz}=+1$. Thus the time-odd conductivity is
symmetric in its spatial indices, while the time-even conductivity is
antisymmetric, with one independent coefficient in each channel. Pyrochlore
\ch{FeF3} thus provides a simple $\vec{k}=0$ non-coplanar example in which both
time-even and time-odd spin conductivities are allowed while the anomalous Hall
conductivity is forbidden.

The leading momentum-space spin texture is a
three-component $d$-wave. Up to a global choice of spin frame
$\vhat{N}_1,\vhat{N}_2,\vhat{N}_3$ the leading form of $\vec{\st}(\vec{q})$
transforms as $q_yq_z\vhat{N}_1+q_zq_x\vhat{N}_2+q_xq_y\vhat{N}_3$, the standard
$T_{2g}$ quadrupole [\cref{fig:textures}(c)]. Odd-rank contributions are
forbidden since $\Gamma\otimes\det\Gamma=T_{1g}$ is even under inversion.

\subsection{\CoTaS{}}
\label{sec:co13tas2}
The material \CoTaS{} crystallizes in space group $P6_322$ (\#182),
with the Co ions forming triangular layers~\cite{park2022field}. The
low-temperature magnetic state, below $T_{N2} \approx
26.5$~K~\cite{park2022field}, is a non-coplanar tetrahedral triple-$Q$ order
involving the three symmetry-related $M$
points~\cite{park2023tetrahedral,takagi2023spontaneous}
\begin{align}
\label{eq:co13tas2-wavevectors}
\vec{Q}_1 &= (\tfrac{1}{2},0,0),&
\vec{Q}_2 &= (0,\tfrac{1}{2},0),&
\vec{Q}_3 &= (\tfrac{1}{2},-\tfrac{1}{2},0).
\end{align}
Choosing three orthogonal spin directions $\vhat{N}_1$, $\vhat{N}_2$, and
$\vhat{N}_3$, the tetrahedral state may be written in the form
\begin{equation}
\label{eq:co13tas2-3q}
\vec{S}(\vec{r})
=
\frac{S}{\sqrt{3}}
\sum_{n=1}^{3}
\vhat{N}_n \cos(\vec{Q}_n\cdot\vec{r}),
\end{equation}
with the origin chosen on a Co site. The second Co layer of the unit cell
follows from the first by the $6_3$ screw, with scalar spin chirality of the same
sign in both layers, the ferro-chiral stacking argued for by
\citet{park2023tetrahedral}. We use the equal-amplitude tetrahedral state of
\cref{eq:co13tas2-3q} as an ideal representative. Recent work reports
coexisting chirality and nematicity in the low-temperature, low-field phase,
consistent with non-equilateral triple-$Q$ order~\cite{kirstein2026tunable}; we
return to such distortions below.

\begin{figure}[t]
\includegraphics[width=\columnwidth]{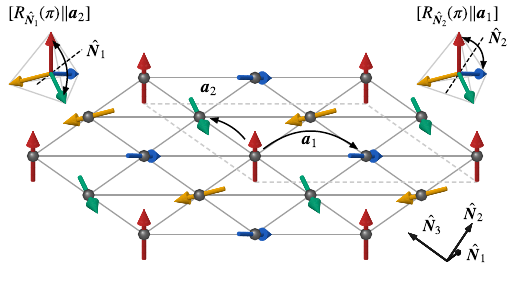}
\caption{The idealized tetrahedral triple-$Q$ order of \CoTaS{},
\cref{eq:co13tas2-3q}, on one triangular Co layer, with its four moment
directions in the insets and the spin frame $\vhat{N}_1,\vhat{N}_2,\vhat{N}_3$
at lower right.
Translation by $\vec{a}_1$ takes each site to one whose moment differs by
the spin rotation $R_{\vhat{N}_2}(\pi)$, so the order is invariant under
$\protect\ssop{R_{\vhat{N}_2}(\pi)}{\vec{a}_1}$ (right inset), giving
$\Gamma([\mat{1}|\vec{a}_1])=\mathrm{diag}(-1,+1,-1)$. Likewise
$\protect\ssop{R_{\vhat{N}_1}(\pi)}{\vec{a}_2}$ (left inset) gives
$\Gamma([\mat{1}|\vec{a}_2])=\mathrm{diag}(+1,-1,-1)$. The gray dashed outline
marks the $2\times2$ magnetic cell.}
\label{fig:co13tas2}
\end{figure}

Since each $\vec{Q}_n$ is a time-reversal invariant $M$ point, translations act on the
three spin components only by signs. For primitive translations $\vec{a}_1$ and
$\vec{a}_2$, a convenient choice of basis gives
\begin{subalign} 
\label{eq:co13tas2-t1}
\Gamma([\mat{1}|\vec{a}_1])
&=
{\rm diag}(-1,+1,-1),\\
\label{eq:co13tas2-t2}
\Gamma([\mat{1}|\vec{a}_2])
&=
{\rm diag}(+1,-1,-1).
\end{subalign}
The four possible sign patterns generated by these translations give the four
spin directions 
\begin{equation}
\label{eq:co13tas2-tetrahedron}
(+,+,+),\qquad
(-,+,-),\qquad
(+,-,-),\qquad
(-,-,+),
\end{equation}
which are the vertices of a regular tetrahedron (\cref{fig:co13tas2}). This is a version of the
tetrahedral triple-$Q$ order discussed in
Refs.~\cite{martin2008itinerant,messio2011} on the two-dimensional triangular
lattice.

The point-group operations act by permuting the three arms of the $M$-point star
and hence simultaneously permuting the three spin directions $\vhat{N}_n$. In
particular, the six-fold operation cyclically permutes the three wave vectors,
while an in-plane two-fold operation exchanges two of them. Together with the
translation matrices above, this defines the real three-dimensional \emph{star}
representation associated with the three $M$ points, rather than a
three-dimensional representation at a single wave vector. In the conventional
Miller--Love notation~\cite{miller1967tables} it is ${}^*M_1$, the
representation induced from the one-dimensional little-group irrep $M_1$ at
$\vec{k}=M$. For this symmetric
triple-$Q$ state, permutations of the three equally weighted star arms are
implemented by orthogonal transformations in spin space, allowing the full paramagnetic space group to be retained as the parent, in contrast to the single-$\vec{k}$ orders of \ch{MnO} and
\ch{Gd3(Ru_{0.95}Rh_{0.05})4Al12} below.

\begin{figure}[t]
\includegraphics[width=\columnwidth]{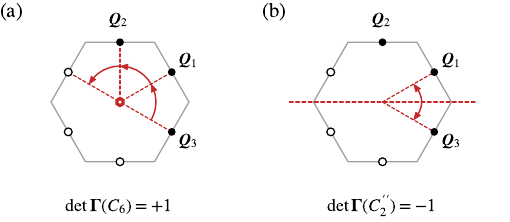}
\caption{The $M$-point arms $\vec{Q}_n$ of \CoTaS{} (filled) and
their equivalent opposites (open). For the ferro-chiral stacking,
$\det\Gamma$ is the sign of the permutation of the arms. (a)~The six-fold rotation cycles them, so
$\det\mat{\Gamma}(C_6)=+1$. (b)~The two-fold about the dashed axis exchanges
$\vec{Q}_1$ and $\vec{Q}_3$ and takes $\vec{Q}_2$ to its opposite, so
$\det\mat{\Gamma}(C_2'')=-1$.}
\label{fig:co13tas2-arms}
\end{figure}

Its determinant can be fixed without constructing the full representation
matrices: for this stacking, $\det\mat{\Gamma}(g)$ is simply the parity of the
permutation of the three arms by $g$ (\cref{fig:co13tas2-arms}). Translations
permute no arms, and indeed have determinant $+1$, since
\begin{equation}
\label{eq:co13tas2-qsum}
\vec{Q}_1+\vec{Q}_2+\vec{Q}_3
\equiv \vec{0}
\end{equation}
modulo a reciprocal lattice vector. An operation that keeps each Co layer in
place can be written as a symmetry fixing the Co site chosen as the origin,
followed by a lattice translation. The site symmetry simply permutes the three
cosine modes, and hence the $\vhat{N}_n$, while the translation has determinant
$+1$. The determinant of the combined operation is therefore the parity of the
arm permutation. Every other operation is one of these composed with the $6_3$
screw, which permutes the arms cyclically, so the rule holds throughout provided
$\det\mat{\Gamma}(6_3)=+1$. This last sign is fixed by the stacking. The $6_3$
screw is a rotation about $\vhat{c}$ that carries each triangle of one layer to
a triangle of the other with the same orientation. Since the scalar chirality
transforms with $\det\mat{\Gamma}(g)$ [\cref{eq:chirality-det-gamma}],
$\det\mat{\Gamma}(6_3)=+1$ exactly when the two layers have chirality of the
same sign, as here. A six-fold rotation, a three-cycle, therefore has
determinant $+1$, whereas an in-plane two-fold rotation exchanges two arms and
has determinant $-1$. These are precisely the characters of $A_2$ in the $D_6$
point group, and hence
\begin{equation}
\label{eq:co13tas2-det}
\Lambda^3\Gamma = \det\Gamma = A_2.
\end{equation}

Since $\Gamma$ is built entirely from finite-wave-vector components at the $M$
point, the order is compensated, with no linear piezomagnetism or time-odd spin
conductivity (\cref{sec:magnetization}). For the time-even spin conductivity the
relevant representation is $\Lambda^2\Gamma=\Gamma\otimes\det\Gamma={}^*M_2$. Since $\det\Gamma=A_2$ is
a $\vec{k}=0$ representation, tensoring by it does not change the wave-vector
content of $\Gamma$, and therefore the time-even spin conductivity vanishes as
well.

The anomalous Hall conductivity however is different. The order is non-coplanar,
so it is not forbidden by a spin-only symmetry, and its symmetry is controlled
by the top exterior power $\Lambda^3\Gamma = \det\Gamma$. In $D_6$ the $A_2$
irrep is the out-of-plane component of the axial-vector representation. Since
$\det\Gamma=A_2$, an anomalous Hall conductivity in the plane, or equivalently a
Hall pseudovector along the $c$ axis, is allowed even in the non-relativistic
limit, consistent with the large spontaneous Hall conductivity $\sigma_{xy}$
measured in the triple-$Q$
phase~\cite{park2023tetrahedral,takagi2023spontaneous}. In contrast, the
momentum-space spin texture vanishes identically. The representation $\Gamma$
contains only the finite-$M$ components, while $\Lambda^2\Gamma =
\Gamma\otimes\det\Gamma$ has the same wave-vector content, since
$\det\Gamma=A_2$ lives at $\vec{k}=0$, so neither even- nor odd-rank momentum
multipoles are allowed despite the finite anomalous Hall response. This allowed
Hall response and the vanishing spin magnetization and spin conductivities were
also obtained from spin-group symmetry by \citet{watanabe2024symmetry}, who
noted that the Hall response occurs without non-relativistic spin--momentum
locking.

The precise stacking matters only for the Hall response. Reversing every moment of the
second layer gives a tetrahedral order with the same parent space group, but
with scalar chirality of opposite sign in the two layers. This reverses
$\mat{\Gamma}(g)$ for every operation that exchanges the layers, replacing
$\Gamma$ by $\Gamma\otimes B_1={}^*M_4$, where $B_1$ is the irrep of $D_6$ that
is $-1$ on those operations. The exterior square $\Lambda^2\Gamma={}^*M_2$ is
unchanged, so this order too is compensated without spin conductivities or a
spin texture. But since $\det\Gamma=A_2\otimes B_1=B_2$ does not occur in
$\Gamma_1^+=A_2\oplus E_1$, its anomalous Hall response vanishes. The
measured $\sigma_{xy}$ thus also selects the ferro-chiral stacking.

This separation is not specific to this tetrahedral order. More generally, for any
three-dimensional $\Gamma$ with no $\vec{k}=0$ component but with $\det\Gamma$
trivial under translations, $\Lambda^2\Gamma = \Gamma\otimes\det\Gamma$ also has
no $\vec{k}=0$ component. Both spin conductivities and the momentum-space spin
texture therefore vanish, while an anomalous Hall response remains possible when
$\det\Gamma$ overlaps with $\Gamma_1^+$. In particular, these translation-enforced
prohibitions do not require equal amplitudes. If the three components in
\cref{eq:co13tas2-3q} have unequal but nonzero amplitudes and remain mutually
orthogonal, the translations still act as in
\cref{eq:co13tas2-t1,eq:co13tas2-t2}, so both spin conductivities and the
momentum-space spin texture still vanish.

\subsection{\ch{Cr2O3}}
\label{sec:cr2o3}

\begin{figure}[t]
\includegraphics[width=\columnwidth]{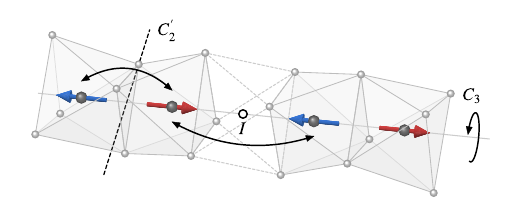}
\caption{A column of octahedra along the three-fold axis $C_3$ of \ch{Cr2O3},
drawn horizontally. Pairs of face-sharing \ch{CrO6} octahedra alternate with empty ones
(dashed), and the moments alternate along the axis. The two-fold axis $C_2'$ in
the shared face exchanges the two Cr of a pair, and the inversion center $I$ in
the empty octahedron exchanges Cr across it. Both join opposite moments, so
$\Gamma(C_2')=\Gamma(I)=-1$.}
\label{fig:cr2o3}
\end{figure}

The material \ch{Cr2O3} crystallizes in the corundum structure with space group
$R\bar{3}c$ (\#167)~\cite{newnham1962refinement}, and has a collinear
antiferromagnetic order at
$\vec{k}=0$~\cite{brockhouse1953antiferromagnetic,corliss1965magnetic} below the
N\'eel temperature $T_N \approx 307$~K~\cite{mcguire1956antiferromagnetism}. The
Cr moments lie along the rhombohedral axis and alternate in sign between the two
magnetic sublattices. The one-dimensional representation follows directly from
how the generators act on these sublattices. A three-fold rotation about the
rhombohedral axis preserves them, while the in-plane two-fold rotation and the
inversion each exchange them. These two act on different pairs of Cr sites (\cref{fig:cr2o3}): the
two-fold axes of $R\bar{3}c$ lie at $z=\tfrac{1}{4}$ rather than through the
origin, so $C_2'$ relates $(0,0,z)$ to $(0,0,\tfrac{1}{2}-z)$, the near neighbor
across the shared octahedral face, whereas the inversion relates it to
$(0,0,-z)$. Thus
\begin{subalign}
\label{eq:cr2o3-c3}
\Gamma(C_3) &= +1, \\
\label{eq:cr2o3-c2}
\Gamma(C_2') &= -1, \\
\label{eq:cr2o3-inversion}
\Gamma(I) &= -1.
\end{subalign}
These characters identify the representation as
\begin{equation}
\label{eq:cr2o3-rep}
\Gamma=A_{2u}
\end{equation}
of the $D_{3d}$ point group, with $\Gamma(S_6)=-1$ and $\Gamma(\sigma_d)=+1$ for
the remaining improper operations.

Since $\Gamma=A_{2u}$ does not contain the trivial irrep $A_{1g}$, the order is
compensated. The quadrupole representation in the $D_{3d}$ point group is
\begin{equation}
\label{eq:cr2o3-quad}
\Gamma_2^+ = A_{1g} \oplus 2E_g,
\end{equation}
which has no overlap with $A_{2u}$. Thus there is no linear piezomagnetism in
the non-relativistic limit. Similarly, the axial-vector representation is
\begin{equation}
\label{eq:cr2o3-axial}
\Gamma_1^+ = A_{2g} \oplus E_g,
\end{equation}
so $A_{2u}$ has no overlap with $\Gamma_0^+$, $\Gamma_1^+$, or $\Gamma_2^+$. The
time-odd spin conductivity therefore vanishes, while the time-even spin
conductivity and anomalous Hall conductivity vanish because the order is
collinear. Thus \ch{Cr2O3} provides a simple example of a compensated collinear
antiferromagnet that, unlike \ch{MnF2}, is not piezomagnetic in the
non-relativistic limit. The momentum-space spin texture also vanishes in the
non-relativistic limit. Collinearity requires it to be even in momentum, while
$\Gamma=A_{2u}$ is odd under inversion, so $\Gamma$ cannot occur in any of the
allowed even-rank momentum multipoles. The same inversion parity does permit the
tensor structure of a time-odd Edelstein response: the polar-vector
representation of $D_{3d}$ is $\Gamma_1^-=A_{2u}\oplus E_u$, which contains
$\Gamma$. This is the condition of \cref{sec:edelstein}, and equivalently that
for the equilibrium linear magnetoelectric effect, for which \ch{Cr2O3} is the
archetypal
material~\cite{dzyaloshinskii1960magnetoelectrical,astrov1960magnetoelectric}.
In the non-relativistic limit the uniform Edelstein response vanishes in simple
models in which the carrier spin is conserved, such as Bloch electrons in a fixed
collinear exchange field, but is generically allowed once the carriers can
exchange spin with the ordered moments (\cref{sec:edelstein}). The equilibrium
magnetoelectric coefficient is not restricted in this way. For example, exchange striction and thermal spin
fluctuations produce a finite longitudinal equilibrium magnetoelectric response
in \ch{Cr2O3} without spin-orbit coupling~\cite{mostovoy2010temperature}.

\subsection{\ch{MnO}}
\label{sec:mno}

\begin{figure}[t]
\includegraphics[width=\columnwidth]{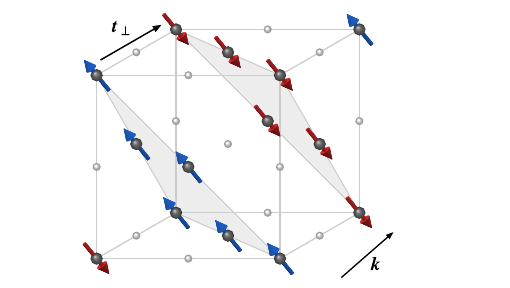}
\caption{The unit cell of \ch{MnO}, with Mn dark and O light.
The moments are ferromagnetic within $(111)$ planes, two of which are shaded,
and alternate between them. The propagation vector $\vec{k}=L$ is normal to the
planes, and a translation $\vec{t}_{\perp}$ between neighboring planes reverses
the moments,
\cref{eq:mno-layer-translation}.}
\label{fig:mno}
\end{figure}

The material \ch{MnO} crystallizes in the rocksalt structure with space group
$Fm\bar{3}m$ (\#225)~\cite{kuriyama1962xray} and develops the type-II collinear
antiferromagnetic
order~\cite{shull1951neutron,roth1958magnetic,goodwin2006magnetic} with
propagation vector
\begin{equation}
\label{eq:mno-k}
\vec{k}=L=(\tfrac{1}{2},\tfrac{1}{2},\tfrac{1}{2}).
\end{equation}
The magnetic structure (\cref{fig:mno}) consists of ferromagnetic $(111)$ planes whose moments
alternate in sign between neighboring planes below the N\'eel temperature $T_N
\approx 118$~K~\cite{roth1958magnetic,morosin1970exchange}, accompanied by a
rhombohedral contraction along the selected $\langle 111 \rangle$
axis~\cite{morosin1970exchange}.

Unlike the previous examples, for a single $\vec{k}$-domain the parent space group is not $G_0$. The cubic point group relates the
four arms of the $L$-point star,
\begin{equation}
\label{eq:mno-l-star}
(\tfrac{1}{2},\tfrac{1}{2},\tfrac{1}{2}),\quad
(-\tfrac{1}{2},\tfrac{1}{2},\tfrac{1}{2}),\quad
(\tfrac{1}{2},-\tfrac{1}{2},\tfrac{1}{2}),\quad
(\tfrac{1}{2},\tfrac{1}{2},-\tfrac{1}{2}),
\end{equation}
up to reciprocal lattice vectors. A single-$Q$ domain selects one of these arms
and therefore breaks the operations that permute it with the other three. The
subgroup of $Fm\bar{3}m$ that preserves the selected $\vec{k}=(1/2,1/2,1/2)$ is
the index-four rhombohedral subgroup $G=R\bar{3}m$ (\#166). Keeping
$Fm\bar{3}m$ would instead require the four-dimensional star representation,
which acts on the whole order-parameter space spanned by the four arms, including
multi-$\vec{k}$ states, rather than on any one ordered state.

Within this reduced parent group the collinear order is described by a real
one-dimensional representation at the finite wave vector $\vec{k}$. Its
translation character follows directly from the modulation of the
antiferromagnetic state,
\begin{equation}
\label{eq:mno-translation}
\Gamma([\mat{1}|\vec{t}])
=
e^{i\vec{k}\cdot\vec{t}}.
\end{equation}
Since the $L$ point is a time-reversal invariant wave vector, $2\vec{k}\equiv
0$, these characters are real and take only the values $\pm1$. In particular, a
translation connecting neighboring ferromagnetic $(111)$ planes changes the sign
of the order parameter,
\begin{equation}
\label{eq:mno-layer-translation}
\Gamma([\mat{1}|\vec{t}_{\perp}])=-1,
\end{equation}
which encodes the alternating stacking of the type-II antiferromagnet. In the
notation of the $R\bar{3}m$ group this is the $T$ point and the associated
representation is the one-dimensional representation $T^+_1$ in the Miller--Love
convention~\cite{miller1967tables}, with no star since the star of $T$ in $R\bar{3}m$ has a single arm.

Several physical properties then follow without requiring the detailed
point-group content of $\Gamma$. Since the representation is at finite
wave vector, the order is compensated, with no linear piezomagnetism or
time-odd spin conductivity (\cref{sec:magnetization}).
The time-even spin conductivity vanishes identically because the order is
collinear. Thus all of these conclusions follow directly from the dimensionality
and wave-vector content of the representation, without needing to assign a spin
space group label. The spin texture also vanishes identically for \ch{MnO}. The
defining one-dimensional representation lives at the finite $T$ point and
therefore cannot overlap any $\vec{k}=0$ momentum multipole, while the collinear
spin-only symmetry excludes the odd-momentum sector altogether.
 
\subsection{\ch{Gd3(Ru_{0.95}Rh_{0.05})4Al12}}
\label{sec:gdru}
The material \ch{Gd3(Ru_{0.95}Rh_{0.05})4Al12} is a Rh-substituted variant of
the breathing-kagome magnet \ch{Gd3Ru4Al12}~\cite{hirschberger2019skyrmion}. It
crystallizes in space group $P6_3/mmc$ (\#194)~\cite{gladyshevskii1993structure}
and hosts a commensurate coplanar spin helix below $T_{N2} \approx 9$~K with
propagation vector~\cite{yamada2025metallic}
\begin{equation}
\label{eq:gdru-k-hex}
\vec{k}=(\tfrac{1}{6},\tfrac{1}{6},0).
\end{equation}
A single-$\vec{k}$ helix selects one of the three symmetry-related pairs
$\{\vec{k},-\vec{k}\}$ in the hexagonal Brillouin zone. The full $P6_3/mmc$
space group therefore cannot be the parent space group of its spin
space group: acting with the full point group generates all three pairs and
hence a six-dimensional star representation, as the full cubic group does for
\ch{MnO} (\cref{sec:mno}). Restricting instead to the subgroup
that preserves the selected pair gives the index-three orthorhombic subgroup
\begin{equation}
\label{eq:gdru-parent}
G=Cmcm
\end{equation}
(\#63), in whose reciprocal-lattice coordinates the propagation vector is
$\vec{k}=(\tfrac{1}{3},0,0)$.

The origin of the two-dimensional representation is particularly transparent for
a helix. Let $\hat{\vec N}_1$ and $\hat{\vec N}_2$ span the spin rotation plane.
The magnetic structure may be written as
\begin{equation}
\label{eq:gdru-helix}
\vec{S}(\vec{r})
=
S\left[
\hat{\vec N}_1\cos(\vec{k}\cdot\vec{r})
+
\hat{\vec N}_2\sin(\vec{k}\cdot\vec{r})
\right].
\end{equation}
We have idealized the experimentally refined structure by this circular
single-harmonic helix, neglecting its small distortions and the accompanying
magnetization, $\approx 2\%$ of the saturated
moment~\cite{yamada2025metallic}. The pair of functions
$\{\cos(\vec{k}\cdot\vec{r}), \sin(\vec{k}\cdot\vec{r})\}$ therefore furnishes a
real two-dimensional representation of $Cmcm$. Under a translation $\vec{t}$
this basis transforms as
\begin{equation}
\label{eq:gdru-translation-rep}
\mat{\Gamma}([\mat{1}|\vec{t}])
=
\begin{pmatrix}
\cos(\vec{k}\cdot\vec{t})
&
-\sin(\vec{k}\cdot\vec{t})
\\
\sin(\vec{k}\cdot\vec{t})
&
\cos(\vec{k}\cdot\vec{t})
\end{pmatrix}.
\end{equation}
Thus the spatial phase advance of the helix is represented directly as a
rotation in spin space. Since the phase of the helix is $\vec{k}\cdot\vec{r}$
itself, an operation $[\mat{R}|\vec{t}]$ with $\mat{R}\vec{k}=\vec{k}$ acts in
the same way as its translation alone, so the little group of $\vec{k}$ carries
the irrep in which every point operation acts trivially. In the Miller--Love
notation~\cite{miller1967tables} this is $\Sigma_1$, with $\Sigma=(u,0,0)$ the
line of $Cmcm$ containing $\vec{k}$, and $\Gamma$ is the two-dimensional star
representation ${}^*\Sigma_1$ induced from it.

\begin{figure}[t]
\includegraphics[width=\columnwidth]{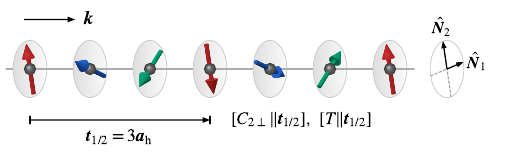}
\caption{The helix of \ch{Gd3(Ru_{0.95}Rh_{0.05})4Al12}, \cref{eq:gdru-helix},
on a row of Gd sites related by $\vec{a}_{\mathrm{h}}$, drawn against position
along $\vec{k}$. The moments turn in the plane normal to $\vec{k}$ (disks). The
phase advances by $\pi/3$ per $\vec{a}_{\mathrm{h}}$, so the translation
$\vec{t}_{1/2}=3\vec{a}_{\mathrm{h}}$ reverses every moment. This gives the
$p$-wave symmetries $\protect\ssop{C_{2\perp}}{\vec{t}_{1/2}}$ and
$\protect\ssop{T}{\vec{t}_{1/2}}$.}
\label{fig:gd-helix}
\end{figure}

For the commensurate state the magnetic unit cell is six chemical unit cells, so
the translation $\vec{t}_{1/2}=3\vec{a}_{\mathrm{h}}$ by three hexagonal lattice
vectors has $\vec{k}\cdot\vec{t}_{1/2}=\pi$, and
$\mat{\Gamma}([\mat{1}|\vec{t}_{1/2}])=-\mat{1}$, which in three-dimensional
spin space is a two-fold rotation about the normal to the ordering plane
(\cref{fig:gd-helix}). Combined with the spin-only symmetry of coplanar order,
this gives the half-period symmetries $\ssop{C_{2\perp}}{\vec{t}_{1/2}}$ and
$\ssop{T}{\vec{t}_{1/2}}$ used to characterize $p$-wave
magnets~\cite{hellenes2023p,yamada2025metallic}. The results below do not rely
on them: they follow from $\Lambda^2\Gamma$ for any coplanar
$\{\vec{k},-\vec{k}\}$ order, commensurate or not (\cref{sec:coplanar-odd}).

\begin{figure}[t]
\includegraphics[width=\columnwidth]{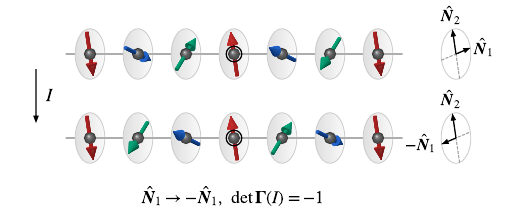}
\caption{The helix of \cref{fig:gd-helix} (top) and its image under inversion
about the circled site (bottom), each moment carried with its site. The image is
the original helix with $\vhat{N}_1\to-\vhat{N}_1$ (frames at right), a
reflection in spin space, so $\det\mat{\Gamma}(I)=-1$.}
\label{fig:gd-inversion}
\end{figure}

Several consequences also follow directly from the representation. Since
$\Gamma$ is constructed entirely from the finite wave vectors $\vec{k}$ and
$-\vec{k}$, the order is compensated, with no linear piezomagnetism or time-odd
spin conductivity (\cref{sec:magnetization}). The time-even spin conductivity
instead depends on the one-dimensional representation $\det\Gamma$, which can be
determined without any further work. In the basis $\{\cos(\vec{k}\cdot\vec{r}),
\sin(\vec{k}\cdot\vec{r})\}$, any operation $g = [\mat{R}|\vec{t}]$ with
$\mat{R}\vec{k} \equiv -\vec{k}$ acts as
\begin{equation}
\label{eq:gdru-kflip}
\mat{\Gamma}(g)
=
\begin{pmatrix}
\cos{(\vec{k}\cdot\vec{t})} & \sin{(\vec{k}\cdot\vec{t})} \\
\sin{(\vec{k}\cdot\vec{t})} & -\cos{(\vec{k}\cdot\vec{t})}
\end{pmatrix}
\end{equation}
and therefore has $\det\mat{\Gamma}(g) = -1$, independently of the phase
$\vec{k}\cdot\vec{t}$ carried by its translation part. Since $Cmcm$ is
centrosymmetric and inversion sends $\vec{k} \rightarrow -\vec{k}$, we have
\begin{equation}
\label{eq:gdru-det-inversion}
\det\mat{\Gamma}(I) = -1,
\end{equation}
so $\det\Gamma$ is odd under inversion (\cref{fig:gd-inversion}). Thus an
operation that preserves $\vec{k}$ acts as a rotation and has determinant $+1$,
while one that reverses $\vec{k}$ acts as a reflection and has determinant $-1$.
The determinant is therefore the one-dimensional representation recording
whether an operation preserves or reverses $\vec{k}$, which is precisely how the
polar-vector component along $\vec{k}$ transforms. Since $\vec{k} \parallel
\vec{a}$, we have $\det\Gamma = B_{3u}$ of $D_{2h}$, transforming as $q_x$ in
the convention with $x \parallel \vec{a}$ and $z \parallel \vec{c}$. As
$\Gamma_0^+$, $\Gamma_1^+$ and $\Gamma_2^+$ are all even, there is no overlap
and the time-even spin conductivity vanishes as well. The anomalous Hall
conductivity vanishes since the order is coplanar, so the uniform magnetization,
the piezomagnetic response, the spin conductivities and the anomalous Hall
conductivity considered above are all forbidden in the non-relativistic limit.

The momentum-space spin texture, controlled by $\Lambda^2\Gamma = \det\Gamma$,
is purely perpendicular to the ordering plane and begins at $p$-wave order. With
$\det\Gamma = B_{3u}$ transforming as $q_x$, this gives $\vec{\st}(\vec{q})$
transforming as $q_x(\vhat{N}_1\times\vhat{N}_2)$ to leading order
[\cref{fig:textures}(d)], while the even in-plane texture vanishes. Being the
rank-one case, this also permits a time-even Edelstein response, by
\cref{sec:edelstein}. The contrast with the time-even spin conductivity is
instructive: both are controlled by $\det\Gamma$, but the former needs an
inversion-odd multipole and the latter an inversion-even one, so the
inversion-odd $\det\Gamma$ allows one and forbids the other.

\subsection{Survey of the MAGNDATA Database}
\label{sec:magndata}
The steps followed in each of these examples are mechanical. The parent space
group and the defining representation are extracted from the magnetic structure
(\cref{app:constructing-gamma}), and each condition of \cref{tab:summary} is an
average, over the operations of the parent space group, of a product of
characters. We have implemented these steps in a
short program, published with this paper~\cite{zenodo-code}, which accepts any
commensurate magnetic structure in the magCIF format, and have applied it to the 2318
commensurate magnetic structures in the MAGNDATA
database~\cite{gallego2016magndata,aroyo2006bilbao}, of which 1390 are
collinear, 601 coplanar and 327 non-coplanar. The results are tabulated in the
Supplemental Material~\cite{supplemental}, which also describes the method in detail. Each
structure is analyzed as it is reported, so the entries for the compounds that
\cref{tab:examples} treats in an idealized form can differ from it, as the
Supplemental Material describes.

\section{Discussion}

Having illustrated the approach on specific materials, we now place it in a
broader context. The same representation-first viewpoint applies to conventional
magnetic space groups, each of which is encoded in a real one-dimensional
representation of its parent space group (\cref{sec:msg}). For collinear order,
the defining representation is the Landau order-parameter irrep of the magnetic
transition when that irrep is one-dimensional (\cref{sec:landau}). We then use
the exterior powers to characterize odd-parity magnets in general, including
their wave-vector content and inversion parity (\cref{sec:oddparity}), and
compare the resulting texture conditions with existing spin space group
classifications (\cref{sec:textureclass}). We close with the scope and
limitations of the approach (\cref{sec:scope}).

\subsection{Magnetic Space Groups}
\label{sec:msg}
A similar representation-first viewpoint can be advocated for the usual magnetic
space groups (MSGs), excluding the type II (grey) groups, which do not describe
spontaneous breaking of time-reversal symmetry. Recall that each MSG defines a
real one-dimensional irrep of the parent space group $G$,
\begin{equation}
\label{eq:msg-sign}
\Phi: G \rightarrow \mathbb{Z}_2,
\end{equation}
with the sign indicating the presence or absence of time-reversal symmetry.
These are subgroups of $\mathbb{Z}_2 \times G$ instead of the $\ot{} \times G$
we have for the spin space groups. The corresponding operation for $g \equiv
[\mat{R}|\vec{t}]$ takes the form (in the spin space group notation)
\begin{equation}
\label{eq:msg-as-ssg}
\ssop{\Phi(g)\det(\mat{R})\mat{R}}{g}.
\end{equation}
This is the three-dimensional axial-vector representation $\Gamma_1^+$ times the
one-dimensional irrep $\Phi$, which thus can be interpreted as a spin space
group with $\Gamma=\Phi \otimes \Gamma^+_1$. As there is no spin-only group this
should be interpreted within the non-coplanar framework. The usual
classification of non-trivial magnetic space groups into types I, III, IV is
then recovered by the following conditions on $\Phi$. Type I groups have $\Phi =
\Gamma^+_0$, the trivial irrep. Type III groups have $\Phi \neq \Gamma^+_0$ but
$\vec{k}=0$, while type IV groups have $\vec{k}\neq
0$~\cite{litvin2013magnetic}.
  
\begin{table*}[t]
\centering
\footnotesize
\setlength{\tabcolsep}{4pt}
\renewcommand{\arraystretch}{1.25}
\begin{tabular}{@{}l l c l l c c c@{}}
\toprule
& & & & & \multicolumn{3}{c@{}}{MSG type} \\
\cmidrule(l){6-8}
\multicolumn{2}{@{}l}{Response} & Symbol
  & Spin space group ($\Gamma$) & Magnetic space group ($\Phi$)
  & I & III & IV \\
\midrule
\multicolumn{2}{@{}l}{Net spin magnetization} & $\vec{M}$
  & $\Gamma \cap \Gamma_0^+$
  & $\Phi \cap \Gamma_1^+$
  & \maybe & \maybe & \no \\
\multicolumn{2}{@{}l}{Piezomagnetism} & $\vec{C}_{\mu\nu}$
  & $\Gamma \cap (\Gamma_0^+ \oplus \Gamma_2^+)$
  & $\Phi \cap (\Gamma_1^+ \oplus \Gamma_2^+ \oplus \Gamma_3^+)$
  & \maybe & \maybe & \no \\
\multirow{2}{*}{Spin conductivity}
  & Time-odd & $\Sigma^-$
  & $\Gamma \cap (\Gamma_0^+ \oplus \Gamma_1^+ \oplus \Gamma_2^+)$
  & $\Phi \cap (\Gamma_0^+ \oplus \Gamma_1^+ \oplus \Gamma_2^+ \oplus \Gamma_3^+)$
  & \yes & \maybe & \no \\
  & Time-even & $\Sigma^+$
  & $\Lambda^2\Gamma \cap (\Gamma_0^+ \oplus \Gamma_1^+ \oplus \Gamma_2^+)$
  & Always
  & \yes & \yes & \yes \\
\multicolumn{2}{@{}l}{Anomalous Hall conductivity} & $\sigma^-$
  & $\Lambda^3\Gamma \cap \Gamma_1^+$
  & $\Phi \cap \Gamma_1^+$
  & \maybe & \maybe & \no \\
\multirow{2}{*}{Edelstein effect}
  & Time-odd & $\alpha^-$
  & $\Gamma \cap \Gamma_1^-$
  & $\Phi \cap (\Gamma_0^- \oplus \Gamma_1^- \oplus \Gamma_2^-)$
  & \maybe & \maybe & \no \\
  & Time-even & $\alpha^+$
  & $\Lambda^2\Gamma \cap \Gamma_1^-$
  & $\Gamma_1^+ \cap \Gamma_1^-$
  & \maybe & \maybe & \maybe \\
\midrule
\multirow{2}{*}{Spin texture}
  & Even rank & $\vec{\st}(\vec{q})$
  & $\Gamma \cap \Gamma_\ell^+$
  & $(\Phi\otimes\Gamma_1^+) \cap \Gamma_\ell^+$
  & \maybe & \maybe & \no \\
  & Odd rank & $\vec{\st}(\vec{q})$
  & $\Lambda^2\Gamma \cap \Gamma_\ell^-$
  & $\Gamma_1^+ \cap \Gamma_\ell^-$
  & \maybe & \maybe & \maybe \\
\bottomrule
\end{tabular}
\caption{Symmetry conditions for each observable, stated in the spin space group
language of \cref{sec:observables} and translated into the magnetic space group
language of this section. A magnetic space group corresponds to $\Gamma = \Phi
\otimes \Gamma_1^+$. Piezomagnetism is listed in its general form, which reduces
to the $\Gamma\cap\Gamma_2^+$ of \cref{tab:summary} for compensated order. Types
are as defined in the text, $\Phi = \Gamma_0^+$ for type I, $\Phi \neq
\Gamma_0^+$ at $\vec{k}=0$ for type III, and $\vec{k}\neq 0$ for type IV. The
last three columns record whether the response is allowed for every magnetic
space group of that type (\yes), forbidden for that type whatever the point
group (\no), or allowed subject to the condition in the preceding column
(\maybe).}
\label{tab:msg}
\end{table*}

Our results can be translated immediately to this language, and are collected in
\cref{tab:msg}. A magnetic space group is compensated if and only if $\Phi$ does
not overlap with $\Gamma^+_1$. It is piezomagnetic if and only if $\Phi$
overlaps with the pseudovector, quadrupole or octupole representations since
$\Gamma^+_1 \otimes \Gamma^+_2 = \Gamma^+_1 \oplus \Gamma^+_2 \oplus
\Gamma^+_3$, and the scalar part $\Gamma_0^+$ of the strain contributes only a
further $\Gamma_1^+$. For either quantity to be finite requires that
$\vec{k}=0$, i.e. that the magnetic space group is type I or type III. This
condition is less prescriptive than it appears, since the sum of the three
irreps $\Gamma^+_1 \oplus \Gamma^+_2 \oplus \Gamma^+_3$ contains every (even)
one-dimensional irrep of the point group except for the cubic groups $O$, $T_d$
and $O_h$ where the trivial irrep is missing. Thus only type IV MSGs, type I
MSGs with point group $O$, $T_d$ or $O_h$, and centrosymmetric type III MSGs
where $\Phi$ is inversion odd can fail to be piezomagnetic.

Next we can consider transport. Since $\Phi^2$ and $\det\Gamma_1^+$ are trivial,
$\det\Gamma = \det(\Phi\otimes\Gamma_1^+) = \Phi^3\det\Gamma_1^+ = \Phi$, so the
time-even part of the spin conductivity is always allowed: $\Gamma \otimes \det
\Gamma = \Gamma^+_1$. This is expected, as the time-even spin conductivity is
allowed by symmetry even in free space so long as relativistic effects are
included~\cite{dyakonov1971possibility,dyakonov1971current,sinova2015spin}. The
time-odd part has a non-trivial, but fairly relaxed condition that $\Phi$
overlaps with $\Gamma^+_0\oplus\Gamma^+_1\oplus\Gamma^+_2\oplus\Gamma^+_3$. This
requires $\Phi$ to lie at $\vec{k}=0$, so that the group is of type I or III,
and for every crystallographic point group it holds whenever $\Phi$ is even
under inversion. Only type IV groups, and type III groups with inversion-odd
$\Phi$, can therefore lack a time-odd spin conductivity.

Like in the spin space group case, the symmetric part of the thermal or charge
conductivity is always allowed. The antisymmetric Hall part is allowed if
$\det\Gamma = \Phi$ overlaps with $\Gamma^+_1$, which is the same as the
condition for lack of compensation. Note that $\Gamma^+_1$ is purely
three-dimensional for the cubic point groups $T$, $T_h$, $O$, $T_d$ and $O_h$,
and so contains no one-dimensional $\Phi$; no magnetic space group with a cubic
point group can therefore be uncompensated or exhibit an anomalous Hall
response. This is a key difference from the spin space group case, where the
antisymmetric part of the charge conductivity can be allowed even for
compensated magnets.

Substituting $\Gamma=\Phi\otimes\Gamma_1^+$ and $\det\Gamma=\Phi$ into the
momentum-space spin-texture condition gives
\begin{equation}
\Gamma\otimes(\det\Gamma)^\ell
=
\Gamma_1^+\otimes\Phi^{\ell+1}.
\end{equation}
Since $\Phi^2=\Gamma_0^+$, the conditions for even- and odd-rank textures
separate. For even $\ell \geq 2$, a contribution is allowed when
$\Phi\otimes\Gamma_1^+$ overlaps with $\Gamma_\ell^+$, or equivalently when
$\Phi$ overlaps with
$\Gamma_{\ell-1}^+\oplus\Gamma_\ell^+\oplus\Gamma_{\ell+1}^+$. For odd $\ell$,
by contrast, the dependence on $\Phi$ drops out entirely and the condition
reduces to an overlap between $\Gamma_1^+$ and $\Gamma_\ell^-$. Since
$\Gamma_1^+$ is even and $\Gamma_\ell^-$ odd under inversion, this overlap is
possible only in a point group without inversion. Thus even-in-momentum spin
textures probe the magnetic representation $\Phi$, while odd-in-momentum
textures are controlled entirely by the crystallographic symmetry.

The lowest ranks reproduce familiar response conditions. At $\ell=0$ the
condition is precisely that for a net magnetization, while at $\ell=2$ it
becomes
\begin{equation}
\Phi
\subseteq
\Gamma_1^+\oplus\Gamma_2^+\oplus\Gamma_3^+,
\end{equation}
which is exactly the condition for piezomagnetism. Thus an MSG admits a $d$-wave
spin texture if and only if it admits piezomagnetism, and higher even ranks
extend this same magnetic-multipole hierarchy~\cite{radaelli2024tensorial}.

For type IV magnetic space groups, $\Phi$ lives at finite wave vector, so all
even-rank contributions are forbidden and only odd-in-momentum spin textures can
occur. In a centrosymmetric parent group the spin texture therefore vanishes
identically.

The Edelstein conditions reduce in the same way. The time-odd component requires
$\Phi$ to be contained in $\Gamma_0^- \oplus \Gamma_1^- \oplus \Gamma_2^-$, and
thus to live at $\vec{k}=0$, so only types I and III can exhibit it; in a
centrosymmetric group $\Phi$ must in addition be odd under inversion. The
time-even component is the $\ell=1$ case of the odd-rank texture condition, and
so loses its dependence on $\Phi$ entirely, reducing to $\Gamma_1^+ \cap
\Gamma_1^- \neq 0$. This condition is independent of the MSG and requires a
non-centrosymmetric point group, but not every such group qualifies:
$\Gamma_1^-$ and $\Gamma_1^+$ share no irrep in $C_{3h}$, $D_{3h}$ and $T_d$, as
\cref{tab:vector-quadrupole} shows. The remaining eighteen non-centrosymmetric
point groups are the gyrotropic ones, and for those the time-even Edelstein
response is permitted for every type.

\subsection{Collinear Magnets and Landau Theory}
\label{sec:landau}
For collinear order the representation-first description is closely tied to an
object that is already standard: the Landau order-parameter irrep of the
magnetic transition, with which it coincides when that irrep is one-dimensional.
This equivalence was implicit in \citet{mcclarty2024landau} and demonstrated
explicitly in \citet{schiff2025collinear}. \Citet{schiff2026landau} extend the
Landau description to multi-dimensional irreps and to coplanar and non-coplanar
order. We summarize the collinear argument here.

As set out in \cref{eq:collinear-gamma}, a collinear spin space group is fixed
entirely by a single real one-dimensional representation $\Gamma$ of the parent
space group $G$. The same representation appears if one instead writes down a
Landau theory for the transition. In the non-relativistic limit the free energy
is invariant under $\ot{} \times G_0$, and the order parameter has one component
for each dimension of the Landau irrep of $G_0$, each a vector in spin space.
For collinear order these all lie along a common unit vector $\vhat{N}$. Spin
rotations act only on $\vhat{N}$, so the entire spatial transformation law is
carried by the amplitudes of the components, which are scalars under spin
rotations. If the Landau irrep is one-dimensional there is a single amplitude
$\phi$, every operation of $G_0$ maps $\phi$ to $\pm\phi$, and so $G=G_0$ and
the Landau irrep is $\Gamma$ itself. If it is multi-dimensional, as for a
wave-vector star with more than one arm or a two- or three-dimensional irrep at
$\vec{k}=0$, the ordered state selects a direction in the space of amplitudes, and
$\phi$ is the amplitude along that direction. The parent space group $G<G_0$ is
then the subgroup that maps $\phi$ to $\pm\phi$, and $\Gamma$ is the sign with
which it does so. For \ch{MnO} the Landau irrep is the four-dimensional star
representation of the $L$ point of $Fm\bar{3}m$, and $\Gamma$ is the
one-dimensional $T_1^+$ of $R\bar{3}m$ (\cref{sec:mno}).

The practical consequence is that the input required by the results of
\cref{sec:mnf2,sec:mnte,sec:cr2o3,sec:mno} is precisely the output of a
conventional representation analysis of the ordered state, as routinely
performed when refining a magnetic structure~\cite{perezmato2015symmetry}.
Decomposing the permutation representation of the magnetic sites into irreps of
$G_0$, the real one-dimensional irreps are the candidate regular collinear
orders, with $G=G_0$, and each is a candidate $\Gamma$. This is the route taken
in \cref{sec:mnf2,sec:mnte}, where the permutation representations $A_{1g}
\oplus B_{2g}$ of \ch{MnF2} and $A_{1g} \oplus B_{2g}$ of \ch{MnTe} identify
$\Gamma$ immediately, with the trivial irrep in each case corresponding to the
ferromagnet. The systematic development of these Landau theories, and their
associated physical consequences for collinear altermagnets, is given
in~\citet{schiff2025collinear}.

Seen this way, many of the boxed criteria of \cref{sec:observables} are
statements about invariants of the Landau free energy. Compensation is the
requirement that no term linear in the net magnetization is allowed, and the
piezomagnetic condition that $\Gamma$ overlaps $\Gamma_2^+$ is the existence of
a trilinear invariant coupling the order parameter, the strain, and the
magnetization along $\vhat{N}$ -- which is why the allowed coupling in \ch{MnF2}
takes the form $\vec{M} \propto \epsilon_{xy} \vhat{N}$. For collinear magnets,
then, no spin space group need be constructed at any stage: the Landau irrep one
already has gives the representation the construction requires, directly when it
is one-dimensional and by restriction to $G$ otherwise. For coplanar or
non-coplanar order arising from a single irrep, however, that irrep is
necessarily multi-dimensional, and the relation to the underlying spin space
group is more delicate, as developed in \citet{schiff2026landau}.

\subsection{Odd-Parity Magnets}
\label{sec:oddparity}
We define odd-parity magnets (including the so-called $p$-wave
magnets~\cite{hellenes2023p,brekke2024minimal}) as magnets whose momentum-space
spin texture is purely odd under momentum inversion.~\footnote{The $p$- and
$f$-wave labels are not used uniformly in the literature; some authors define
them by the parity of the band splitting rather than by that of the spin
polarization~\cite{hirschmann2026nodal}.} Textures of this kind have appeared in
other contexts, notably in the theory of Fermi-liquid instabilities, where an
odd-parity order parameter in the particle--hole spin channel spin-splits the
Fermi surface without relativistic spin-orbit
coupling~\cite{wu2007,fradkin2010nematic,kiselev2017}. Our representation-first
approach allows a systematic characterization of this class of odd-parity
magnets, in particular their wave-vector content and inversion parity, directly
from the properties of the defining representation $\Gamma$.

Based on the results of \cref{sec:spintexture}, collinear order cannot support
any odd-rank spin texture, whatever the parent space
group~\cite{hellenes2023p,song2026}, in the absence of other forms of
time-reversal symmetry breaking~\cite{leeb2026collinear,zeng2026oddparity}. For
coplanar and non-coplanar order alike the even-rank texture is controlled by
$\Lambda^1\Gamma$ and the odd-rank texture by $\Lambda^2\Gamma$, so a purely
odd-parity texture requires
\begin{equation}
\Lambda^1\Gamma \cap \Gamma^+_{\ell} = 0
\end{equation}
for all even $\ell$, while
\begin{equation}
\Lambda^2\Gamma \cap \Gamma^-_{\ell} \neq 0
\end{equation}
for at least one odd $\ell$. The smallest odd $\ell$ for which the second
condition is satisfied defines the wave character of the magnet: $\ell=1$ gives
a $p$-wave magnet, $\ell=3$ an $f$-wave magnet, and so on through the odd
multipoles. These conditions place strong constraints on the representation
$\Gamma$ and hence on the spin space group. Which irreps can appear in a
momentum multipole of given rank and parity is established in
\cref{app:parity-multipoles}.

The rank-one condition, $\Lambda^2\Gamma \cap \Gamma^-_1 \neq 0$, is the
condition for a time-even linear Edelstein response found in
\cref{sec:edelstein}. Among odd-parity magnets, the $p$-wave ones are thus
exactly those in which an electric field can generate a time-even spin
polarization at linear order, a non-relativistic charge-to-spin
conversion~\cite{chakraborty2025highly}. Those whose lowest odd rank is $f$-wave
or higher admit no such response, as found for the $h$-wave texture of coplanar
order in the Fe-based superconductors~\cite{dsouza2026odd}.

\subsubsection{Coplanar Odd-Parity Magnets}
\label{sec:coplanar-odd}
Consider coplanar order, for which the defining representation $\Gamma$ is
two-dimensional and
\begin{equation}
    \Lambda^2\Gamma=\det\Gamma.
\end{equation}
A purely odd-parity spin texture requires that $\Gamma$ have no overlap with any
even-rank momentum multipole, while $\det\Gamma$ must overlap with at least one
odd-rank multipole. Since $\Lambda^2\Gamma=\det\Gamma$ is also the
representation carried by the vector spin chirality $\vec{S}_1\times\vec{S}_2$,
this second condition connects directly with the odd-parity order parameter of
\citet{yu2025odd}, and with the microscopic result that the spin polarization
normal to the ordering plane is proportional to the real-space cross product of
the moments~\cite{mitscherling2026microscopic}.

We first constrain the wave-vector content. To have an odd-parity texture
$\Lambda^2 \Gamma = \det \Gamma$ must have a $\vec{k}=0$ component and so the
wave-vector content must be $\{\vec{k},-\vec{k}\}$ including the special case of
two components at the same TRIM, $\vec{k} \equiv -\vec{k} \equiv \vec{K}$. The
case $\vec{k}=0$ cannot produce a purely odd-parity texture. If the parent point
group does not contain inversion, every irrep occurs in some even-rank momentum
multipole, so $\Gamma$ necessarily has an even-rank overlap. If inversion is
present, absence of every even-rank contribution requires every component of
$\Gamma$ to be odd under inversion. But for a two-dimensional representation
this makes $\det\Gamma$ even under inversion, and hence unable to overlap any
odd-rank multipole $\Gamma_\ell^-$. If $\vec{k} \neq 0$ then the even-rank
overlap is automatically forbidden, independent of the presence of inversion,
and the appearance of the odd-parity texture is controlled by the point group.
In short, a coplanar purely odd-parity magnet must have wave-vector content
$\{\vec{k},-\vec{k}\}$ with $\vec{k} \neq 0$.

The remaining condition is a point-group one:
\begin{equation}
    \det\Gamma \cap \Gamma_\ell^- \neq 0
\end{equation}
for at least one odd $\ell$. For the non-centrosymmetric case ($P$ does not
contain inversion) only the wave-vector content is relevant, since every irrep
appears in some odd-rank multipole and in some even-rank multipole. This is the
case for the brownmillerite
\ch{Ca2Cr2O5}~\cite{arevalolopez2015crystal,gallego2016magndata}, whose $\Gamma
= 2X_1$ of $Ima2$ has $\Lambda^2\Gamma = \det\Gamma = A_1$, the polar-axis
component of $\Gamma_1^-$, and so a $p$-wave texture. In the centrosymmetric
case ($P$ contains inversion) the point-group condition is non-trivial. For a
pair $\{\vec{k},-\vec{k}\}$ where $\vec{k} \not\equiv -\vec{k}$ inversion only
swaps $\vec{k}$ and $-\vec{k}$ (it is not in the little group of $\vec{k}$) and
so $\det \mat{\Gamma}(I) = -1$ and hence it overlaps some odd-rank multipole.
For two components at the same TRIM $\{\vec{K},\vec{K}\}$ inversion can act
non-trivially and allow or forbid the odd-rank overlap depending on the
inversion parity of $\det \Gamma$. Both outcomes can occur. For example, in
\ch{GeNi2O4}~\cite{matsuda2008frustrated,gallego2016magndata} the two components
of $\Gamma = T_1^+ \oplus T_2^-$ of $R\bar{3}m$ have opposite inversion parity,
so that $\Lambda^2\Gamma = \det\Gamma = A_{2u}$ and the texture is $p$-wave,
whereas in \ch{LiFeGe2O6}~\cite{redhammer2009magnetic,gallego2016magndata}
$\Gamma = Y_1^+ \oplus Y_2^+$ of $P2_1/c$ has $\Lambda^2\Gamma = \det\Gamma =
B_g$ and no spin texture at all.~\footnote{Inversion can also act within a
single two-dimensional irrep at a TRIM: for
\ch{LaFeAs_{0.55}P_{0.45}O}~\cite{stadel2022multiple,gallego2016magndata},
$\Gamma = M_2$ of $P4/nmm$ has $\Lambda^2\Gamma = \det\Gamma = A_{1u}$, whose
lowest odd multipole has $\ell = 5$, giving the $h$-wave texture found for
coplanar order in the Fe-based superconductors~\cite{dsouza2026odd}.}

To summarize, a coplanar order with $\{\vec{k},-\vec{k}\}$, $\vec{k}\neq0$, has
a purely odd-parity texture except at a repeated TRIM in a centrosymmetric
parent, where $\det\Gamma$ must also be odd under inversion. We collect this
result, together with the non-coplanar results below, in \cref{tab:odd-parity}.

Further, for a non-TRIM pair we can obtain a stronger result, as the wave
character is fixed by $\vec{k}$ alone. An operation that preserves $\vec{k}$
acts on the pair as a rotation and one that reverses it as a reflection (see
\cref{sec:gdru}), so $\det\Gamma$ records only whether
$\mat{R}\vec{k}\equiv\vec{k}$ or $\mat{R}\vec{k}\equiv-\vec{k}$, irrespective of
the irrep at $\vec{k}$ the order transforms as. For $\vec{k}$ in the interior of
the Brillouin zone, no reciprocal lattice vector is needed in these conditions:
$\mat{R}\vec{k}$ and $\pm\vec{k}$ are then both interior points, which cannot
differ by a nonzero reciprocal lattice vector, so $\mat{R}\vec{k}=\pm\vec{k}$
exactly. Hence $\det\Gamma$ transforms identically to $\vec{k}\cdot\vec{q}$. The
magnet is thus $p$-wave and admits a time-even Edelstein response. Its leading
texture is $(\vec{c}\cdot\vec{q})\,\vhat{N}_1\times\vhat{N}_2$, where $\vec{c}$
lies in the subspace of polar vectors transforming as $\det\Gamma$. This
subspace always contains $\vec{k}$. When $\det\Gamma$ occurs only once in
$\Gamma_1^-$, as for $B_{3u}$ of $D_{2h}$ in \ch{Gd3(Ru_{0.95}Rh_{0.05})4Al12},
$\vec{c}$ is therefore parallel to $\vec{k}$. For a parent with point group
$C_i$, by contrast, $\det\Gamma = A_u$ occurs three times in $\Gamma_1^-$, and
$\vec{c}$ can point in any direction. Higher odd waves therefore require
$\vec{k}$ on the zone boundary: at the $K$ point of a hexagonal crystal with
point group $D_{6h}$, for instance, $\det\Gamma$ is an inversion-odd $B$ irrep
and the leading texture is $f$-wave. This is the case for the $120^\circ$ order
of the triangular antiferromagnet \ch{CsMnBr3}, for which $\det\Gamma =
B_{2u}$~\cite{eibschutz1973magnetic,gallego2016magndata}, in agreement with the
$f$-wave assignment obtained from its spin space group~\cite{bai2026time}.
These rules are consistent with first-principles calculations for
triangular-lattice helimagnets, which find $p$-wave textures with nodal lines
normal to the ordering vector at interior wave vectors and $f$-wave symmetry
for order at the $K$ point~\cite{larsen2026odd}.
\begin{cframed}
\textbf{Coplanar Odd-Parity Magnets:} A coplanar order whose wave-vector content
is a non-TRIM pair $\{\vec{k},-\vec{k}\}$ has a purely odd-parity spin texture,
as does one with two components at the same nonzero TRIM, provided that in a
centrosymmetric parent $\mat{\Gamma}(I)$ has eigenvalues of both signs. For a
non-TRIM $\vec{k}$ in the interior of the Brillouin zone, the texture is
$p$-wave and a time-even Edelstein response is allowed. Higher odd waves, such
as $f$-wave, require $\vec{k}$ on the zone boundary or a repeated TRIM.
\end{cframed}
 
This construction includes the original $p$-wave construction of
\citet{hellenes2023p}, but does not in general require the ordering wave vector
to admit a half-period lattice translation. Such a translation exists when there
is a lattice vector $\vec{t}_{1/2}$ satisfying
\begin{equation}
    e^{i\vec{k}\cdot\vec{t}_{1/2}}=-1.
\end{equation} 
Every nonzero TRIM has this property, as do commensurate non-TRIM wave vectors
whose translation character has even order. Certain incommensurate wave vectors
may also admit such a translation, but its existence is not guaranteed.
Odd-parity magnets defined through the resulting effective time-reversal
symmetry have accordingly been taken to require commensurate
order~\cite{lee2026incommensuration}. The
representation criterion above, by contrast, only requires the pair
$\{\vec{k},-\vec{k}\}$ and therefore applies more
generally~\cite{larsen2026odd}, including to experimentally realized
incommensurate odd-parity magnets such as
\ch{NiI2}~\cite{song2025switching} and
\ch{BiFeO3}~\cite{shen2026room}.~\footnote{For an incommensurate spiral there
is no finite magnetic unit cell and hence no ordinary magnetic Bloch momentum.
The momentum-space statement should then be understood either in a
generalized-Bloch description~\cite{larsen2026odd} or through a sequence of commensurate
approximants.}

\begin{table}[t]
\centering
\footnotesize
\setlength{\tabcolsep}{4pt}
\renewcommand{\arraystretch}{1.3}
\begin{tabular}{@{}l l l l@{}}
\toprule
Order & Content & Without $I$ & With $I$ \\
\midrule
Coplanar & $\{\vec{k},-\vec{k}\}$ & always & always \\
 & $\{\vec{K},\vec{K}\}$ & always & both parities \\
\midrule
Non-coplanar & $\{\vec{k},-\vec{k},\vec{K}\}$ & $\vec{K}\neq\vec{0}$
  & odd at $\vec{K}=\vec{0}$ \\
 & $\{\vec{K}_1,\vec{K}_1,\vec{K}\}$ & $\vec{K}\neq\vec{0}$
  & both parities at $\vec{K}_1$, \\
 & & & odd at $\vec{K}=\vec{0}$ \\
 & $\{\vec{K},\vec{K},\vec{K}\}$ & always & both parities \\
\bottomrule
\end{tabular}
\caption{Conditions on $\Gamma$ for a purely odd-parity spin texture, by its
wave-vector content, for parent point groups without and with inversion $I$.
Here $\vec{k}$ is not a TRIM, $\vec{K}$ and $\vec{K}_1$ are distinct TRIMs, and
a repeated TRIM is nonzero. ``Both parities'' requires $\mat{\Gamma}(I)$ to have
eigenvalues of both signs on the repeated components. ``Odd at
$\vec{K}=\vec{0}$'' requires any component at $\vec{K}=\vec{0}$ to be odd under
$I$. No other wave-vector content admits a purely odd-parity texture.}
\label{tab:odd-parity}
\end{table}

\subsubsection{Non-Coplanar Odd-Parity Magnets}
\label{sec:noncoplanar-odd}
Consider non-coplanar order, for which the defining representation $\Gamma$ is
three-dimensional. Odd-parity spin textures arising from non-coplanar magnetic
order have also been identified directly from spin-group
symmetry~\cite{luo2025odd,luo2026spin}. A purely odd-parity spin texture again requires that
$\Gamma$ have no overlap with any even-rank momentum multipole, while
$\Lambda^2\Gamma$ must overlap with at least one odd-rank multipole.

We first constrain the wave-vector content. For $\Lambda^2\Gamma$ to have a
$\vec{k}=0$ component $\Gamma$ must have wave-vector content of the form
$\{\vec{k},-\vec{k},\vec{K}\}$, where $\vec{K}$ is a TRIM (see
\cref{sec:exterior-prelim}). The case $\vec{k}=0$ cannot produce a purely
odd-parity texture. If inversion is absent, the two $\vec{k}=0$ components
necessarily produce an even-rank overlap. If inversion is present, removing this
overlap requires both to be odd under inversion. Their exterior product is then
even. The remaining parts of $\Lambda^2\Gamma$, which involve the third
component, either lie at finite wave vector, if $\vec{K}\neq0$, or are also
even, if $\vec{K}=0$, since that component must then be odd as well. Either way
no odd-rank texture results.

We thus assume that $\vec{k}\neq0$. For a non-centrosymmetric parent, the
remaining condition is simply $\vec{K}\neq0$, since a component at $\vec{K}=0$
would overlap some even-rank multipole. All components of $\Gamma$ then lie at
finite wave vector, while $\Lambda^2\Gamma$ has a $\vec{k}=0$ component, the
exterior square of the canceling pair. Since every irrep appears in some
odd-rank multipole, this is sufficient for a purely odd-parity texture.

For a centrosymmetric parent, a component of $\Gamma$ at $\vec{K}=0$ is allowed
provided it is odd under inversion. Such a component can also overlap
$\Gamma_1^-$, so that, unlike an order built entirely from finite wave vectors,
the magnet may then show a time-odd Edelstein response. The canceling pair
$\{\vec{k},-\vec{k}\}$ must also produce an inversion-odd component of
$\Lambda^2\Gamma$. For a non-TRIM pair $\vec{k}\not\equiv-\vec{k}$ this is
automatic, since inversion exchanges the two sectors. If instead the canceling
pair consists of two components at the same nonzero TRIM $\vec{K}_1$, those two
components must have opposite inversion parity.

There is one additional special case when all three components occur at the same
nonzero TRIM, so that the wave-vector content is $\{\vec{K},\vec{K},\vec{K}\}$.
Then every pair contributes at $\vec{k}=0$ to $\Lambda^2\Gamma$. An odd-rank
texture is allowed provided inversion does not act with the same parity on all
three components: equivalently, $\mat{\Gamma}(I)$ must contain both $+1$ and
$-1$ eigenvalues.\footnote{In this case the odd texture can have more than one
spin component. In a centrosymmetric parent at most two can be odd, since the
components of $\Lambda^2\Gamma$ have inversion parities $s_2s_3$, $s_1s_3$ and
$s_1s_2$, with $s_i$ those of $\Gamma$. A three-component odd texture, type III
in the terminology of \citet{luo2025odd}, therefore requires a
non-centrosymmetric parent.}

The wave character then follows as in the coplanar case. Outside the case
$\{\vec{K},\vec{K},\vec{K}\}$, the $\vec{k}=0$ part of $\Lambda^2\Gamma$ is the
exterior square of the canceling pair alone, so the odd texture points along
the spin direction of the $\vec{K}$ component and its rank is set by that
one-dimensional representation. For a non-TRIM $\vec{k}$ inside the Brillouin
zone it is therefore again $p$-wave, with a time-even Edelstein response, while
higher odd waves require $\vec{k}$ on the zone boundary or a canceling pair at
a TRIM.

These conditions are collected, with those for coplanar order, in
\cref{tab:odd-parity}.

\subsection{Spin Texture Classifications}
\label{sec:textureclass}
Momentum-space textures of the kind derived in \cref{sec:spintexture} have been
analyzed directly using spin space groups, including the symmetry of
non-relativistic electronic bands in \citet{liu2022spin}, the even-wave spin
splitting of collinear altermagnets~\cite{smejkal2022beyond}, and the odd-parity
texture perpendicular to the ordering plane of coplanar $p$-wave
magnets~\cite{hellenes2023p}. Criteria for odd-parity magnets have been
formulated in the same way, as conditions on the spin-group operations
present~\cite{luo2025odd,luo2026spin,bai2026time}. \citet{song2026} have given a
systematic spin space group classification of these textures. The parity relations of
\cref{eq:texture-collinear-parity,eq:texture-coplanar-x,%
eq:texture-coplanar-y,eq:texture-coplanar-z} appear there, together with a
classification of the collinear and coplanar wave hierarchies. For a texture
that is constrained to a fixed spin axis, they define $\ell$ as the lowest
polynomial degree whose representation contains the one-dimensional
representation associated with that texture. A related classification is
given in Ref.~\cite{luo2026unconventional}.

The distinction from the present viewpoint is therefore not the
representation-theoretic character of this criterion, but where the required
representation is obtained. In the spin space group approach it is determined
from an identified spin space group, while here it is already determined by the
defining representation, as $\Lambda^1\Gamma$ for collinear order and
$\Lambda^1\Gamma$ or $\Lambda^2\Gamma$ for coplanar and non-coplanar order.
Moreover, because no reduction to a fixed spin axis is required, the same
overlap condition applies directly to vector-valued spin textures, with
$\Lambda^1\Gamma$ controlling the even and $\Lambda^2\Gamma$ the odd ranks. Thus
the momentum-space multipole content follows from the parent space group and its
defining representation, with the alternation of even and odd ranks between
$\Lambda^1\Gamma$ and $\Lambda^2\Gamma$ visible directly.

\subsection{Scope and Limitations}
\label{sec:scope}
Three important limitations bound the core results of this work. 

The first is the non-relativistic limit itself: couplings that lock spin to
space are invisible here by construction, which is why the anomalous Hall
effects observed in \ch{MnTe}~\cite{betancourt2023spontaneous} and
\ch{Mn3Ge}~\cite{nayak2016large,kiyohara2016giant} require relativistic
interactions or extrinsic perturbations that break the protecting spin-only
symmetry. Our conditions are therefore statements about what survives as
relativistic effects are switched off at fixed magnetic structure, and are most
useful as a leading-order account to which relativistic corrections are added.
When relativistic effects are included in full, the symmetry is instead that of
a magnetic space group, which the same representation-first approach also
covers, though less cleanly, through $\Gamma=\Phi\otimes\Gamma_1^+$
(\cref{sec:msg}).

The second is the kind of question asked. The exterior-power conditions derived
here are overlaps of a power of $\Gamma$ with a representation of the point
group, and so fix the symmetry-allowed form of a response: which components of a
tensor may be nonzero, or which multipoles a momentum-space texture may contain.
A tensor allowed in this sense may still vanish in particular models, through a
conservation law that the full problem need not share. The uniform time-odd
Edelstein response of a collinear magnet, for example, vanishes when the carrier
spin is conserved, as for Bloch electrons in a fixed exchange field, but not
once the carriers exchange spin with the ordered moments (\cref{sec:edelstein}).
Nor is every equilibrium property an overlap of this kind. 

The anisotropy of the static spin susceptibility, at fixed orientation of the
order in spin space, is governed instead by the invariant symmetric bilinear
forms on $\Gamma$, that is, by the number of times $\Gamma_0^+$ occurs in the
symmetric square ${\rm sym}^2(\Gamma)$, together with the spin-only symmetry.
For collinear and coplanar orders, the spin-only symmetry singles out the N\'eel
axis or the ordering plane, allowing anisotropic susceptibilities. By contrast,
a non-coplanar order whose $\Gamma$ is irreducible, as for \ch{FeF3} and
\CoTaS{}, has an isotropic susceptibility enforced by Schur's lemma. For such
properties the symmetric square plays the role that the exterior powers play for
the responses treated here. 

These conditions also do not fix the symmetry labels of individual states. Where
the electronic or magnon bands are degenerate, how those labels connect across
the Brillouin zone, and which of their topological invariants symmetry can
diagnose are constrained by the little co-groups of the spin space group at each
crystal momentum, including its antiunitary elements and the translations they
carry; answering them means constructing at least those
groups~\cite{chen2024enumeration,song2025irreps,xiao2024,%
zhang2026irssg,watanabe2018structure}, and a character overlap in $G$ does not
suffice. 

The third is that every quantity treated is uniform, and every response linear.
The responses are unit-cell averages, and although the spin texture depends on
$\vec{q}$, its multipole moments [\cref{eq:spin-texture-multipoles}] average it
over the Brillouin zone, so that each, like the responses, transforms under a
representation of $G$ at $\vec{k}=\vec{0}$. Sublattice-resolved or locally
staggered responses, such as a local Edelstein effect, can transform under other
representations of $G$ and lie outside the present analysis. Nonlinear
responses, by contrast, are mostly still uniform and should follow in the same
way from overlaps of powers of $\Gamma$ with the appropriate multipole
representations. We leave them to future work.

\section{Conclusions}
\label{sec:conclusions}
We have argued that the symmetry consequences of a magnetic order in the
non-relativistic limit are best read from the real representation $\Gamma$ of
its parent space group, rather than from the spin space group it generates. This
spares assembling and identifying the group, but its main value is structural.
The responses considered here reduce to overlaps between the three exterior
powers of $\Gamma$ and low-rank multipole representations, with the power fixed
by the presence of a spin index and by time-reversal parity, and the spin-only
symmetries of collinear and coplanar order entering through the vanishing of the
higher powers. The same powers impose a hierarchy of wave-vector selection rules
(one ordering wave vector vanishing, two canceling, or all three summing to
zero) and control the even- and odd-rank momentum-space spin textures, of which
the time-even Edelstein effect is the rank-one case. They also delimit
odd-parity magnets: a purely odd-parity texture requires a canceling pair of
nonzero wave vectors, and in coplanar order an interior non-TRIM pair always
gives a $p$-wave texture.
 
The examples of \cref{sec:mnf2,sec:mnte,sec:mn3ge,sec:fef3,sec:co13tas2,%
sec:cr2o3,sec:mno,sec:gdru} show how little is often needed. For \ch{MnO} the
dimension and wave-vector content of the representation settle every response
without any point-group label at all: a representation at finite $\vec{k}$
cannot overlap anything at $\vec{k}=0$, and collinearity disposes of the rest.
For \ch{Gd3(Ru_{0.95}Rh_{0.05})4Al12} the time-even responses follow from
identifying $\det\Gamma$ with the polar-vector component along the propagation
direction, without needing the full irrep label of $\Gamma$. More generally,
neither a decomposition of $\Gamma$ into irreducible representations nor any
label of its components is ever needed, since each condition of
\cref{tab:summary} is an average of a product of characters. The multipole
representations $\Gamma_\ell^\pm$, being those of $\mathrm{O}(3)$, have the same
characters for every parent space group, fixed by each operation's rotation angle
and inversion parity. This is what makes the analysis straightforward to
automate (\cref{sec:magndata}). Where the
point-group content does matter it separates materials that are otherwise alike:
\ch{MnF2} and \ch{Cr2O3} are both compensated collinear antiferromagnets, but
only the first is piezomagnetic, and the distinction is visible immediately as
$B_{2g} \subset \Gamma_2^+$ against $A_{2u} \not\subset \Gamma_2^+$. The
examples also show where care is needed: when the order selects part of a wave-vector star, the
parent group need not be the paramagnetic space group, and using the latter would
force a star representation of the wrong dimension. 

The same construction is not peculiar to spin space groups. A magnetic space
group is encoded by a real one-dimensional representation $\Phi$ of its parent,
from which the conventional type I, III and IV classification and all of the
response conditions follow in the same way, and for collinear order in the
non-relativistic limit $\Gamma$ is the Landau order-parameter irrep of the
transition when that irrep is one-dimensional, and otherwise follows from it by
restriction to the parent space group of the spin space group. The representation the
construction requires is therefore one that a standard analysis of an ordered
structure already produces. This is the practical content of the distinction we
have drawn between classification and inference. A classification remains
necessary to enumerate symmetry types, compare settings, and assign canonical
labels. Going from a magnetic structure to its symmetry-enforced properties can
instead be framed directly in terms of the representation, where the connections
between different responses, and their origin in the ordered state, are made
explicit.
 
\begin{acknowledgments}
We thank Paul McClarty, Judit Romhányi, and Hana Schiff for their contributions
during the early stages of this work, as well as in related collaborations. Work
at the University of Windsor was funded by the Natural Sciences and Engineering
Research Council of Canada (NSERC) (Funding Reference No. RGPIN-2020-04970).
This work benefited from the Visiting Fellows Program of the School of Physical
Sciences at the University of California, Irvine, funded by a grant from the
University of California Office of the President (UCOP).

The author acknowledges the use of generative artificial intelligence tools
(Anthropic Claude Opus 5.5 and Fable 5.2) for language editing, for checking the
manuscript for consistency, and for assistance in writing the code used to
generate the tables and figures, which the author reviewed. All symmetry
analysis, derivations, and physical arguments were carried out and verified by
the author, who takes full responsibility for the content of this work.
\end{acknowledgments}

\appendix

\section{Exterior Powers of a Direct Sum}
\label{app:exterior-sum}

The exterior powers of a reducible representation are fixed by those of its
summands. Let $\Gamma = \Gamma_a\oplus\Gamma_b$ act on $V = V_a\oplus V_b$,
where $\Gamma_a$ and $\Gamma_b$ are themselves representations of $G$. A wedge
of $n$ vectors of $V$ expands into terms carrying $p$ factors from $V_a$ and
$q=n-p$ from $V_b$, and $\mat{\Gamma}(g)$ preserves this grading, so
\begin{equation}
\label{eq:exterior-sum}
\Lambda^n\Gamma
=
\bigoplus_{p+q=n}
\Lambda^p\Gamma_a\otimes\Lambda^q\Gamma_b
\end{equation}
as representations of $G$~\cite{fulton1991representation}. Iterating this gives
\begin{equation}
\label{eq:exterior-sum-many}
\Lambda^n{\left(\bigoplus_i \Gamma_i\right)}
=
\bigoplus_{\sum_i p_i = n}\
\bigotimes_i \Lambda^{p_i}\Gamma_i.
\end{equation}
Each $\Gamma_i$ must be a representation of $G$ in its own right; splitting $V$
into subspaces that the group mixes does not decompose $\Lambda^n\Gamma$.

Since $\Lambda^p\Gamma_i$ vanishes for $p$ greater than the dimension of
$\Gamma_i$, few terms survive in low dimensions. When every summand is
one-dimensional only $p_i = 0,1$ contribute and
\begin{equation}
\label{eq:exterior-sum-1d}
\Lambda^n{\left(\bigoplus_{i=1}^{d} \Gamma_i\right)}
=
\bigoplus_{i_1<\cdots<i_n}
\Gamma_{i_1}\otimes\cdots\otimes\Gamma_{i_n},
\end{equation}
the sum of the products of all $n$ distinct summands.

The reducible two- and three-dimensional cases follow, with $\dG{d}$ denoting a
$d$-dimensional irreducible representation as in \cref{sec:hall}. For
$\Gamma = \dG{1}\oplus\dG{1}'$,
\begin{subalign}
\label{eq:exterior-sum-11}
\Lambda^2\Gamma &= \dG{1}\otimes\dG{1}', \\
\Lambda^3\Gamma &= 0.
\end{subalign}
For $\Gamma = \dG{1}\oplus\dG{2}$,
\begin{subalign}
\label{eq:exterior-sum-12}
\Lambda^2\Gamma &= {\left(\dG{1}\otimes\dG{2}\right)}\oplus \det(\dG{2}), \\
\Lambda^3\Gamma &= \dG{1}\otimes\det(\dG{2}).
\end{subalign}
For $\Gamma = \dG{1}\oplus\dG{1}'\oplus\dG{1}''$,
\begin{subalign}
\label{eq:exterior-sum-111}
\Lambda^2\Gamma &=
{\left(\dG{1}\otimes\dG{1}'\right)}
\oplus
{\left(\dG{1}\otimes\dG{1}''\right)}
\oplus
{\left(\dG{1}'\otimes\dG{1}''\right)}, \\
\Lambda^3\Gamma &= \dG{1}\otimes\dG{1}'\otimes\dG{1}''.
\end{subalign}
In each case the top power reproduces $\Lambda^n\Gamma = \det\Gamma$, as it must
for any $n$-dimensional representation, reducible or not.

\section{Parity Content of Momentum Multipoles}
\label{app:parity-multipoles}

Here we justify a fact used in \cref{sec:oddparity}. The decomposition of
spherical harmonics into irreducible representations of point groups is
standard~\cite{altmann1957}; we reproduce the argument here in the form we
require. Let $P$ be a crystallographic point group and choose a generic
direction $\vhat{q}$ whose stabilizer in $P$ is trivial. Its orbit $P\vhat{q}$
therefore has one point for each element of $P$, so arbitrary functions on this
orbit transform as the regular representation and contain every irrep of $P$.

Suppose first that inversion is not contained in $P$. For a generic choice of
$\vhat{q}$, the orbit $P\vhat{q}$ and its antipodal copy $-P\vhat{q}$ are
disjoint. An even function on their union is specified by arbitrary values on
$P\vhat{q}$, with the values on the antipodal orbit fixed to be equal. An odd
function is specified in exactly the same way, with the antipodal values fixed
to have the opposite sign. Thus both the even and odd functions contain the
regular representation of $P$, and hence every irrep of $P$. Since every
function on a finite set of directions is the restriction of a finite sum of
spherical harmonics, with even and odd functions requiring only even and odd
ranks respectively, every irrep occurs in some even-rank multipole and in some
odd-rank multipole.

If inversion $I$ is contained in $P$, the antipodal point already belongs to the
same orbit. Inversion commutes with every element of the point group, so each
irrep has a definite inversion parity. Splitting the functions on $P\vhat{q}$
into those that are even or odd under $\vhat{q}\rightarrow-\vhat{q}$ therefore
splits the regular representation into its inversion-even and inversion-odd
parts. Consequently, every inversion-even irrep occurs in some even-rank
multipole, while every inversion-odd irrep occurs in some odd-rank multipole.

\section{Extracting the Defining Representation}
\label{app:constructing-gamma}
For a commensurate magnetic structure, choose a supercell on which the
operations of $G_0$ act by permuting its $N$ magnetic sites. Express the moments
in an orthonormal basis of the spin subspace they span, and collect them as the
columns of a matrix $\mat{S}\in\mathbb{R}^{n\times N}$ of rank $n$, where
$n=1,2,3$ for collinear, coplanar and non-coplanar order.

For an operation $g$, let $\mat{P}_g$ be the permutation matrix such that column
$i$ of $\mat{S}\mat{P}_g$ is the moment at site $g(i)$. A compensating
orthogonal spin transformation exists if and only if $g$ preserves every
pairwise scalar product of the moments, that is, the Gram matrix
$\trp{\mat{S}}\mat{S}$,
\begin{equation}
\label{eq:gram-condition}
\trp{\mat{P}_g}\trp{\mat{S}}\mat{S}\mat{P}_g=\trp{\mat{S}}\mat{S}.
\end{equation}
Since the columns of $\mat{S}$ span the occupied spin subspace, this transformation
is then unique,
\begin{equation}
\label{eq:gamma-from-structure}
\mat{\Gamma}(g)
=
\mat{S}\mat{P}_g\trp{\mat{S}}{\left(\mat{S}\trp{\mat{S}}\right)}^{-1},
\end{equation}
and it is orthogonal and satisfies $\mat{S}\mat{P}_g=\mat{\Gamma}(g)\mat{S}$.
The operations satisfying \cref{eq:gram-condition} form the parent space group
$G$, and their matrices $\mat{\Gamma}(g)$ give its defining representation. For
$n<3$ the spin-only symmetries are added as described in
\cref{sec:definitions}. For a structure known only to finite precision,
$\mat{\Gamma}(g)$ is taken instead as the orthogonal matrix that best carries the
moments onto their images, which coincides with \cref{eq:gamma-from-structure}
when \cref{eq:gram-condition} holds exactly; see the Supplemental
Material~\cite{supplemental}.

The response analysis then proceeds by restricting $\Gamma$ and its exterior
powers to the translations, identifying their components at $\vec{k}=0$, and
testing the overlaps listed in \cref{tab:summary}.

\bibliography{draft,sm/sm}

\end{document}